\documentclass{aa}  
\usepackage{booktabs}
\usepackage{ulem}
\usepackage{graphicx}
\usepackage{txfonts}
\usepackage{xcolor}
\usepackage{enumitem}
\usepackage{adjustbox}
\usepackage{tabularx}

\begin{document}

    \title{JWST reveals extended stellar disks for ALMA-bright dusty star-forming galaxies in the Spiderweb protocluster}
    \titlerunning{Extended stellar disks in the Spiderweb protocluster}
    
   \author{Y. H. Zhang\inst{1,2,3,4}
	\and H. Dannerbauer\inst{3,4}
	\and J. M. Pérez-Martínez\inst{3,4}
	\and Y. Koyama\inst{5,6}
	\and X. Z. Zheng\inst{7}
	\and R. Calvi\inst{8}
	\and Z. Chen\inst{5}
	\and K. Daikuhara\inst{9}
	\and C. De Breuck\inst{10}
	\and C. D'Eugenio\inst{3,4,11,12}
	\and B. H. C. Emonts\inst{13}
	\and S. Jin\inst{14,15}
	\and T. Kodama\inst{16}
	\and M. D. Lehnert\inst{17}
   \and J. Nadolny\inst{18,3,4}
	\and A. Naufal\inst{6, 5}
	\and P. G. Pérez-González\inst{19}
          }

   \institute{Purple Mountain Observatory, Chinese Academy of Sciences, 10 Yuanhua Road, Nanjing, 210023, China \\	\email{yuheng@nju.edu.cn}
	\and School of Astronomy and Space Science, University of Science and Technology of China, Hefei, Anhui 230026, China
	\and Instituto de Astrofísica de Canarias (IAC), E-38205 La Laguna, Tenerife, Spain
	\and Universidad de La Laguna, Dpto. Astrofísica, E-38206 La Laguna, Tenerife, Spain
	\and National Astronomical Observatory of Japan, 2-21-1 Osawa, Mitaka, Tokyo 181-8588, Japan
	\and Graduate University for Advanced Studies (SOKENDAI), 2-21-1 Osawa, Mitaka, Tokyo 181-8588, Japan
	\and Tsung-Dao Lee Institute and State Key Laboratory of Dark Matter Physics, Shanghai Jiao Tong University, Shanghai, 201210, China
	\and INAF–Osservatorio Astronomico di Capodimonte, Salita Moiariello 16, 80131 Napoli, Italy
	\and Institute of Space and Astronautical Science, Japan Aerospace Exploration Agency, 3-1-1, Yoshinodai, Chuou-ku, Sagamihara, Kanagawa 252-5210, Japan
	\and European Southern Observatory, Karl–Schwarzschild–Straße 2, D-85748 Garching bei München, Germany
	\and Institute de Physique du Globe de Paris, 5, Rue Jussieu, Paris, France
	\and CEA-Saclay, IRFU, DAp, AIM, 91191, Gif-sur-Yvette, France
	\and National Radio Astronomy Observatory, 520 Edgemont Road, Charlottesville, VA 22903, USA
	\and Cosmic Dawn Center (DAWN), Denmark
	\and DTU Space, Technical University of Denmark, Elektrovej 327, DK-2800 Kgs. Lyngby, Denmark
	\and Astronomical Institute, Tohoku University, 6-3 Aramaki, Aoba-ku, Sendai 980-8578, Japan
	\and Université Lyon 1, ENS de Lyon, CNRS UMR5574, Centre de Recherche Astrophysique de Lyon, F-69230 Saint-Genis-Laval, France
	\and Astronomical Observatory Institute, Faculty of Physics and Astronomy, Adam Mickiewicz University, ul.~S{\l}oneczna 36, 60-286 Pozna{\'n}, Poland 
	\and Centro de Astrobiología (CAB), CSIC-INTA, Ctra. de Ajalvir km 4, Torrejón de Ardoz, E-28850, Madrid, Spain
}

   \date{Received September XX, 2025; accepted March XX, 2025}

 
  \abstract
   {We present JWST/NIRCam imaging of dusty star-forming galaxies (DSFGs) detected by Atacama Large Millimeter/submillimeter Array (ALMA) in the Spiderweb protocluster at $z=2.16$. We identify 22 DSFGs detected by both ALMA and JWST, 10 of which are spectroscopically confirmed as protocluster members. 
   This is the first systematic analysis of a statistical DSFG sample in $z\sim2$ protocluster environments using JWST/NIRCam data.  Most of the DSFG members exhibit very red colors and reside in the dusty star-forming region of the rest-frame $UVJ$ diagram, indicating strong dust obscuration. The Gini-$M_{20}$ diagram suggests that most DSFGs in this protocluster are late-type disks, with a significant fraction displaying clumpy and disturbed rest-frame UV/optical morphologies, but few showing clear merger signatures. 
   The DSFG members exhibit relatively large stellar disks and effective radii with a median stellar mass of log$(M/M_{\odot}) = 10.8 \pm 0.3$, placing them above coeval field DSFGs and typical protocluster galaxies in the size–mass relation at both rest-frame optical and near-infrared wavelengths. These sizes are comparable to those of more evolved field DSFGs at $z\sim1$–$2$, indicating accelerated structural growth in dense environments.
   Moreover, these DSFG members show a decreasing trend in stellar size from shorter to longer wavelengths, with a moderately steep slope comparable to coeval field DSFGs.
   These results may support an inside-out growth scenario for protocluster evolution, in which massive galaxies near the center are more evolved and more strongly affected by AGN feedback and environmental effects, e.g., ram-pressure stripping. We propose that the cold gas accretion at the protocluster outskirts drives intense star formation and stellar disk growth in ALMA-detected DSFGs, which are expected to evolve into massive elliptical galaxies at later stages.
 
}

   \maketitle
%

\nolinenumbers

\section{Introduction}

More than half of the cosmic star formation rate (SFR) has been heavily obscured by dust since redshift 4. This obscured fraction increases to approximately 80\%  \citep{Bouwens2020,Dudzeviciute2020,Zavala2021} at the peak epoch of cosmic star formation \citep{Madau2014} and active galactic nucleus (AGN) activity \citep{Hopkins2006,Zheng2009}, commonly known as ``cosmic noon''.
Dusty star-forming galaxies (DSFGs) are a key population contributing to the majority of cosmic SFR density during cosmic noon \citep{Casey2014,Hodge2020}, which are expected to evolve into massive elliptical galaxies in the local Universe \citep{Lutz2001,Smail2002,Swinbank2006,Toft2014,Gullberg2019}.
DSFGs are preferentially reside in the most massive dark matter halos \citep{Blain2004,Hickox2012,Wilkinson2017}, making them an ideal tracer of protoclusters—the largest large-scale structures at high redshift and progenitors of local mature galaxy clusters \citep{Overzier2016}.
This association is further confirmed by observations of DSFG overdensities within protoclusters across various redshifts \citep{Greve2007,Tamura2009,Dannerbauer2014,Casey2016,AB2018,Wang2021,Zhang2022,Calvi2023,ZhouL2024,Zhang2024,ZhouD2024}. Therefore, investigating DSFGs within protocluster environments is essential for understanding the formation and evolution of massive galaxies and their connections to large-scale structures.

Over the past decades, DSFGs have been extensively studied and are characterized by high stellar masses \citep{Wardlow2011,Simpson2014}, intense star formation rates \citep{Blain2002,Chapman2005,Swinbank2014}, and elevated AGN activity \citep{Chapman2005,Wang2013}. 
High-resolution observations with Atacama Large Millimeter/submillimeter Array \citep[ALMA;][]{Wootten2009} at sub-arcsecond scales have revealed that some DSFGs exhibit disk-like structures in their dust components \citep{Hodge2016,Gullberg2019,Chen2020,Cochrane2021}, with potential arm and bar features \citep{Hodge2019}. 
The discovery of rotating disks confirmed by kinematic analyses, further supports the presence of these structures in a substantial portion of the DSFG population \citep{Hodge2012,Chen2017,Lelli2021,Rizzo2021,Birkin2024,Venkateshwaran2024,Amvrosiadis2025}.
Previous studies have shown that the stellar components of most DSFGs exhibit disturbed morphologies with clumpy features \citep{Swinbank2010,Chen2015,Gomez2018,Cowie2018,Zavala2018,Lang2019,Ling2022} and have stellar size measurements larger than that of their dust components observed at far-infrared wavelengths \citep{Hodge2016,Gullberg2019,Chen2022,Cochrane2021}. 
Major mergers have been proposed as a main driver of the starburst activity in DSFGs \citep{Chen2015,Cowie2018}; however, this conclusion is primarily based on HST observations that probe the rest-frame UV and optical bands of these galaxies, which are heavily affected by dust extinction \citep{Hainline2011,da_Cunha2015,Simpson2017,Lang2019,Dudzeviciute2020,Popping2022}. Some studies, however, suggest that mergers do not play a more important role in driving star formation in DSFGs compared to general galaxy populations \citep{Swinbank2010,Targett2013,Ren2025}, as indicated by cosmological hydrodynamical simulations \citep{McAlpine2019}.

Recently, the James Webb Space Telescope (JWST) has revealed  disk-like stellar components with disturbed and clumpy structures in DSFGs at $z\sim2-4$ \citep{Cheng2022,Cheng2023,Gillman2023,Fujimoto2025,Wu2023,Rujopakarn2023, Boogaard2024,Lebail2024,Liu2024}, along with a notable prevalence of bar and bulge features \citep{Chen2022,Smail2023,Gillman2024,McKinney2025,EspejoSalcedo2025,Hodge2025} which are invisible in previous HST observations. The broad wavelength coverage ($1-5\,\mu{\rm m}$) of JWST has further confirmed a decreasing trend in stellar sizes from shorter to longer wavelengths \citep{Cheng2022,Gillman2023,Price2025,Boogaard2024}, which are apparently larger than the compact dust structures \citep{Chen2022,Tadaki2023,Hodge2025}. The similar merger fraction of DSFGs compared to that of general field galaxies also suggests that major mergers do not play an primary role in triggering their intense star formation \citep{Gillman2023,McKay2025,Ren2025}.  Moreover, larger stellar disks in individual galaxies have been reported in several protoclusters compared to coeval field galaxies \citep{Colina2023,Crespo2024,Wang2025,Umehata2025a}, suggesting accelerated size growth in overdense environments as seen in simulations \citep{Esposito2025}. However, due to the limited sample size, the origins of DSFGs with stellar disks and intense star formation, particularly in overdense environments, remain to be explored. 

The Spiderweb protocluster is one of the most massive galaxy protoclusters at $z\sim2-3$ and has been extensively studied for over twenty years, following the initial discovery of its central radio galaxy \citep{Roettgering1994,Pentericci1997,Miley2006} and surrounding overdense environments \citep{Kurk2000,Pentericci2000}. 
Several hundred hours of observations, spanning from X-ray to radio wavelengths, have spectroscopically confirmed over one hundred member galaxies and established the overdensity of various galaxy populations, including X-ray sources, Ly$\alpha$ emitters (LAEs), H$\alpha$ emitters (HAEs), Pa$\beta$ emitters, CO(1-0) emitters, submillimeter galaxies (SMGs), DSFGs, and quiescent galaxies, \citep{Pentericci2000,Pentericci2002,Kurk2000,Kurk2004,Kurk2004b,Croft2005,Kuiper2011,Koyama2013,Tanaka2013,Rigby2014,Dannerbauer2014,Shimakawa2014,Shimakawa2018,Shimakawa2024b, Zeballos2018,Jin2021,Tozzi2022a, Tozzi2022b,Daikuhara2024,Zhang2024,Naufal2024}. 
Considerable efforts have been dedicated to examining the environmental effects of the surrounding overdensity on member galaxies from multiple perspectives, such as UV morphologies \citep{Naufal2023}, star formation and gas-phase metallicity \citep{Perez2023}, dust extinction \citep{Perez2024b}, AGN activity \citep{Tozzi2022a, Shimakawa2024a, Shimakawa2025}, and the cold molecular gas content \citep{Dannerbauer2017,Emonts2016,Emonts2018,Chen2024,Perez2025}. The recent discovery of a nascent intracluster medium (ICM) \citep{Mascolo2023} and a diffuse hot halo \citep{Tozzi2022b,Lepore2024} confirms that the Spiderweb protocluster will ultimately evolve into a mature galaxy cluster in the local Universe, making this structure an ideal laboratory for investigating the nature of DSFG members and their relationship with the surrounding overdense environment.

In this paper, we utilize newly acquired JWST data \citep{Shimakawa2024b,Perez2024b} to identify ALMA-detected DSFGs and construct the first statistical sample of DSFGs in $z\sim2$ protocluster environments. In Section \ref{sec: sample}, we describe the observations and results of counterpart identification of these DSFGs. Section \ref{sec: results} presents the results of both non-parametric and parametric morphological analyses of our DSFG sample, followed by a discussion of their origins and evolutionary trajectories in Section \ref{sec: discussion}. We summarize our findings in Section \ref{sec: summary}. 
For simplicity, we refer all sources detected by submillimeter/millimeter facilities as DSFGs, including the traditional SMGs detected by single-dish telescopes.
Throughout this paper, we assume a flat $\Lambda$CDM cosmology with $H_0=71$\,kms\,$^{-1}$\,Mpc$^{-1}$, $\Omega _{\rm \Lambda}$=0.7 \citep{Spergel2003,Spergel2007}, which corresponds to a scale of 0.490\,physical\,Mpc (pMpc) or 1.549\,comoving\,Mpc (cMpc)\,per\,arcmin at $z=2.16$.

\begin{figure*}
	\centering
	\includegraphics[width=\textwidth]{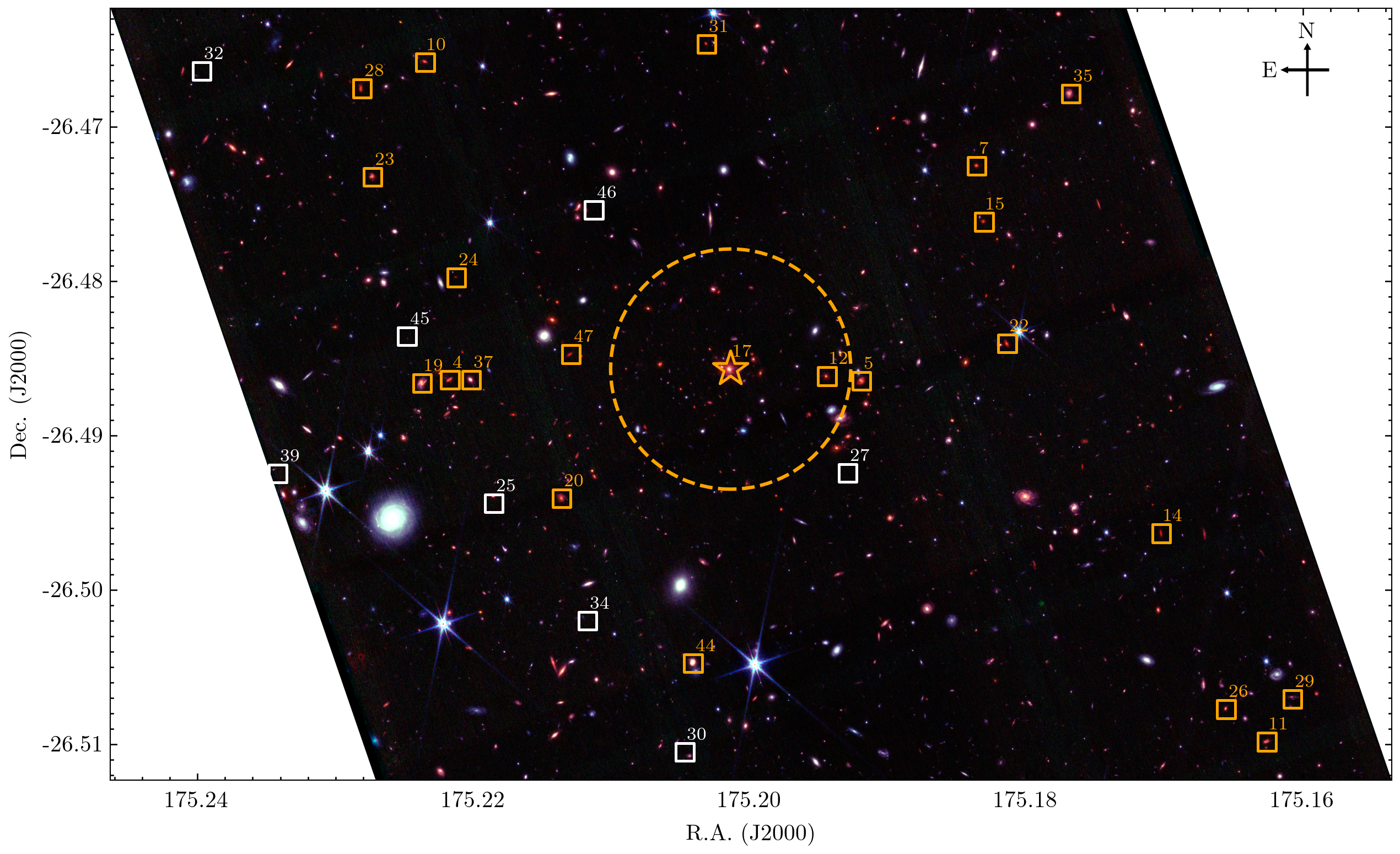}
	\caption{JWST color-composite (F410M/F182M/F115W for RGB) image of the Spiderweb protocluster. The 22 ALMA sources detected by JWST and the 8 undetected ones are marked with orange and white squares, respectively, with their IDs labeled nearby. The Spiderweb galaxy is marked as a star, which represents the center of the protocluster. The dashed circle shows the $r_{500}=229$\,kpc derived from the recently detected Sunyaev–Zeldovich signal \citep{Mascolo2023}. }
	\label{fig:fov_map}
\end{figure*}

\section{Datasets}
\label{sec: sample}

\subsection{ALMA observations}
\label{subsec: alma data}

The ASW$^2$DF is a 1.2\,mm ALMA survey covering $\sim$20\,arcmin$^2$ across six contiguous fields centered on the Spiderweb galaxy. The observations were conducted between December 2021 and April 2022 in Cycle 8 (Project ID: 2021.1.00435.S; PI: Y. Koyama), with a total on-source time of 12.84 hrs. After the data reduction with Common Astronomy Software Applications \citep[CASA version 6.2.1.7;][]{CASA2022}, the final images reach a sensitivity range of $40.3-57.1\,\mu$Jy at a spatial resolution of $0\farcs5-0\farcs9$.  Source catalogs were constructed using three independent tools to ensure detection robustness, identifying 47 reliable ALMA sources (\textgreater4$\sigma$) with an estimated 90\% purity in the main catalog. We refer the reader to \citet{Zhang2024} for more details about the observations and source extractions.

\begin{table*}
	\centering
	\caption{Physical properties of the ten DSFG members in the Spiderweb protocluster.} 
	\label{tab: dsfgs_phys}
		\begin{tabular}{ccccccccccc}
			\toprule
			ID & R.A. & Dec. & $z$ & log(M$_{*}$) & SFR$_{\rm 1.2\,mm}$ & LABOCA & HAE & CO emitter & Pa$\beta$ emitter & X-ray  \\
			& & & & (M$_\odot$) & (M$_\odot$yr$^{-1}$) & ID  & ID & ID & ID & ID \\
			\midrule
			04 & 11:40:53.21 & -26:29:11.1 & 2.188 & $10.94\pm0.08$ & $447\pm66$ & 02 & -- & -- & -- & -- \\
			05 & 11:40:46.06 & -26:29:11.3 & 2.149 & $11.37\pm0.12$ & $481\pm74$ & -- & 902 & 01$_{\rm ext}$ & -- & -- \\
			12 & 11:40:46.66 & -26:29:10.2 & 2.166 & $10.90\pm0.11$ & $214\pm38$ & -- & 880 & 15$_{\rm ext}$ & -- & 87 \\
			14 & 11:40:40.86 & -26:29:46.8 & 2.163 & $10.29\pm0.28$ & $307\pm50$ & -- & -- & 23$_{\rm ext}$ & -- & -- \\
			17 & 11:40:48.34 & -26:29:08.5 & 2.162 & $12.42\pm0.07$ & $378\pm65$ & 07 & 73 & 02$_{\rm ext}$ & 30 & 58 \\
			20 & 11:40:51.27 & -26:29:38.6 & 2.151 & $10.90\pm0.10$ & $285\pm59$ & -- & 1300 & 09 & -- & -- \\
			23 & 11:40:54.56 & -26:28:23.8 & 2.161 & $10.73\pm0.10$ & $179\pm42$ & -- & 1162 & 45 & -- & -- \\
			28 & 11:40:54.74 & -26:28:03.1 & 2.152 & $10.83\pm0.09$ & $151\pm36$ & 15 & 1181 & -- & -- & -- \\
			29 & 11:40:38.58 & -26:30:25.5 & 2.121 & $10.05\pm0.15$ & $113\pm36$ & -- & -- & 34 & -- & -- \\
			47 & 11:40:51.11 & -26:29:05.1& 2.146 & $10.46\pm0.26$ & $182\pm44$ & -- & -- & 38$_{\rm ext}$ & -- & -- \\
			\bottomrule
		\end{tabular}
	\tablefoot{Stellar masses are adopted from \citet{Perez2023,Shimakawa2024a} or derived from NIR luminosities measured by JWST/NIRCam (see Sec. \ref{subsubsec: mass_size}). SMG, HAE, and CO emitter IDs follow \citet{Dannerbauer2014}, \citet{Perez2023,Shimakawa2024a}, and \citet{Jin2021}, respectively, with extended gas reservoirs marked in boldface \citep{Chen2024}. Pa$\beta$ and X-ray counterparts are cross-matched from \citet{Shimakawa2024b} and \citet{Tozzi2022a}.	 }
\end{table*}

\subsection{JWST/NIRCam imaging}
\label{subsec: jwst data}
The JWST/NIRCam \citep{Rieke2005} observations of the Spiderweb protocluster covered an area of $3\times6$\,arcmin$^2$ using four filters to trace rest-frame $U$, $V$, $J$ band continuum (F115W, F182M, F410M), and Pa$\beta$ emission (F405N) at $z=2.16$ \citep[ID: 1572,][]{Dannerbauer2021}. With integration times ranging from 21 to 63 minutes, the reduced images achieve median 5$\sigma$ depths of 25.4, 24.3, 23.3 and 24.6 mag in a 1\farcs5 diameter aperture for the F115W, F182M, F405N and F410M filters, respectively \citep{Perez2024b}. Sources were extracted using {\sc SExtractor} in dual-image mode, with a median-stacked detection image. We refer the reader to \citet{Shimakawa2024b} and \citet{Perez2024b} for more details about the data and source selection.  Astrometric calibration using 18 Gaia EDR3 stars resulted in an alignment within 0\farcs01 of HST/ACS images.

\begin{figure}
	\centering
	\includegraphics[width=\linewidth]{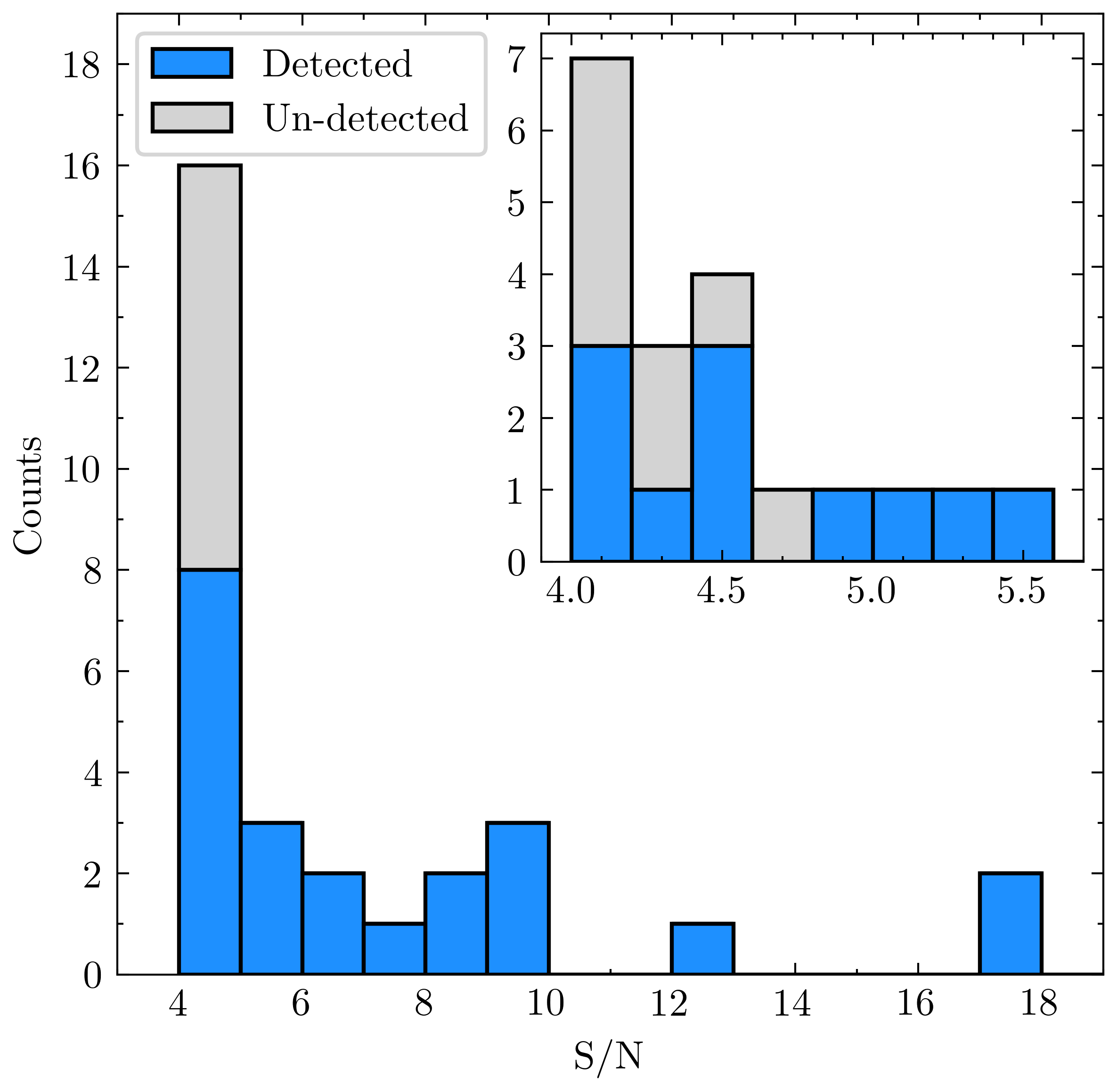}
	\caption{The signal-to-noise ratio distribution of the 30 ALMA sources within the JWST/NIRCam footprint. The 22 sources detected by JWST are indicated in blue, while the eight undetected sources are shown in gray.}
	\label{fig:03_snr_hist}
\end{figure}

\begin{figure*}
	\centering
	\includegraphics[width=0.49\textwidth]{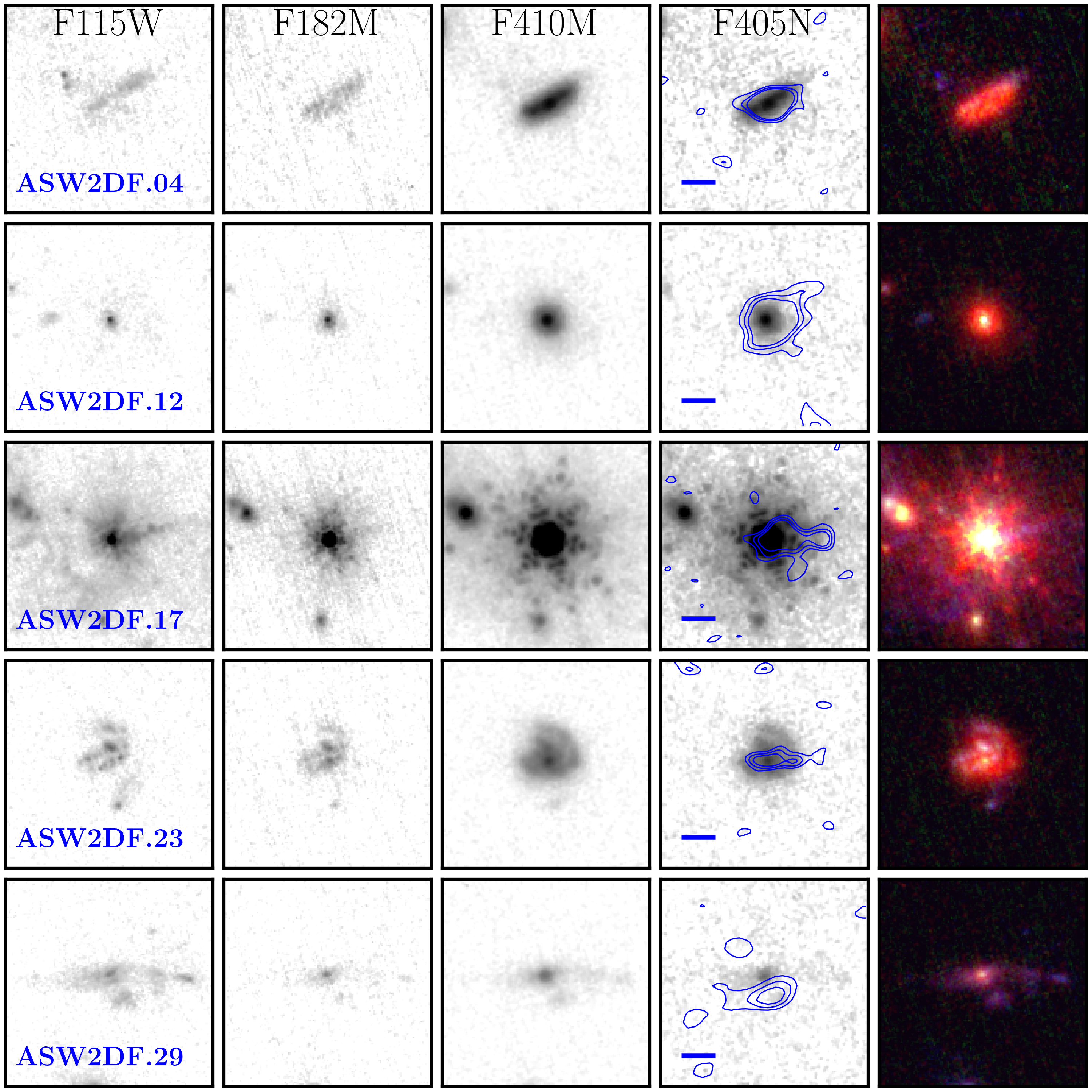}
	\includegraphics[width=0.49\textwidth]{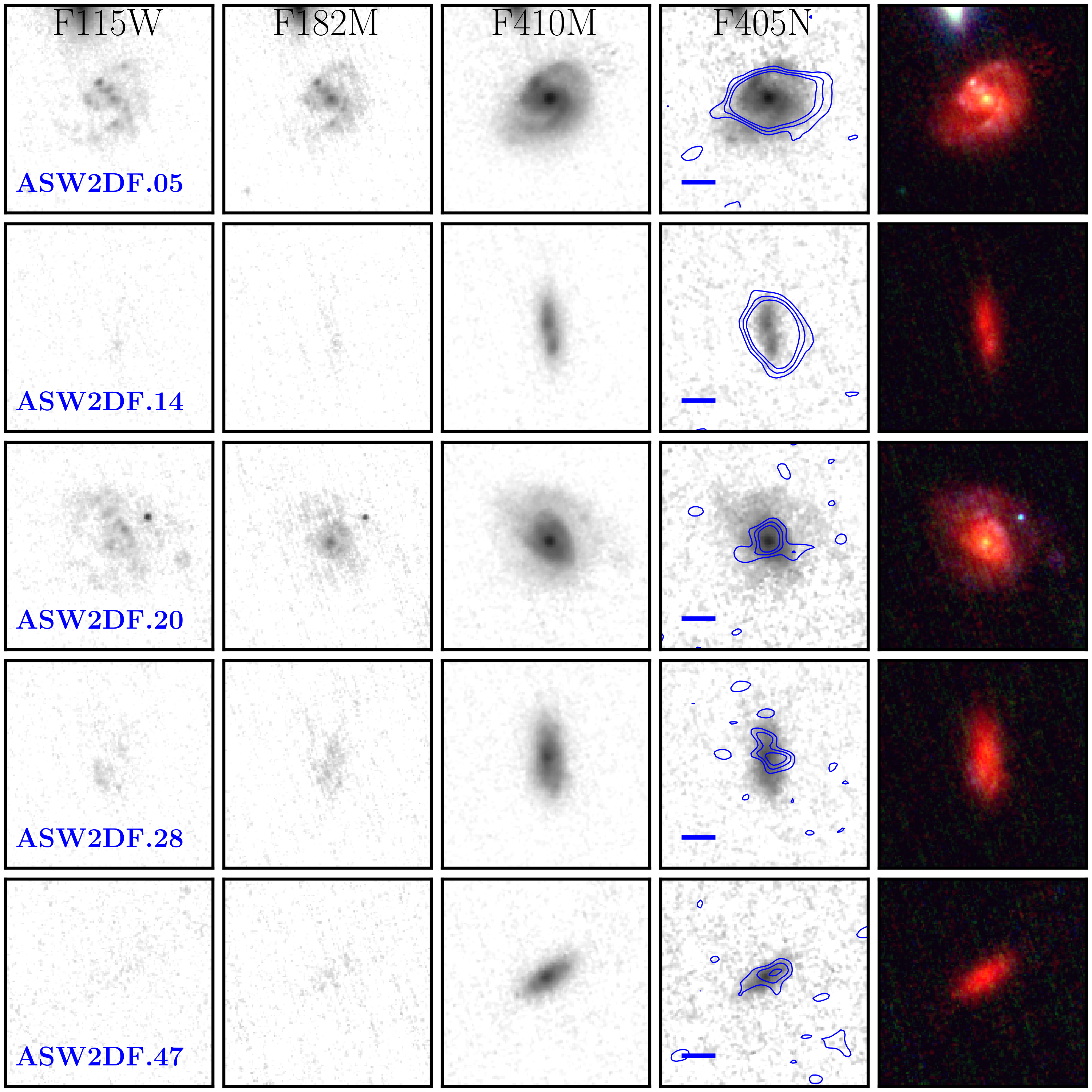}
	\caption{The gallery of ten protocluster members detected by both ALMA and JWST/NIRCam, with a cutout size of 4$\times$4\,arcsec. The names of the four filters are shown in the first row.  The ALMA 1.2\,mm contours at [2$\sigma$, 3$\sigma$, 4$\sigma$] are overlaid on the F405N image in the fourth column for each source, with a blue bar showing a physical distance of 5\,kpc. The RGB-composite (F410M/F182M/F115W) color images are shown in the fifth column. The source IDs are positioned in the lower-left corner of each group of panels, consistent with the numbering in Fig.~\ref{fig:fov_map}.} 
	\label{fig: cutouts}
\end{figure*}

\section{Results}
\label{sec: results}

\subsection{JWST+ALMA counterparts}
\label{subsec: cpt}

The overlaping region between the ALMA and JWST footprints encompass $3.9 \times 3.3$\,arcmin$^2$, covering approximately 60\% of the JWST/NIRCam field of view \citep{Shimakawa2024b}. By cross-matching the ALMA and JWST/NIRCam catalogs with a 0\farcs5 radius, we identified 30 ALMA sources located within the JWST coverage. Among these, 22 are detected in more than one NIRCam filters, corresponding to a detection rate of 73\%. All of the 30 ALMA sources are overlaid on the composite color image in Figure~\ref{fig:fov_map}.

Figure~\ref{fig:03_snr_hist} shows the signal-to-noise ratio (S/N) distribution of all 30 ALMA sources. Notably, all sources with S/N~$\textgreater5$ are detected by JWST, supporting the reliability of our ALMA source extraction. We note that eight ALMA sources are not detected by JWST in any of the F115W, F182M, F405N, or F410M filters above $1.5\sigma$ at the limiting magnitudes of $23.3$–$25.4$, and their cutouts are presented in Appendix~\ref{sec:appendix_undetections}. 
Most of the JWST–undetected sources are located on the eastern side of the ALMA field (Figure~\ref{fig:fov_map}), where the imaging has higher angular resolution and therefore a larger number of synthesized beams \citep{Zhang2024}, increasing the likelihood of spurious detections. Based on the estimated image purity \citep{Zhang2024}, approximately five of the 30 ALMA sources within the JWST coverage are expected to be spurious, which is somewhat fewer than the eight JWST–undetected sources identified above.
The nature of these additional JWST–undetected sources remains uncertain. One plausible explanation is that they represent extremely dust–obscured systems, such as NIRCam–dark galaxies \citep{Perez-Gonzalez2024, SunF2025}.

We cross-match the 22 JWST-detected ALMA sources with known protocluster members to assess their membership status. Six sources (ASW2DF.05, 12, 17, 20, 23, and 28) are confirmed as H$\alpha$ emitters (HAEs) \citep{Shimakawa2014, Dannerbauer2017, Perez2023, Shimakawa2024a}, while eight (ASW2DF.05, 12, 14, 17, 20, 23, 29, and 47) have been identified as CO emitters \citep{Jin2021}, five of which host extended molecular gas reservoirs \citep{Dannerbauer2017,Chen2024}. 

Additional redshift constraints are obtained from our HST/WFC3 G141 grism observations \citep[ID: 17117;][]{Koyama2022}, which confirm ASW2DF.04 as a protocluster member \citep{Naufal2024}. In summary, 10 of the 22 ALMA sources detected by JWST are confirmed members of the Spiderweb protocluster, while the remaining sources lack spectroscopic redshift information are presented in Appendix~\ref{sec:appendix_nonmembers}.
Table~\ref{tab: dsfgs_phys} summarizes the positions, redshifts, stellar masses, ALMA-derived SFRs, and associations with various galaxy populations of the 10 confirmed DSFG members. We emphasize that this represents the first systematic study of a statistical DSFG sample in $z \sim 2$ protocluster environments using JWST/NIRCam data.

\subsection{Colors}
\label{subsec: color}

Figure \ref{fig: cutouts} shows cutout images of the 10 DSFG members in four JWST/NIRCam filters and corresponding color composites. Most sources are faint  in the short-wavelength (SW) filters (F115W and F182M) but appear significantly brighter in the long-wavelength (LW) filters (F405N and F410M). The color composites further reveal that the majority of sources are distinctly red, indicative of strong dust obscuration.

Moreover, the galaxy morphologies vary significantly across different filters. Given the redshift of the Spiderweb protocluster ($z = 2.16$), the F410M filter probes the rest-frame near-infrared (NIR) and effectively traces the stellar components of the associated DSFGs. In F410M, most sources display smooth disk morphologies, with several exhibiting clear spiral features, suggesting that mergers have not recently disturbed their stellar morphology.
In contrast, in the bluer filters, which trace ongoing star formation and younger stellar populations, many galaxies show irregular shapes and clumpy structures within their disks; a few even exhibit multiple components, indicative of clumpy star formation, inhomogenous dust distribution, or potential merger activity. In such cases, the ALMA-detected dust emission might be spatially offset from the stellar continuum traced by JWST (e.g., a 0.5$^{\prime\prime}$ offset for ASW2DF.29). Overall, the sample reveals a population of very red, disk-like galaxies undergoing intense, dust-obscured star formation, consistent with recent findings that DSFGs are predominantly disk galaxies hosting vigorous star formation \citep{Cheng2023,Liu2023,Magnelli2023,Wu2023,Rujopakarn2023,Gillman2024,Fujimoto2025}. As shown in Figure~\ref{fig: cutouts}, the central Spiderweb galaxy (ASW2DF.17) is significantly affected by PSF contamination across all JWST/NIRCam filters, prominently due to its intense AGN activity. Therefore, we exclude it from the subsequent morphological analysis.

\begin{figure}
	\centering
	\includegraphics[width=\linewidth]{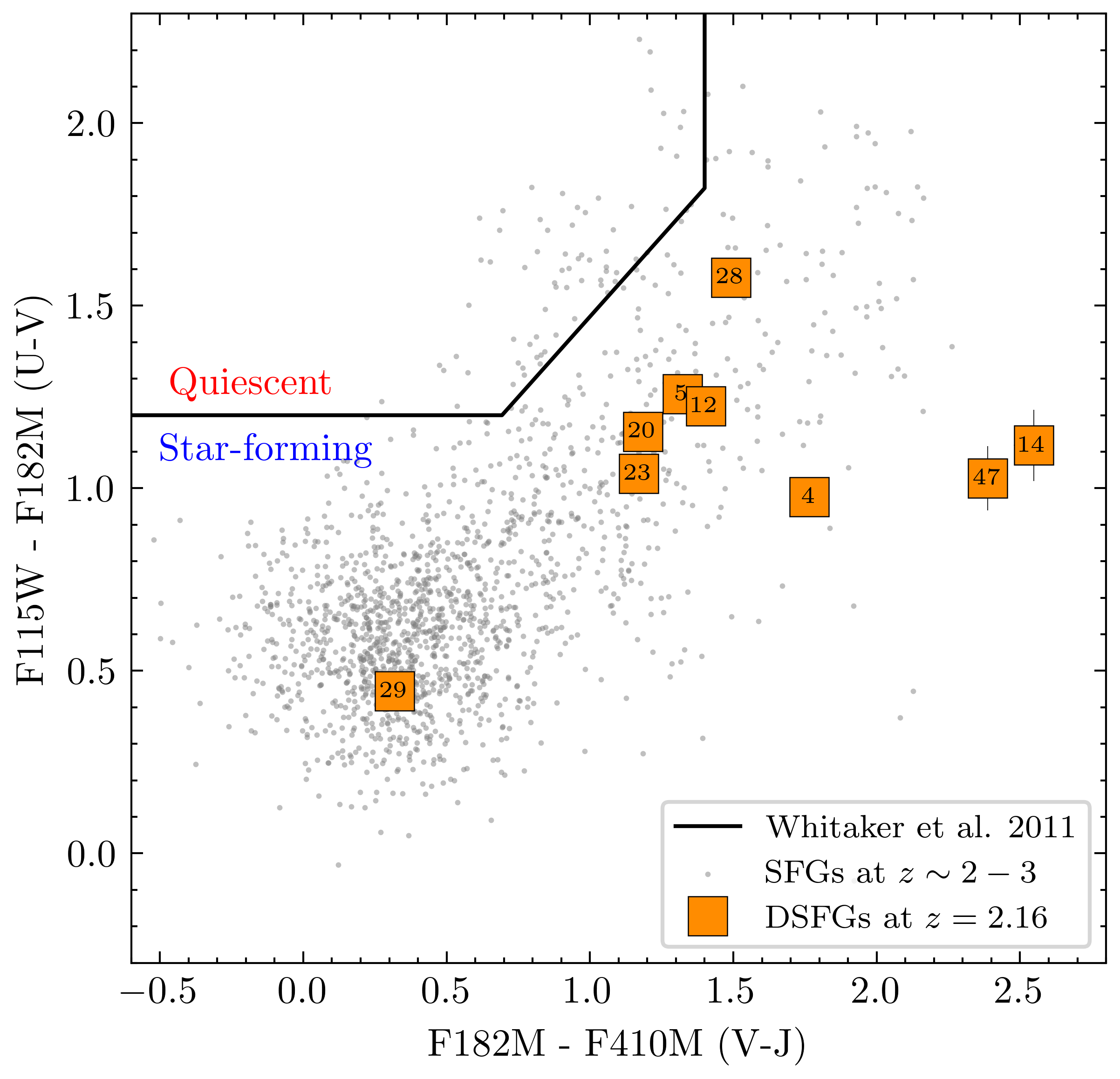}
	\caption{Rest-frame $UVJ$ (F410M, F182M, F115W) diagram of the nine JWST-detected DSFG members,  marked with their IDs. The criteria to separate the quiescent and star-forming regions are adopted from \citet{Whitaker2011}. The normal star-forming galaxies and quiescent galaxies at $z\sim2-3$ from JWST/CEERS and JADES surveys are shown with gray dots \citep{Ren2024}.}
	\label{fig: uvj}
\end{figure}

The $UVJ$ diagram \citep{Williams2009, Brammer2011, Whitaker2011} has been widely used to classify high-redshift galaxies into star-forming and quiescent populations in the JWST era \citep{Ren2024, Ji2024, Ito2024, Martorano2024, Polletta2024}. 
We use the total magnitudes measured in F115W, F182M, and F410M filters to trace the rest-frame $U$, $V$, and $J$ colors of the nine DSFG members. 
The resulting rest-frame $UVJ$ diagram is shown in Fig.~\ref{fig: uvj}, with the criteria to separate star-forming and quiescent galaxies according to \citet{Whitaker2011}. We also present the normal star-forming galaxies and quiescent galaxies at $z\sim2-3$ from JWST/CEERS and JADES surveys in Fig.~\ref{fig: uvj} for a comparison \citep{Ren2024}. It is evident that all ALMA sources are located in the star-forming region of the $UVJ$ diagram. Most are concentrated in the upper-right region, clearly offset from the bulk of normal star-forming galaxies in the field, highlighting their dusty star-forming nature.

\subsection{Non-parametric morphologies}
\label{subsec: morph_n}

Thanks to the high spatial resolution and the broad wavelength coverage up to 4\,$\mu$m provided by JWST/NIRCam, we could study the morphologies of the stellar components (traced by rest-frame NIR) in our DSFG sample for the first time. 
Significant efforts have already been made on structural studies of ALMA-selected DSFGs with newly acquired JWST data \citep{Boogaard2024,Chen2022,Cheng2022,Cheng2023,Colina2023,Crespo2024,Fujimoto2025,Gillman2023,Gillman2024,Hodge2025,Kamieneski2023,Lebail2024,Liu2024,Price2025,Polletta2024,Rujopakarn2023,Smail2023,Tadaki2023,Wu2023,Umehata2025a,Umehata2025b},  which could be utilized in this paper to compare DSFGs in field and protocluster environments. 

Two approaches are commonly-used to measure the morphological properties of galaxies: non-parametric and parametric ways. In this subsection we primarily focus on the non-parametric analysis, which is advantageous for capturing the diverse and  complex shapes of galaxies without assuming a specific parametric form for the galaxy profiles.

The non-parametric morphological analysis are performed by the code \texttt{statmorph} \citep{Rodriguez-Gomez2019}, which could return various quantitative parameters including Gini $(G)$ and $M_{20}$ coefficients \citep{Abraham2003,Lotz2004}, and the concentration $(C)$, asymmetry $(A)$, and smoothness/clumpiness $(S)$ indicators \citep[CAS;][]{Bershady2000,Conselice2003,Lotz2004}. We use the background-subtracted images to make a cutout in a size of $201\times201$ pixels ($6^{\prime\prime}\times6^{\prime\prime}$) for each source. The {\sc Python} package \texttt{Photutils} is employed to create a segmentation of each target based on the F410M image, with a detection threshold of 1.5$\sigma$. 
Subsequently, we run \texttt{statmorph} in all three images (F115W, F182M, and F410M), assessing the results with a flag and S/N per pixel for each source in each filter. The flag indicates the quality of the basic morphological measurements, which can be zero to five from good to catastrophic.
We only retain robust results with a flag equal zero \citep{Rodriguez-Gomez2019}, or an S/N per pixel greater than two \citep{Lotz2004}. 
Finally, we obtain robust non-parametric morphological measurements for all nine DSFG members in the F410M image, and for three and two members in the F182M and F115W images, respectively. Among them, two sources (ASW2DF.05 and ASW2DF.23) yield reliable measurements across all three filters.

We mainly employ the Gini $(G)$ and $M_{20}$ coefficients to classify our DSFGs in this work. The Gini coefficient assesses the distribution of light across the pixels within a galaxy. A value of one indicates that all the flux is concentrated in a single pixel, while $G=0$ means a perfectly uniform surface brightness. The $M_{20}$ coefficient quantifies the second-order moment of the pixels contributing 20\% of the total flux in a galaxy. A more negative $M_{20}$ value suggests that the light is more centrally concentrated, and vice versa. Specifically, we adopt the `bulge statistic' and `merger statistic' \citep{Lotz2008b,Snyder2015a, Snyder2015b} criteria, in which the former distinguishes early-type (E/S0/Sa) and late-type (Sb/Sc/Irr) galaxies, while the latter separates the mergers and non-mergers in the Gini-$M_{20}$ plane.

\begin{figure}
	\includegraphics[width=\linewidth]{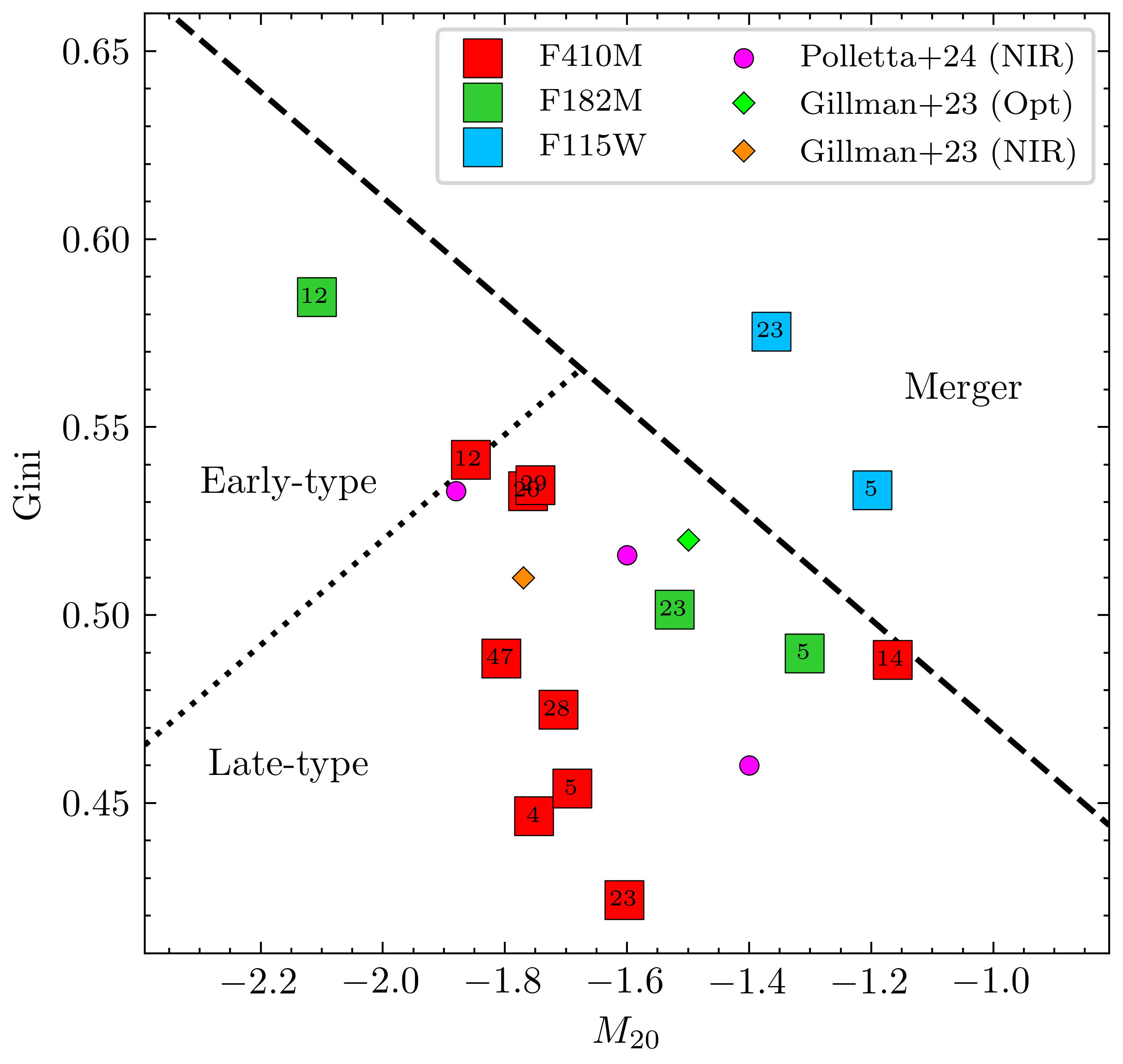}
	\caption{The Gini-$M_{20}$ diagram of nine DSFG members across three filters—F410M (red), F182M (green), and F115W (blue)—is presented. The medians of the field DSFG sample are depicted as orange and lime diamonds in the rest-frame optical and NIR bands \citep{Gillman2024}. Three extremely dusty starburst at cosmic noon are also shown in magenta circles for a comparison \citep{Polletta2024}. }
	\label{fig: gini-m20}
\end{figure}

Fig.~\ref{fig: gini-m20} shows the Gini-$M_{20}$ diagram of our DSFG members. Sources with reliable measurements in any of the three filters are shown in squares, color coded by the filters (RGB for F410M/F182M/F115W). 
The same measurements for a sample of DSFGs in the CEERS field \citep{Gillman2023}, and three massive dusty starbursts which is possibly related to a proto-structure at cosmic noon \citep{Polletta2024} are also shown in Fig.~\ref{fig: gini-m20} for a comparison.

The majority of our sample lies in the late-type galaxy region, exhibiting a broad range of Gini and $M_{20}$ values, consistent with our visual classifications and previous studies of field DSFGs \citep{Ling2022, Gillman2023}. Specifically, eight out of nine DSFG members are classified as late-type galaxies in the F410M filter. 
The exception, ASW2DF.12, is categorized as early-type in both F410M and F182 filters. This is due to the presence of a strong central AGN as previously confirmed \citep{Tozzi2022a,Shimakawa2024a}. 
Two members are classified as mergers in the F115W filter. ASW2DF.23 is likely a minor merger system, consisting of a northern main component and a southern blue companion. In contrast, ASW2DF.05 exhibits a clear disk morphology in the other two filters, suggesting that its merger classification in F115W is likely due to strong dust attenuation affecting its rest-frame UV emission. 
These results suggest that, while localized starburst activity is common in these ALMA-selected DSFGs,  their overall star formation is more likely driven by secular processes rather than major mergers. \citep{Huang2025}.

Among the sources (ASW2DF.05 and 23) with robust morphological measurements across all three filters, we observe a clear trend in their measured values, shifting from the late-type region toward the merger region at shorter wavelengths. A similar trend has been found for field DSFGs based on median measurements in the rest-frame optical and NIR bands \citep{Gillman2023}. This wavelength-dependent shift reflects the underlying stellar populations: rest-frame NIR emission, tracing older and redder stars, reveals a more stable and centrally concentrated morphology, while rest-frame UV emission, tracing recent star formation, appears significantly more clumpy and disturbed. This is further supported by the smoothness ratios in our sample, with median values of $S_{U}/S_{V}=2.8$ and $S_{U}/S_{J}=2.4$, indicating that ongoing star formation is substantially more clumpy than the stellar distribution. These results are consistent with the inside-out galaxy evolution scenario, in agreement with previous studies on field DSFGs \citep{Chen2022, Gillman2023, Kamieneski2023}. However, we note that the boundaries in the Gini–$M_{20}$ diagram were originally defined at rest-frame 0.65\,$\mu$m and are therefore strictly applicable to the F182M filter. As a result, the apparent trend toward the merger region at shorter wavelengths may be subtle, since classifications could differ in the F115W and F410M filters, and the sample sizes in the F115W and F182M filters are limited.

\begin{figure}
	\centering
	\includegraphics[width=\linewidth]{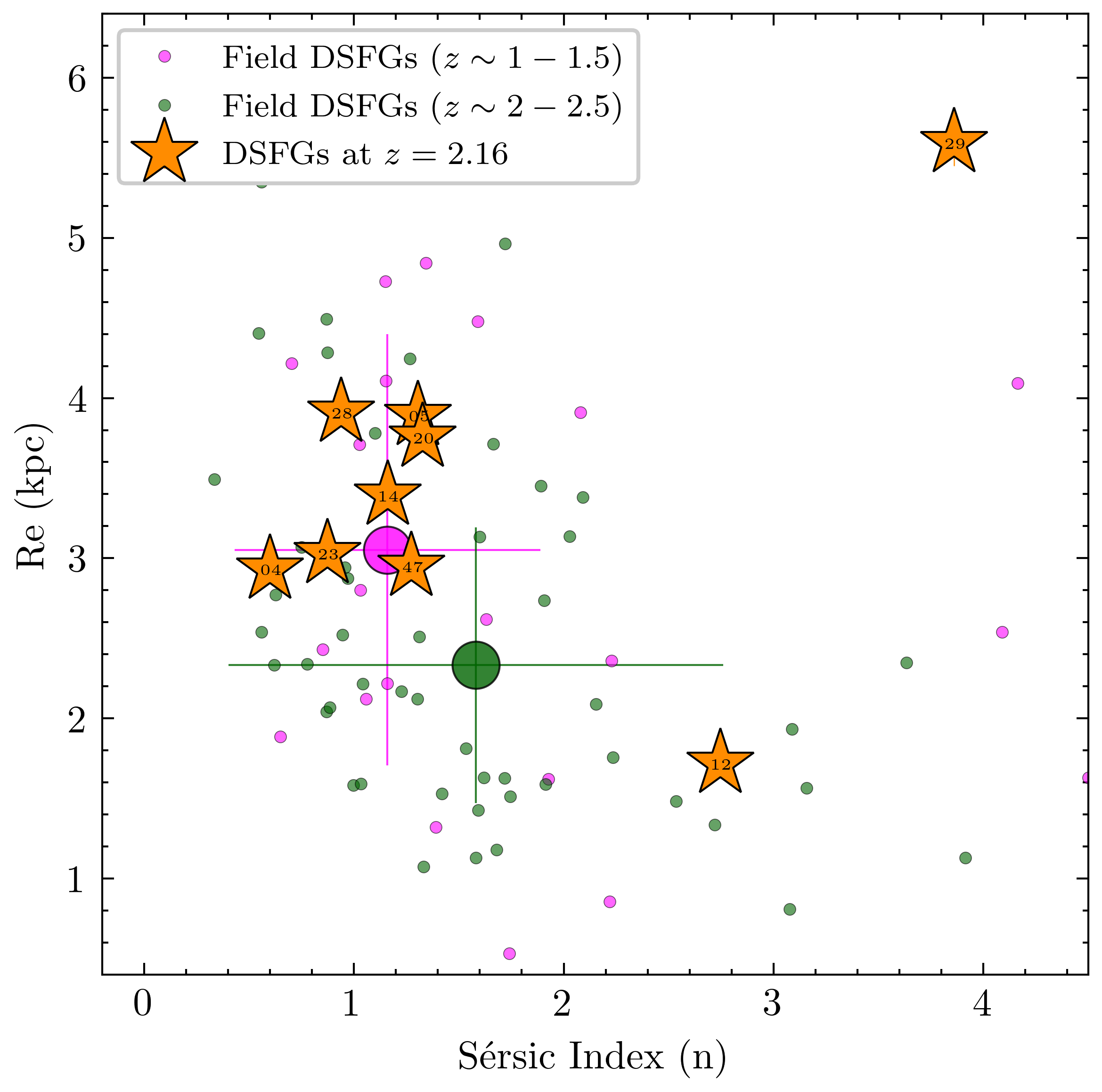}
	\caption{Rest-frame near-infrared effective radii, measured from the F410M filter, are plotted against Sérsic indices for the nine DSFG members in the Spiderweb protocluster.  For comparison, 27 (55) field DSFGs at $z\sim1-1.5$ ($z\sim2-2.5$) from the literature \citep{Cheng2022,Cheng2023,Boogaard2024,Price2025,Ren2025} are shown as violet and pink circles, respectively. The median values for each redshift bin, along with their median absolute deviations, are indicated by larger markers of the same color with corresponding error bars.
	}
	\label{fig: re_vs_n}
\end{figure}

\subsection{Parametric morphologies}
\label{subsec: morph_p}

In addition to the non-parametric morphological analysis described above, we obtain parametric measurements of the DSFG members using \texttt{Galight} \citep{Ding2020,Ding2021}, which has been widely adopted on JWST data \citep{Yang2022,Ding2022,Ding2023,Williams2023,Liu2024,Zhuang2024,Casey2024} and demonstrated to be robust \citep{Kawinwanichakij2021,Yang2021,Casey2024} compared to the commonly used \texttt{GALFIT} \citep{Peng2002}. We employ \texttt{Galight} to select several unsaturated point sources in each of the three filters and generate an empirical PSF through a median stacking approach. We note that the Full Width Half Maximum (FWHM) of our PSFs are overall larger than the theoretical PSF generated by \texttt{WebbPSF} \citep{Perrin2012,Perrin2014}, while comparable with the empirical PSFs extracted from archival JWST/NIRCam data \citep{Finkelstein2023} within 5\% uncertainty, confirming the robustness of our PSF extraction. 

\begin{figure*}
	\centering
	\includegraphics[width=0.49\textwidth]{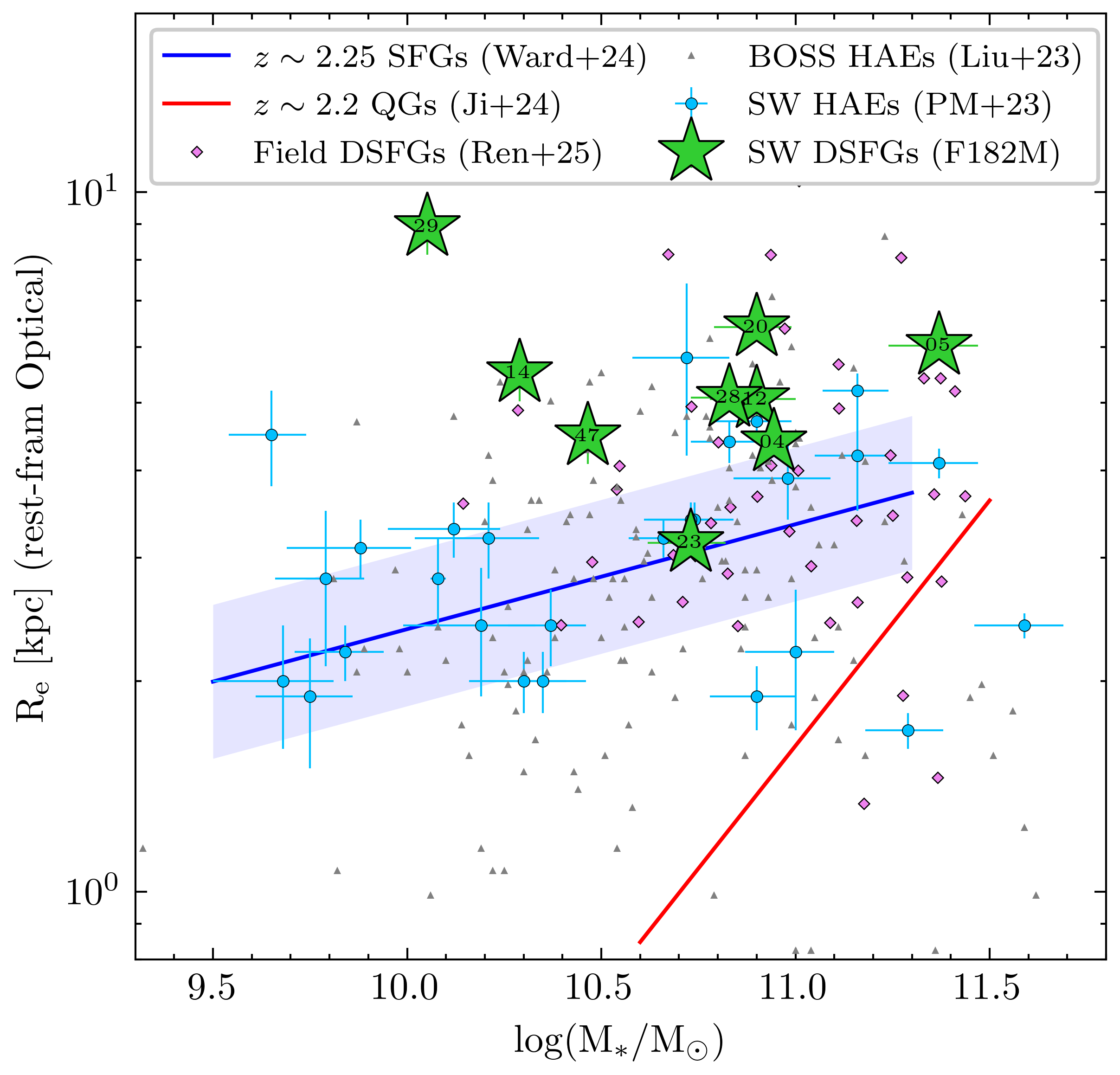}
	\includegraphics[width=0.49\textwidth]{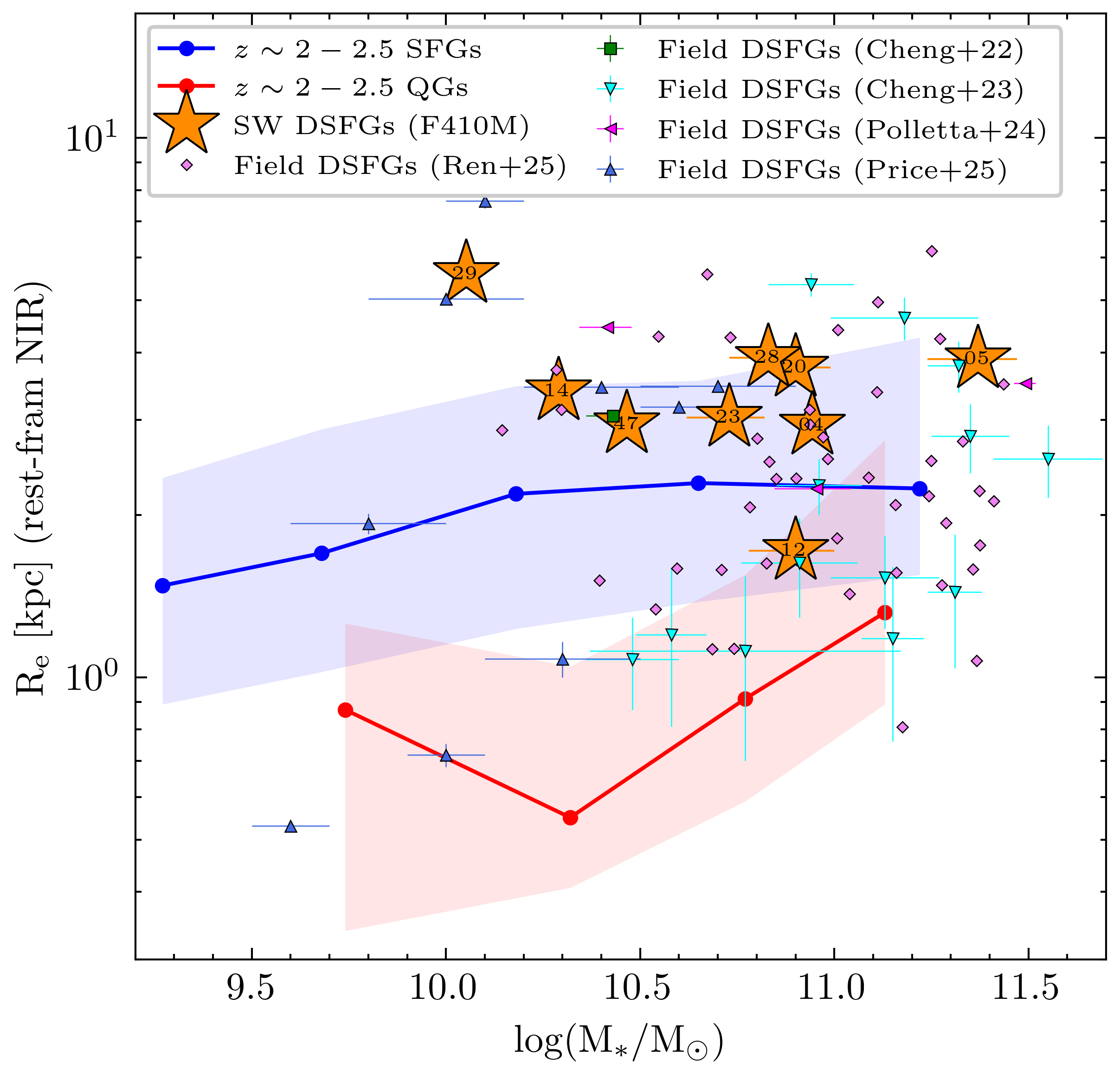}
	\caption{Left: Rest-frame optical size-mass relation of the nine DSFG members in the Spiderweb protocluster, with effective radii measured using the F182M filter. For comparison, HAEs from the BOSS1244/1542 \citep{Liu2023} and Spiderweb protoclusters \citep{Perez2023}, as well as field DSFGs at similar redshifts \citep{Ren2025}, are shown. The optical size-mass relations of field star-forming and quiescent galaxies at $z\sim2.2$ are also included \citep{Ward2024, Ji2024}. Right:   Rest-frame NIR size-mass relation of the nine DSFG members, measured using the F410M filter. Archival DSFG samples from recent JWST studies are presented for comparison \citep{Cheng2022, Cheng2023, Polletta2024, Price2025, Ren2025}, along with the NIR (1.5\,$\mu$m) size-mass relations of star-forming and quiescent galaxies at $z\sim2-2.5$ \citep{Martorano2024}.
	}
	\label{fig: size-mass relation}
\end{figure*}

Similarly, we create a background-subtracted cutout for each source in each filter, sized at $201\times201$ pixels ($6\times6$\,arcsec). The detection thresholds are adjusted adaptively to ensure that the segmentation encompasses the entire galaxy, particularly in bluer filters (e.g., F115W), where the galaxy may exhibit multiple clumps. We individually verify the segmentation for each sources  to maintain consistency across all three filters and alignment with the segmentation previously generated using \texttt{statmorph}. We fit the target galaxy along with any possible nearby companions with single Sérsic model, and mask out other sources within each cutout. We determine the uncertainty for each parameter by deriving the 1$\sigma$ interval of the posterior distribution. Finally, we obtain robust size measurements for all the nine DSFG members across F115W, F182M and F410M filters. Interestingly, \texttt{Galight} provides a function that compiles all data for each source and feeds it into \texttt{statmorph} to obtain non-parametric morphological measurements. We confirm the reliability of our results by validating the consistency of \texttt{statmorph} outputs across both approaches.

\subsubsection{Effective radii}
\label{subsubsec: eff_radii}

We convert the effective radii of the nine DSFG members into physical scales using their respective redshifts.
In Fig.~\ref{fig: re_vs_n} we present the effective radii as a function of Sérsic indices of these protocluster members derived from the F410M image, which traces the stellar components at rest-frame NIR wavelength.  For comparison, we compile the size measurements at rest-frame near-infrared wavelengths for field DSFGs from the literature \citep{Cheng2022,Cheng2023,Boogaard2024,Price2025,Ren2025}, and divide them into two redshift bins: 27 sources at $z\sim1-1.5$ and 55 sources at $z\sim2-2.5$. These two groups are shown as magenta and green circles in Fig.~\ref{fig: re_vs_n}, respectively. For each bin, the median values and their median absolute deviations are indicated by larger markers of the same color, along with corresponding error bars. We note that the effective radii of field DSFGs at higher redshift are systematicly smaller than those at lower redshift, as observed in several JWST studies \citep{Gillman2023,Ren2025}.

It is clearly seen that, our DSFG sample exhibits systematically larger effective radii relative to field DSFGs at similar redshifts. Two outliers, ASW2DF.12 and ASW2DF.29, deviate from this overall trend. The size measurement of ASW2DF.12 is likely affected by an X-ray AGN at its center \citep{Tozzi2022a,Shimakawa2024a}, while ASW2DF.29 shows an irregular morphology characterized by a bright core, small blue clumps, and a clear spatial offset between the dust (traced by ALMA) and stellar components—indicative of strong asymmetry and heavy dust obscuration. These features suggest ASW2DF.29 is likely a recent merger hosting partially obscured starbursts and a nascent AGN. Despite these two exceptions, the sample as a whole remains significantly larger than coeval field DSFGs.

Moreover, the effective radii and Sérsic indices of our DSFG sample closely match those of field DSFGs at $z \sim 1$–1.5, as shown in Fig.~\ref{fig: re_vs_n}. We attribute this size similarity to accelerated evolution driven by the gas-rich and overdense environment of the protocluster. This environment promotes the growth of more extended star-forming regions and larger stellar disks, causing our DSFG members to exhibit sizes comparable to field DSFGs approximately 2\,Gyr later.  However, given the limited sample size and focus on a single protocluster, larger samples across multiple protoclusters are needed to confirm this scenario.

\subsubsection{Size-mass relation}
\label{subsubsec: mass_size}

The conclusions above are based on a comparison of overall sizes between DSFGs in the Spiderweb protocluster and coeval field DSFGs, without accounting for stellar mass. 
Out of our ten spectroscopically confirmed ALMA-detected sources, we exclude the Spiderweb galaxy (ASW2DF.17) from our analysis given its singular nature. We obtain stellar masses for five of the nine remaining ALMA-detected sources (ASW2DF.05, 12, 20, 23, 28) from \citet{Perez2023}, where they were identified as HAEs (see Table~\ref{tab: dsfgs_phys}). Their stellar masses are based on up to ten-band SED modelling using a \cite{Chabrier03} IMF and covering rest-frame UV to NIR wavelengths (B band to Spitzer/IRAC 3.6 and 4.5 $\mathrm{\mu m}$ in the observed frame), a range that captures the bulk of the stellar emission and yields estimates with typical uncertainties of $\lesssim 0.2$ dex (see \citealt{Perez2023} for more details). The remaining four sources (ASW2DF.04, 14, 29, 47) were not included in that study, and thus no SED-based stellar masses are available. Visual inspection of the multiband imaging confirms that these galaxies are either undetected or fall below the 5$\sigma$ depth in many ground-based bands, leaving SED fitting significantly unconstrained. However, the JWST observations are deeper than the pre-existing datasets, and all four sources are securely detected in the relevant NIRCam bands, enabling us to apply alternative methods. We therefore estimate their stellar masses using their F115W–F182M colors and F410M-based NIR luminosities, adopting the mass-to-light ratio prescriptions from \citet{Bell2003} and correcting for a \cite{Chabrier03} IMF. Similar model-based mass-to-light approaches, albeit with different prescriptions have also been succesfully used in past works up to $\mathrm{z\sim2}$ (e.g., \citealt{Kriek08}; \citealt{Koyama2013}). We validate our approach by confirming that it reproduces, within reasonable uncertainties (0.17 dex standard deviation), the SED-derived stellar for our first five sources, thereby supporting its robustness for the rest of the sample (see Appendix~\ref{sec:stellar masses}).

Figure~\ref{fig: size-mass relation} presents the rest-frame optical and NIR effective radii as a function of stellar mass of our DSFG members. For comparison, we include empirical size–mass relations derived from JWST data for star-forming galaxies and quiescent galaxies at $z\sim2-3$ \citep{Ward2024,Ji2024,Martorano2024}, as well as rest-frame optical sizes of $z\sim2.2$ HAEs in protocluster environments \citep{Liu2023,Perez2023}. Archival measurements of DSFGs at rest-frame NIR wavelengths from recent JWST studies are also shown \citep{Cheng2022,Cheng2023,Polletta2024,Price2025,Ren2025}.

As shown in Fig.~\ref{fig: size-mass relation}, the nine DSFGs in the Spiderweb protocluster are systematically larger than coeval star-forming galaxies at fixed stellar mass, in both rest-frame optical and NIR wavelengths. These galaxies also represent the largest systems among the HAE populations in protocluster environments. It is worth noting that the effective radii of  HAEs  in BOSS protoclusters were measured using HST/F160W imaging, which corresponds to rest-frame 0.5\,$\mu$m at $z\sim2.2$ \citep{Liu2023}. Also, the size measurements of HAEs in the Spiderweb protocluster were made in the $K_{\rm s}$ band,  corresponding to rest-frame 0.7\,$\mu$m at $z\sim2.16$ \citep{Perez2023}. These rest-frame wavelengths differ slightly from the rest-frame 0.57\,$\mu$m traced by F182M for our DSFG members, which may introduce systematic uncertainties of up to $\sim20$–$30\%$ in the measured sizes, following the size–wavelength relation discussed in the next subsection.

Additionally, the F182M image is shallower than the F115W and F410M images, reducing its sensitivity to low surface brightness features and underestimating galaxy sizes compared to measurements in both bluer and redder bands (see Fig.~\ref{fig: cutouts}). When focusing on the rest-frame NIR sizes derived from F410M, our DSFGs are not only larger than field star-forming galaxies at similar redshifts but also rank among the largest DSFGs with comparable stellar masses across both field and protocluster environments. 
While the archival JWST surveys used for comparison employed different filters and depths \citep[e.g.,][]{Cheng2022,Cheng2023,Polletta2024,Price2025,Ren2025}, their samples—characterized by median redshifts of $z \sim 2$–3 and rest-frame NIR measurements—remain well suited for a meaningful comparison with our DSFG sample.
In summary, our sample exhibits larger sizes than both typical HAEs in protoclusters and field DSFGs at similar redshifts. Future   JWST observations across multiple bands and environments will be essential to confirm and extend these findings.

\begin{figure*}
	\centering
	\includegraphics[width=\textwidth]{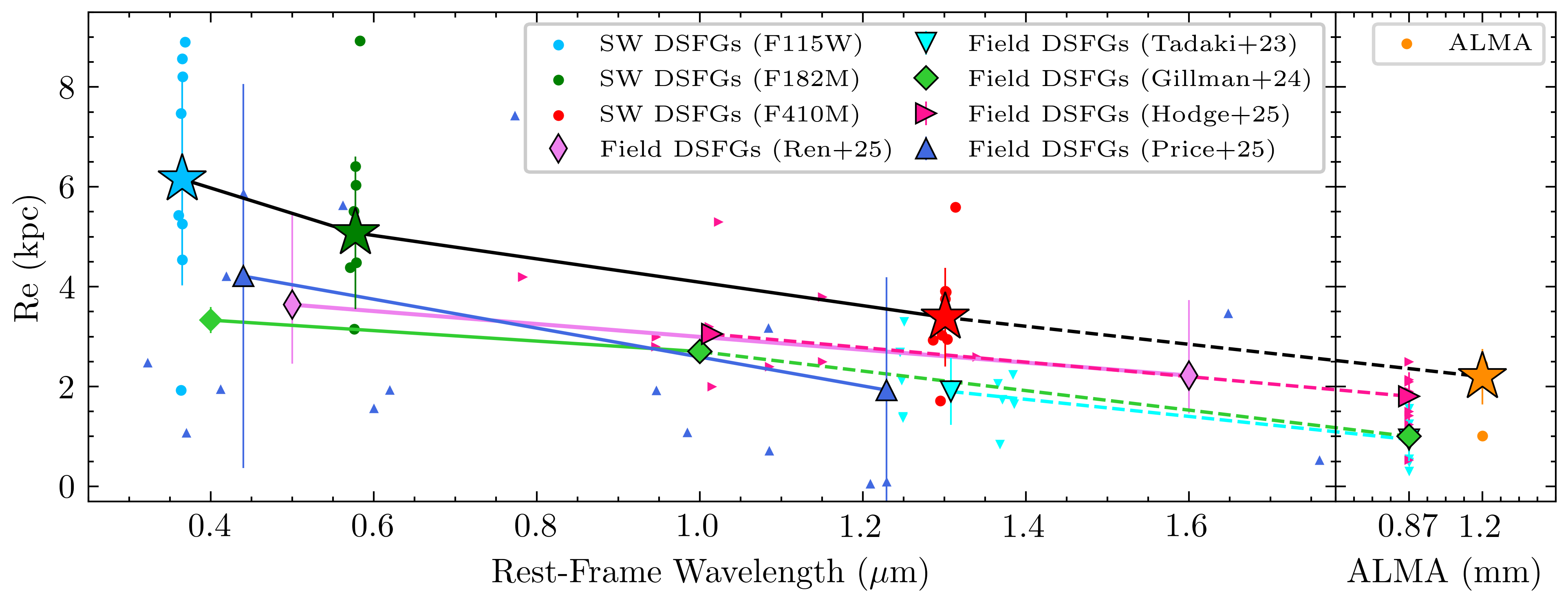}
	\caption{Size-wavelength relation of the nine DSFG members in the Spiderweb protocluster. The individual measurements and median values are marked as circles and stars with different colors for corresponding wavelengths. Field DSFG samples at similar redshifts from recent JWST studies are shown with different markers and colors for a comparison \citep{Tadaki2023,Gillman2024,Price2025,Hodge2025,Ren2025}. The green point at rest-frame 0.4\,$\mu$m represents the extrapolated result from the slope and the size at rest-frame 1\,$\mu$m in \citet{Gillman2024}. Solid lines indicate the decreasing slopes from rest-frame UV to NIR, while dashed lines connect the JWST and ALMA size measurements in each sample. }
	\label{fig: size evolution}
\end{figure*}

\subsubsection{Size-wavelength relation}
\label{subsubsec: size_evolution}

We characterized the stellar morphologies and measured the dust component sizes in these DSFGs using ALMA 1.2 mm observations. The \texttt{imfit} function in CASA was used to model the sources with an elliptical Gaussian profile. Due to varying resolutions and depths across the six fields, some sources could not be deconvolved from the synthesized beam and remained unresolved, while a few failed the fitting process. Overall, six out of nine DSFG members had dusty component size measurements. The sizes were determined by the FWHM along the major and minor axes, and were circularized into effective radii using the following equation \citep{Fujimoto2018}:
\begin{equation}
    R_{\rm e} = r_{\rm e,maj} \times \sqrt{q},
\label{eq: alma_re}
\end{equation}
where $r_{\rm e,maj}$ is the radius along the major axis and $q$ is the axis ratio, respectively.

We note that the sizes of the nine DSFG members were successfully measured across all three JWST/NIRCam filters, with four of them having additional size measurements of their dust components. In Fig.~\ref{fig: size evolution}, we present the physical sizes of the stellar and dust components of our member DSFGs as a function of rest-frame UVJ bands and observed millimeter band. Their median values are shown in star symbols and linked together with black solid lines to demonstrate the size evolution across the four bands. Additionally, the size-wavelength relation of coeval DSFG samples is shown as different markers and colors for a comparison \citep{Tadaki2023,Gillman2024,Price2025,Hodge2025,Ren2025}. It is evident that all samples, whether in protocluster or field environments, follow a similar trend: galaxy sizes systematically decrease with increasing wavelength from rest-frame UV to NIR, and the stellar components are generally larger than the dust emission traced by ALMA \citep{Tadaki2023,Gillman2024,Hodge2025}. This result aligns with previous studies revealing that dust components are smaller than stellar components traced by rest-frame optical observations with HST \citep{Rivera2018,Lang2019,Tadaki2020}, but contrasts with recent findings from hydrodynamical simulations \citep{Cochrane2019,Popping2022}.

Once again, our DSFG sample exhibits consistently larger size measurements across all observed bands compared to field DSFGs at similar redshifts.  
Specifically, the median effective radii of our DSFG sample at rest-frame $UVJ$ and 380\,$\mu$m are $5.8\pm2.7$\,kpc, $5.1\pm2.0$\,kpc, $3.2\pm1.2$\,kpc and $2.2\pm0.6$\,kpc, respectively. 
As the F182M filter corresponds to rest-frame 1.3\,$\mu$m for our DSFG sample, archival measurements at the same wavelength yield sizes of $2.6 \pm 0.2$, $2.5 \pm 0.1$, and $1.8 \pm 0.8$\,kpc for the DSFGs in \citet{Gillman2024}, \citet{Ren2025}, and \citet{Price2025}, respectively, all smaller than the $\sim 3.2$\,kpc measured for DSFGs in the Spiderweb protocluster.
Furthermore, we estimate the median stellar mass of our DSFG sample to be log$(M/M_{\odot})=10.8\pm0.3$, comparable to log$(M/M_{\odot})=11.2\pm0.1$ and log$(M/M_{\odot})=11.0\pm0.3$ in \citet{Gillman2024} and \citet{Ren2025} samples, but higher that the log$(M/M_{\odot})=10.3\pm0.4$ in \citet{Price2025} sample. These comparisons further indicate that our DSFG sample tends to have larger effective radii than their coeval field counterparts. 
The median stellar-to-dust size ratio for our protocluster members is $1.5\pm0.7$, consistent with the range of 1.6–2.7 reported for coeval field DSFGs \citep{Tadaki2023,Gillman2024,Hodge2025}.  It is worth noting that  previous studies measured dust sizes at 870\,$\mu$m, while our measurements are based on 1.2\,mm data, which may introduce systematic differences. 

From rest-frame UV to NIR wavelengths, our DSFGs exhibit a pronounced decrease in size. The corresponding slope is comparable to that of field DSFGs, although field samples show considerable diversity in their measured slopes. For reference, \citet{Gillman2024} reported a size–wavelength slope of $-0.60\pm0.09$ for DSFGs across similar wavelength ranges, which is steeper than the $-0.15\pm0.07$ slope found for normal star-forming galaxies \citep[see also][]{Suess2022}. Following the normalization procedure of \citet{Gillman2024}, we obtain an even steeper slope of $-1.20\pm0.30$ for our DSFG members.
We speculate that two factors may contribute to this trend. First, the rest-frame UV sizes of our sources are likely inflated by centrally concentrated dust obscuration more strongly than in field DSFGs or normal star-forming galaxies. Second, the proximity of these DSFG members to large-scale filaments and extended gas reservoirs \citep{Chen2024} may promote continuous gas accretion, fueling clumpy starbursts in their outskirts. Both effects could naturally lead to a steeper slope in the size–wavelength relation.
For comparison, the slopes derived for the DSFG samples in \citet{Ren2025} and \citet{Price2025} are $-0.94 \pm 0.13$ and $-1.98 \pm 0.52$, respectively.
The substantial differences among these values are likely driven by sample diversity—such as variations in sample size, stellar mass, and selection depth—and should therefore be interpreted with caution.
Overall, samples with larger stellar masses tend to exhibit moderate slopes, as seen in our sample and in the literature \citep{Gillman2024, Ren2025}, whereas less massive DSFGs may show much steeper slopes \citep{Price2025}. A larger and more mass-complete sample will be essential for disentangling these factors and for quantifying the role of stellar mass in shaping the size–wavelength slope across different mass regimes.

\subsection{Notes on individual member}
\label{subsec: individual member}

This subsection presents individual descriptions of the nine DSFG members, detailing their counterpart identifications, morphologies, and any distinguishing structures. We summarize all morphological measurements of the nine DSFG members, as well as their brief descriptions in Table~\ref{tab: dsfgs_morph}. We note that the central Spiderweb galaxy (ASW2DF.17) is significantly affected by PSF contamination across all JWST/NIRCam filters, and thus excluded from the discussion on individual member.

\paragraph{ASW2DF.04} was not identified as an HAE or CO line emitter in previous studies \citep{Shimakawa2018,Jin2021,Perez2023}, but has recently been confirmed as a protocluster member based on HST observations \citep{Naufal2024}. It lies within the LABOCA beam of the single-dish source DKB02 \citep{Dannerbauer2014}, and is likely its dominant flux contributor. In JWST imaging, ASW2DF.04 shows a uniform disk morphology across all filters, although the F115W band reveals a slight central depression in brightness, suggestive of significant centrally concentrated dust attenuation. A small blue feature to the northeast may represent a low-redshift interloper.

\paragraph{ASW2DF.05} has been extensively studied as an HAE \citep{Dannerbauer2017}, and is the brightest CO emitter \citep{Jin2021} with extended molecular gas reservoirs \citep{Chen2024}. It displays a prominent disk morphology with two grand-design spiral arms clearly visible in all JWST filters. From shorter to longer wavelengths, its central bulge becomes increasingly prominent, while the disk appears progressively smoother. Multiple star-forming clumps are evident in the F115W band, consistent with features previously identified in HST imaging \citep{Dannerbauer2017}.

\paragraph{ASW2DF.12} has been identified as an HAE, CO emitter, and X-ray source in previous studies \citep{Jin2021,Tozzi2022a,Perez2023}. Hosting a central AGN \citep{Shimakawa2024a}, it exhibits a compact and centrally concentrated morphology. Based on non-parametric morphological analysis, it is classified as an early-type galaxy, consistent with AGN-driven structural compaction. These features suggest it may represent a transitional phase in the quenching pathway of massive protocluster galaxies.

\paragraph{ASW2DF.14}  is a confirmed CO emitter within the Spiderweb protocluster \citep{Jin2021}, and exhibits extended molecular gas reservoirs \citep{Chen2024}. It is barely visible in the F115W and F182M filters, but becomes clearly visible at longer wavelengths, where it resolves into two distinct components. This morphology strongly suggests that ASW2DF.14 is an ongoing merger.

\paragraph{ASW2DF.20}  has been identified as both an HAE and a CO emitter \citep{Jin2021,Perez2023}. It shows a well-defined disk morphology across all JWST filters, with a compact central core. The diffuse, extended emission observed in the F115W band indicates the presence of widespread star-forming regions in the outer disk. A point-like blue component nearby is likely a foreground, low-redshift interloper.

\paragraph{ASW2DF.23}  is a confirmed HAE and CO emitter \citep{Jin2021, Perez2023}. At shorter wavelengths, it appears fragmented into multiple clumps, while at longer wavelengths it presents a more coherent morphology consistent with a single galaxy. A small blue feature connected to the northern part of the main body suggests a possible minor merger.

\paragraph{ASW2DF.28}  is identified as an HAE \citep{Perez2023}, and lies within the LABOCA beam of the single-dish source DKB15 \citep{Dannerbauer2014}. It is faint in the F115W and F182M filters, but displays a smooth, disk-like morphology at longer wavelengths.

\paragraph{ASW2DF.29}  has the most complicated morphology in our sample. It has been identified as a CO emitter with extended molecular gas reservoirs \citep{Jin2021, Chen2024}. Across all JWST filters, it exhibits two distinct components to the north and south. The northern component features a bright central core, while the southern component aligns with the ALMA detection position. The presence of multiple components and irregular morphology strongly suggests that ASW2DF.29 is undergoing a merger.

\paragraph{ASW2DF.47}   has been identified as a CO emitter \citep{Jin2021}. Similar to ASW2DF.14 and ASW2DF.28, it is barely detected in the F115W and F182M filters, but reveals a clear, smooth disk morphology at longer wavelengths. 

Additionally, we measured the morphological properties of the remaining 12 ALMA sources lacking spectroscopic redshifts. These sources were visually inspected and are described individually below. ASW2DF.07, 10, 11, 15, 22, 24, and 26, 31 exhibit very red colors and generally smaller sizes compared to our DSFG sample, suggesting that they are high-redshift galaxies with significant dust obscuration. ASW2DF.19 is bright in all four filters, displaying a clear stellar disk and prominent starburst clumps, indicative of an ongoing starburst at a relatively lower redshift. Similarly, ASW2DF.35, 37, and 44 show extended stellar disks and strong emission across all four filters, resembling dusty star-forming galaxies at lower redshifts. A summary of the morphological measurements, along with brief notes on each source, is provided in Table~\ref{tab: other_morph}.

\section{Discussion}
\label{sec: discussion}

\subsection{The origin of extended stellar disks}
\label{subsec: disk origin}

\begin{figure}
	\centering
	\includegraphics[width=\linewidth]{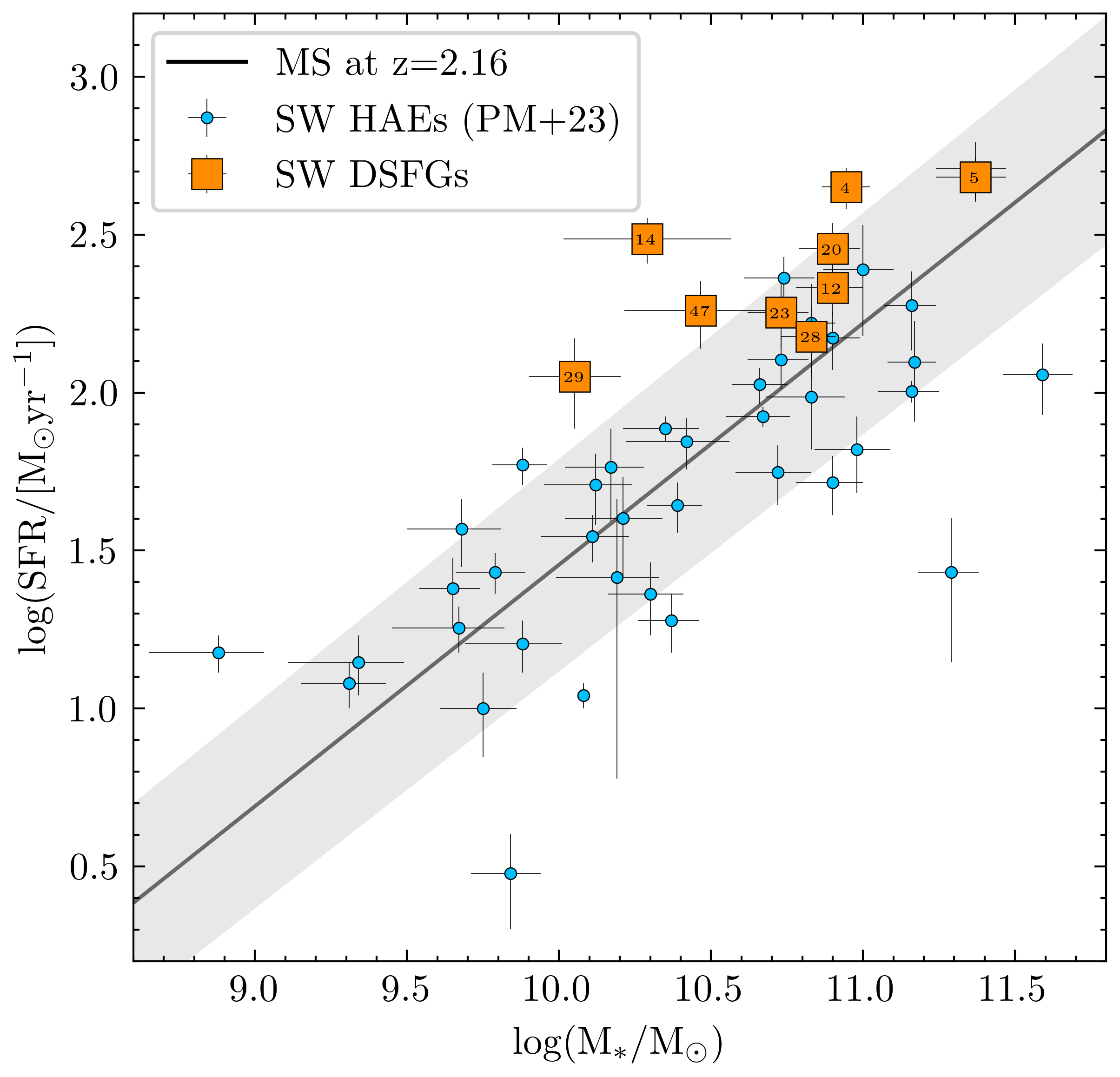}
	\caption{Star-forming main sequence relation for the nine DSFG members.  The reference main sequence relation at $z = 2.16$ is shown in black line and the corresponding 3$\sigma$ scatter are marked with grey shaded region \citep{Speagle2014}.  The blue solid circles represent the HAEs in the Spiderweb protocluster for a comparison \citep{Perez2023}.} 
	\label{fig: ms}
\end{figure}

In this section, we explore the star formation mode and the origin of extended stellar disks in the DSFG members. Stellar masses for the nine protocluster DSFGs are adopted from either the HAE catalog by \citet{Perez2023} or derived from JWST observations, as described previously. We estimate SFRs using ALMA 1.2\,mm fluxes \citep{Scoville2023, Zhang2024}. In Fig.~\ref{fig: ms}, we compare these DSFGs with all HAE members confirmed by \citet{Perez2023}, as well as the main sequence at the same redshift \citep{Speagle2014}. 
While these sources exhibit elevated star formation rates, the majority remain consistent with the high-mass end of the coeval star-forming main sequence within the 3$\sigma$ scatter, showing no clear evidence for starburst activity. The elevated star formation is attributed to localized star-forming clumps, likely triggered by disk instabilities in combination with cold gas accretion and/or minor mergers, as evidenced by their irregular rest-frame UV morphologies.

We visually inspect the DSFG sample to identify  merger systems and assess the role of mergers in driving star formation within the Spiderweb protocluster. ASW2DF.14 is classified as a late-stage merger based on the presence of two distinct components in the F410M image. This classification is supported by a lower reduced $\chi^2$ and cleaner residuals when fitting a dual Sérsic model, compared to a single-component fit. Another ongoing merger is ASW2DF.29, which exhibits an elongated disk morphology with a small blue companion to the south. The southeastern edge of the disk appears disconnected from the central structure, possibly forming a tidal tail due to ongoing interaction. Notably, a significant offset of $\sim$4\,kpc is observed between the dust and stellar components, suggesting a heavily obscured starburst region interacting with its northern companion. This spatial offset is unlikely to be caused by astrometric uncertainty, as the alignment between JWST and ALMA detections is consistent for most sources in our sample \citep{Hodge2025}.

We also identify ASW2DF.23 as a potential merger candidate, given its multiple star-forming clumps in the bluer bands and irregular morphology in all three filters. 
Given the presence of multiple satellites around the Spiderweb galaxy \citep{Miley2006} and a prominent dusty companion to its west (as shown in Fig.~\ref{fig: cutouts}), we also classify the Spiderweb galaxy as a merger candidate.
With ten confirmed DSFG members in our sample, this yields a merger fraction of 20\%, up to 30\% when including the potential candidate. This estimate aligns with recent JWST-based studies of field DSFGs, which report a merger fraction of ~20\%, with an upper limit of 60\% \citep{Gillman2024,McKay2025,Ren2025}. Similar merger fractions have also been observed among HAE populations in protocluster environments at comparable redshifts \citep{Liu2023, Naufal2023}. Earlier studies reported higher merger fractions of $\sim$50\% or more in DSFGs \citep{Chen2015, Stach2019}, largely due to HST observations being limited to rest-frame UV wavelengths, where dust-obscured galaxies often appear morphologically disturbed, potentially biasing merger classifications.

These results suggest that mergers are not the prominent drivers of star formation or disk growth in DSFGs \citep{McAlpine2019,Ren2025}. Instead, intense star formation and the development of extended stellar disks may result from smooth accretion of cold gas along large-scale filaments, as supported by both simulations and observations \citep{Dekel2009, Daddi2021,Umehata2025b}. Indeed, the tight spatial associations between DSFGs and gas filaments traced by Ly$\alpha$ emission have been observed in a few galaxy overdensities \citep{Umehata2019,Daddi2021}. 
In addition to the morphological analysis in the three main filters corresponding to the rest-frame $U$, $V$, and $J$ bands, we also measured sizes in the F405N filter, which is sensitive to ongoing star formation activity. The sizes measured in the F405N and F410M filters are found to be nearly identical, which suggests that the contribution of Pa$\beta$ emission is relatively minor compared to the continuum emission in the F405N filter. This may also provide a natural explanation for why these DSFGs were not identified as Pa$\beta$ emitters in our previous JWST studies \citep{Shimakawa2024b,Perez2024b}.
Moreover, DSFGs in protoclusters tend to exhibit larger stellar disks compared to their field counterparts \citep{Wu2023, Crespo2024, Wang2025, Umehata2025a}, indicating accelerated size evolution in dense environments \citep{Esposito2025}. 

Simulations suggest that coherent angular momentum in inflowing gas can expand rotating disks and enhance stellar rotational velocities \citep{Ceverino2010,Stewart2017,Umehata2025a}. Conversely, counter-rotating or misaligned gas streams may disrupt existing disks and induce localized star-forming clumps, as seen in both this study and prior observations \citep{Lebail2024, Rujopakarn2023, Kalita2024, Wang2025, Umehata2025a}. Nevertheless, we cannot rule out mergers as contributors to the mass and size buildup of DSFGs, with their influence likely varying across different protocluster evolutionary stages \citep{Hopkins2009, Jian2012, Liu2023}. A larger DSFG sample with spatially resolved kinematic measurements will be crucial to disentangle the dominant mechanisms shaping their evolution in overdense environments.

\subsection{The possible fate of  DSFG members}
\label{subsec: evolution fate}

The hypothesis that high-redshift DSFGs are progenitors of massive elliptical galaxies in the local universe has been supported by both simulations \citep{Gonzalez2011,McAlpine2019} and observations \citep{Hainline2011,Simpson2014,Toft2014,Ikarashi2015,Chen2016,Wilkinson2017,Stach2019}, based on their consistent number densities and physical properties. Observational evidence from JWST \citep{Chen2022, Lebail2024, Gillman2024, McKinney2025} and ALMA \citep{Gullberg2019, Hodge2025} has revealed the presence of stellar bulges in high-redshift DSFGs, though these bulges generally do not dominate the total stellar mass compared to the extended disk components.

In our sample, ASW2DF.12—located just 180\,kpc from the Spiderweb galaxy—exhibits a prominent stellar bulge component, as indicated by a best-fitting Sérsic index of 2.7 in F410M filter and supported by an X-ray detection \citep{Tozzi2022a}. We propose that this source may represent an early stage of bulge formation in DSFGs within a protocluster environment. Additionally, five DSFG members (ASW2DF.05, 20, 23, 28, 29) show weak but distinct point-like light concentrations superimposed on their stellar disks (see Fig.~\ref{fig: cutouts}).
Such central concentrations may be driven by accretion of cold gas with counter-rotating or perpendicular angular momentum, leading to angular momentum loss and compact central starbursts \citep{Scannapieco2009, Sales2012, Aumer2013, Dubois2014, Daikuhara2024}. Alternatively, these point-like components might signify early AGN activity. Recent studies of X-ray-detected HAEs in the Spiderweb protocluster reveal similar compact cores as potential indicators of AGN feedback \citep{Shimakawa2025}. Although the central brightness in our sample is less pronounced than in the X-ray-detected HAEs, it suggests that these DSFGs may represent progenitor phases of HAE populations situated closer to the cluster core. Their stellar disks are likely fueled by cold gas accretion, while their centers may host growing black holes that contribute to the observed central light excess.

These findings are consistent with the inside-out evolutionary scenario \citet{Shimakawa2018,SunH2024}, in which the Spiderweb protocluster is in a maturing phase. Its core already hosts several quiescent galaxies \citep{Naufal2024}, along with nascent ICM  \citep{Tozzi2022b,Mascolo2023,Lepore2024} and emerging AGN feedback observed in some HAEs \citep{Shimakawa2024a}. In contrast, DSFGs located at the protocluster outskirts could continue their vigorous star formation, likely fueled by cold gas accretion along surrounding large-scale filaments \citep{Zhang2024}.  This spatial trend is also evident in Fig.~\ref{fig:fov_map}, where ASW2DF.12, located within $r_{500}$, may reflect a more advanced stage of AGN-regulated evolution, while most DSFGs reside in the outer regions of the protocluster—likely experiencing an earlier phase of AGN feedback.

We speculate that the directionality of gas inflows may play a role in shaping galaxy morphology and star formation.  In particular, inflows aligned with the disk could potentially lead to size growth and clumpy disk structures, whereas misaligned or perpendicular inflows might instead trigger compact central starbursts. We note, however, that this interpretation is speculative, as we do not have direct evidence for the direction of gas inflows in our sample.
As these DSFGs gradually migrate toward the protocluster center, their stellar cores are expected to grow further through sustained gas accretion, eventually giving rise to bulge-dominated systems \citep{Tan2024b, Han2024}. On the other hand, enhanced AGN feedback coupled with gas overconsumption may begin to suppress their star formation activities and drive them toward quenching \citep{Perez2025,Remus2025,Kimmig2025}. Moreover, the elevated galaxy density near the protocluster center increases the likelihood of interactions such as mergers and ram-pressure stripping \citep{Dannerbauer2017}, processes that can disrupt disk structures and facilitate the transformation of DSFGs into massive elliptical galaxies, as observed in a forming clusters at similar redshift \citep{XuK2025}.

\section{Summary}
\label{sec: summary}

In this paper, we present JWST/NIRCam imaging of ALMA-detected DSFGs within the Spiderweb protocluster at $z=2.16$. Three filter, F115W, F182M, and F410, which correspond to rest-frame $U$, $V$, and $J$ bands at the redshift of the protocluster, will be used for our analysis. Our main results are summarized as follows:

\begin{enumerate}[]
	
	\item Within the JWST footprint, 30 sources from the ALMA main catalog are covered, of which 22 are successfully detected by JWST, yielding a detection rate of 73\%. By cross-matching with catalogs of various source populations, we identify 10 JWST-detected ALMA sources as confirmed members of the protocluster, including the central Spiderweb radio galaxy.
	
	\item Most of the DSFG members appear very red in both color-composite images and the rest-frame $UVJ$ diagram, highlighting the heavily dust-obscured nature of these ALMA-detected galaxies. Non-parametric morphological analysis places the majority of DSFGs within the late-type disk region of the Gini–$M_{20}$ plane. Median morphological measurements shift toward the merger regime from redder to bluer filters, indicating that the young stellar populations traced by rest-frame UV emission are more clumpy and disturbed.
	
	\item Parametric morphological analysis indicates that DSFG members in the Spiderweb protocluster host larger stellar disks with comparable Sérsic indices relative to coeval DSFG samples. Their sizes lie systematically above the typical size–mass relation in both rest-frame optical and NIR wavelengths, pointing to an accelerated size evolution in overdense environments. Furthermore, we identify a downsizing trend in stellar components from rest-frame UV to NIR, with stellar sizes exceeding those of the dust components traced by ALMA. 
	While our results generally align with recent studies \citep{Gillman2024, Hodge2025}, the moderately steep gradient seen in our DSFG sample may indicate enhanced outer-disk star formation. Yet, this conclusion remains tentative due to the limited sample size and possible selection differences, and will require confirmation with larger, mass-complete datasets.
	
	\item The extended stellar disks observed in these DSFGs may arise from the smooth accretion of cold gas streams with coherent angular momentum from surrounding large-scale filaments, suggesting a secular pathway in their size and mass growth. 
	As these DSFGs migrate toward the protocluster center, We speculate that they may undergo bulge growth possibly driven by counter-rotating or perpendicular cold gas accretion, and could also be affected by AGN feedback, mergers, and ram-pressure stripping \citep{XuK2025}, eventually evolving into massive elliptical galaxies similar to those in the local Universe.

\end{enumerate}

Future JWST/NIRISS and ALMA spectroscopic observations will help confirm additional dusty members in the Spiderweb protocluster and refine our understanding of their morphologies and spatial distributions, further testing environmental effects. 
The combination with Ly$\alpha$ three-dimensional mapping may help clarify how cosmic gas accretion shapes galaxy morphology in the Spiderweb protocluster as a function of radial distance from the central galaxy.

\begin{acknowledgements}

This research is based on observations made with the NASA/ESA James Webb Space Telescope obtained from the Space Telescope Science Institute, which is operated by the Association of Universities for Research in Astronomy, Inc., under NASA contract NAS 5–26555. These observations are associated with the JWST/NIRCam program ID\#1572, PI: H. Dannerbauer. NIRCam was built by a team at the University of Arizona (UofA) and Lockheed Martin’s Advanced Technology Center, led by Prof. Marcia Rieke at UoA.
This paper makes use of the following ALMA data: ADS/JAO.ALMA\#2021.1.00435.S ALMA is a partnership of ESO (representing its member states), NSF (USA) and NINS (Japan), together with NRC (Canada), MOST and ASIAA (Taiwan), and KASI (Republic of Korea), in cooperation with the Republic of Chile. The Joint ALMA Observatory is operated by ESO, AUI/NRAO and NAOJ. 
The National Radio Astronomy Observatory is a facility of the National Science Foundation operated under cooperative agreement by Associated Universities, Inc. 
This work is supported by the National Key Research and Development Program of China (2023YFA1608100); the National Science Foundation of China (NSFC, Grant No. 12233005); the China Manned Space Program (Grant Nos. CMS-CSST-2025-A08 and CMS-CSST-2025-A20); and the Office of Science and Technology, Shanghai Municipal Government (Grant Nos. 24DX1400100, ZJ2023-ZD-001).
YZ acknowledges the support from the China Scholarship Council (202206340048), and the National Science Foundation of Jiangsu Province (BK20231106). 
HD, YZ, JMPM and ZC acknowledges ﬁnancial supports from the Agencia Estatal de Investigación del Ministerio de Ciencia e Innovación (AEIMCINN) under grant (La evolución de los cúmulos de galaxias desde el amanecer hasta el mediodía cósmico) with reference (PID2019105776GB-I00/DOI:10.13039/501100011033). 
In addition, HD, YZ and JMPM acknowledge support from the Agencia Estatal de Investigación del Ministerio de Ciencia, Innovación y Universidades (MCIU/AEI) under grant (Construcción de cúmulos de galaxias en formación a través de la formación estelar oscurecida por el polvo) and the European Regional Development Fund (ERDF) with reference (PID2022-143243NB-I00/10.13039/501100011033).
JMPM acknowledges funding from the European Union’s Horizon-Europe research and innovation programme under the Marie Sklodowska-Curie grant agreement No 101106626.
YK, and TK acknowledge support from JSPS KAKENHI Grant Number 23H01219.
KD acknowledges financial support from JSPS KAKENHI Grant Numbers 25K23411.
CDE acknowledges funding from the MCIN/AEI (Spain) and the “NextGenerationEU”/PRTR (European Union) through the Juan de la Cierva-Formación program (FJC2021-047307-I), as well as funding from LabEx UnivEarthS project N9.
TK acknowledges financial support from JSPS KAKENHI Grant Numbers 24H00002 (Specially Promoted Research by T. Kodama et al.), 22K21349 (International Leading Research by S. Miyazaki et al.), and JSPS Core-to-Core Program (JPJSCCA20210003; M. Yoshida et al.).
J.N.~acknowledge the support of the National Science Centre, Poland through the SONATA BIS grant 2018/30/E/ST9/00208. 
J.N.~acknowledges the support of the Polish National Agency for Academic Exchange (NAWA) Bekker grant BPN/BEK/2023/1/00271, and the kind hospitality of the IAC.
PGP-G acknowledges support from grant PID2022-139567NB-I00 funded by Spanish Ministerio de Ciencia, Innovaci\'on y Universidades MICIU/AEI/10.13039/501100011033, and the European Union FEDER program {\it Una manera de hacer Europa}. 
\end{acknowledgements}

\bibliographystyle{aa}
\bibliography{aa}

@ARTICLE{SunH2024,
	author = {{Sun}, Hanwen and {Wang}, Tao and {Xu}, Ke and {Daddi}, Emanuele and {Gu}, Qing and {Kodama}, Tadayuki and {Zanella}, Anita and {Elbaz}, David and {Tanaka}, Ichi and {Gobat}, Raphael and {Guo}, Qi and {Han}, Jiaxin and {Lu}, Shiying and {Zhou}, Luwenjia},
	title = "{JWST's First Glimpse of a z > 2 Forming Cluster Reveals a Top-heavy Stellar Mass Function}",
	journal = {\apjl},
	year = 2024,
	month = jun,
	volume = {967},
	number = {2},
	eid = {L34},
	pages = {L34},
	doi = {10.3847/2041-8213/ad4986},
	archivePrefix = {arXiv},
	eprint = {2403.05248},
	primaryClass = {astro-ph.GA},
	adsurl = {https://ui.adsabs.harvard.edu/abs/2024ApJ...967L..34S}
}

@ARTICLE{Chabrier03,
       author = {{Chabrier}, Gilles},
        title = "{Galactic Stellar and Substellar Initial Mass Function}",
      journal = {\pasp},
         year = 2003,
        month = jul,
       volume = {115},
       number = {809},
        pages = {763-795},
          doi = {10.1086/376392},
archivePrefix = {arXiv},
       eprint = {astro-ph/0304382},
 primaryClass = {astro-ph},
       adsurl = {https://ui.adsabs.harvard.edu/abs/2003PASP..115..763C}
}

@ARTICLE{Kriek08,
       author = {{Kriek}, Mariska and {van der Wel}, Arjen and {van Dokkum}, Pieter G. and {Franx}, Marijn and {Illingworth}, Garth D.},
        title = "{The Detection of a Red Sequence of Massive Field Galaxies at z \raisebox{-0.5ex}\textasciitilde 2.3 and Its Evolution to z \raisebox{-0.5ex}\textasciitilde 0}",
      journal = {\apj},
         year = 2008,
        month = aug,
       volume = {682},
       number = {2},
        pages = {896-906},
          doi = {10.1086/589677},
archivePrefix = {arXiv},
       eprint = {0804.4175},
 primaryClass = {astro-ph},
       adsurl = {https://ui.adsabs.harvard.edu/abs/2008ApJ...682..896K}
}

@ARTICLE{Lepore2024,
	author = {{Lepore}, M. and {Di Mascolo}, L. and {Tozzi}, P. and {Churazov}, E. and {Mroczkowski}, T. and {Borgani}, S. and {Carilli}, C. and {Gaspari}, M. and {Ginolfi}, M. and {Liu}, A. and {Pentericci}, L. and {Rasia}, E. and {Rosati}, P. and {R{\"o}ttgering}, H.~J.~A. and {Anderson}, C.~S. and {Dannerbauer}, H. and {Miley}, G. and {Norman}, C.},
	title = "{Feeding and feedback processes in the Spiderweb proto-intracluster medium}",
	journal = {\aap},
	year = 2024,
	month = feb,
	volume = {682},
	eid = {A186},
	pages = {A186},
	doi = {10.1051/0004-6361/202347538},
	archivePrefix = {arXiv},
	eprint = {2312.06392},
	primaryClass = {astro-ph.GA},
	adsurl = {https://ui.adsabs.harvard.edu/abs/2024A&A...682A.186L}
}

@ARTICLE{Ceverino2010,
	author = {{Ceverino}, Daniel and {Dekel}, Avishai and {Bournaud}, Frederic},
	title = "{High-redshift clumpy discs and bulges in cosmological simulations}",
	journal = {\mnras},
	year = 2010,
	month = jun,
	volume = {404},
	number = {4},
	pages = {2151-2169},
	doi = {10.1111/j.1365-2966.2010.16433.x},
	archivePrefix = {arXiv},
	eprint = {0907.3271},
	primaryClass = {astro-ph.CO},
	adsurl = {https://ui.adsabs.harvard.edu/abs/2010MNRAS.404.2151C}
}

@ARTICLE{XuK2025,
	author = {{Xu}, Ke and {Wang}, Tao and {Daddi}, Emanuele and {Elbaz}, David and {Sun}, Hanwen and {Chen}, Longyue and {Gobat}, Raphael and {Zanella}, Anita and {Liu}, Daizhong and {Xiao}, Mengyuan and {Cen}, Renyue and {Kodama}, Tadayuki and {Kohno}, Kotaro and {Yang}, Tiancheng and {Zhang}, Zhi-Yu and {Zhou}, Luwenjia and {Valentino}, Francesco},
	title = "{Ram-pressure stripping caught in action in a forming galaxy cluster 3 billion years after the Big Bang}",
	journal = {arXiv e-prints},
	year = 2025,
	month = mar,
	eid = {arXiv:2503.21724},
	pages = {arXiv:2503.21724},
	doi = {10.48550/arXiv.2503.21724},
	archivePrefix = {arXiv},
	eprint = {2503.21724},
	primaryClass = {astro-ph.GA},
	adsurl = {https://ui.adsabs.harvard.edu/abs/2025arXiv250321724X}
}

@ARTICLE{Remus2025,
	author = {{Remus}, Rhea-Silvia and {Kimmig}, Lucas C.},
	title = "{Relight the Candle: What Happens to High-redshift Massive Quenched Galaxies}",
	journal = {\apj},
	year = 2025,
	month = mar,
	volume = {982},
	number = {1},
	eid = {30},
	pages = {30},
	doi = {10.3847/1538-4357/ad8b4b},
	archivePrefix = {arXiv},
	eprint = {2310.16089},
	primaryClass = {astro-ph.GA},
	adsurl = {https://ui.adsabs.harvard.edu/abs/2025ApJ...982...30R}
}

@ARTICLE{Kimmig2025,
	author = {{Kimmig}, Lucas C. and {Remus}, Rhea-Silvia and {Seidel}, Benjamin and {Valenzuela}, Lucas M. and {Dolag}, Klaus and {Burkert}, Andreas},
	title = "{Blowing Out the Candle: How to Quench Galaxies at High Redshift{\textemdash}An Ensemble of Rapid Starbursts, AGN Feedback, and Environment}",
	journal = {\apj},
	year = 2025,
	month = jan,
	volume = {979},
	number = {1},
	eid = {15},
	pages = {15},
	doi = {10.3847/1538-4357/ad9472},
	archivePrefix = {arXiv},
	eprint = {2310.16085},
	primaryClass = {astro-ph.GA},
	adsurl = {https://ui.adsabs.harvard.edu/abs/2025ApJ...979...15K}
}

@ARTICLE{Perez2025,
	author = {{P{\'e}rez-Mart{\'\i}nez}, J.~M. and {Dannerbauer}, H. and {Emonts}, B.~H.~C. and {Allison}, J.~R. and {Champagne}, J.~B. and {Indermuehle}, B. and {Norris}, R.~P. and {Serra}, P. and {Seymour}, N. and {Thomson}, A.~P. and {Casey}, C.~M. and {Chen}, Z. and {Daikuhara}, K. and {De Breuck}, C. and {D'Eugenio}, C. and {Drouart}, G. and {Hatch}, N. and {Jin}, S. and {Kodama}, T. and {Koyama}, Y. and {Lehnert}, M.~D. and {Macgregor}, P. and {Miley}, G. and {Naufal}, A. and {R{\"o}ttgering}, H. and {S{\'a}nchez-Portal}, M. and {Shimakawa}, R. and {Zhang}, Y. and {Ziegler}, B.},
	title = "{COALAS: III. The ATCA CO(1{\textendash}0) look at the growth and death of H{\ensuremath{\alpha}} emitters in the Spiderweb protocluster at z = 2.16}",
	journal = {\aap},
	year = 2025,
	month = apr,
	volume = {696},
	eid = {A236},
	pages = {A236},
	doi = {10.1051/0004-6361/202450785},
	archivePrefix = {arXiv},
	eprint = {2411.12138},
	primaryClass = {astro-ph.GA},
	adsurl = {https://ui.adsabs.harvard.edu/abs/2025A&A...696A.236P}
}

@ARTICLE{Tanaka2013,
	author = {{Tanaka}, Masayuki and {Toft}, Sune and {Marchesini}, Danilo and {Zirm}, Andrew and {De Breuck}, Carlos and {Kodama}, Tadayuki and {Koyama}, Yusei and {Kurk}, Jaron and {Tanaka}, Ichi},
	title = "{On the Formation Timescale of Massive Cluster Ellipticals Based on Deep Near-infrared Spectroscopy at z \raisebox{-0.5ex}\textasciitilde 2}",
	journal = {\apj},
	year = 2013,
	month = aug,
	volume = {772},
	number = {2},
	eid = {113},
	pages = {113},
	doi = {10.1088/0004-637X/772/2/113},
	archivePrefix = {arXiv},
	eprint = {1306.4406},
	primaryClass = {astro-ph.CO},
	adsurl = {https://ui.adsabs.harvard.edu/abs/2013ApJ...772..113T}
}

@ARTICLE{Birkin2024,
	author = {{Birkin}, Jack E. and {Puglisi}, A. and {Swinbank}, A.~M. and {Smail}, Ian and {An}, Fang Xia and {Chapman}, S.~C. and {Chen}, Chian-Chou and {Conselice}, C.~J. and {Dudzevi{\v{c}}i{\={u}}t{\.{e}}}, U. and {Farrah}, D. and {Gullberg}, B. and {Matsuda}, Y. and {Schinnerer}, E. and {Scott}, D. and {Wardlow}, J.~L. and {van der Werf}, P.},
	title = "{KAOSS: turbulent, but disc-like kinematics in dust-obscured star-forming galaxies at z   1.3-2.6}",
	journal = {\mnras},
	year = 2024,
	month = jun,
	volume = {531},
	number = {1},
	pages = {61-83},
	doi = {10.1093/mnras/stae1089},
	archivePrefix = {arXiv},
	eprint = {2301.05720},
	primaryClass = {astro-ph.GA},
	adsurl = {https://ui.adsabs.harvard.edu/abs/2024MNRAS.531...61B}
}

@ARTICLE{SunF2025,
	author = {{Sun}, Fengwu and {Yang}, Jinyi and {Wang}, Feige and {Eisenstein}, Daniel J. and {Decarli}, Roberto and {Fan}, Xiaohui and {Rieke}, George H. and {Ba{\~n}ados}, Eduardo and {Bosman}, Sarah E.~I. and {Cai}, Zheng and {Champagne}, Jaclyn B. and {Colina}, Luis and {D'Eugenio}, Francesco and {Fudamoto}, Yoshinobu and {Li}, Mingyu and {Lin}, Xiaojing and {Liu}, Weizhe and {Lyu}, Jianwei and {Mazzucchelli}, Chiara and {Jin}, Xiangyu and {Jun}, Hyunsung D. and {Wu}, Yunjing and {Zhang}, Huanian},
	title = "{The Identification of Two JWST/NIRCam-Dark Starburst Galaxies at $z=6.6$ with ALMA}",
	journal = {arXiv e-prints},
	year = 2025,
	month = jun,
	eid = {arXiv:2506.06418},
	pages = {arXiv:2506.06418},
	doi = {10.48550/arXiv.2506.06418},
	archivePrefix = {arXiv},
	eprint = {2506.06418},
	primaryClass = {astro-ph.GA},
	adsurl = {https://ui.adsabs.harvard.edu/abs/2025arXiv250606418S}
}

@ARTICLE{Esposito2025,
	author = {{Esposito}, Michela and {Borgani}, Stefano and {Strazzullo}, Veronica and {Pannella}, Maurilio and {Granato}, Gian Luigi and {Ragone-Figueroa}, Cinthia and {Saro}, Alex and {Nonino}, Mario and {Valentini}, Milena},
	title = "{Galaxy populations of protoclusters in cosmological hydrodynamical simulations}",
	journal = {\aap},
	year = 2025,
	month = may,
	volume = {697},
	eid = {A142},
	pages = {A142},
	doi = {10.1051/0004-6361/202453507},
	adsurl = {https://ui.adsabs.harvard.edu/abs/2025A&A...697A.142E}
}

@ARTICLE{EspejoSalcedo2025,
	author = {{Espejo Salcedo}, J.~M. and {Pastras}, S. and {V{\'a}cha}, J. and {Pulsoni}, C. and {Genzel}, R. and {F{\"o}rster Schreiber}, N.~M. and {Jolly}, J.-B. and {Barfety}, C. and {Chen}, J. and {Tozzi}, G. and {Liu}, D. and {Lee}, L.~L. and {Wuyts}, S. and {Tacconi}, L.~J. and {Davies}, R. and {{\"U}bler}, H. and {Lutz}, D. and {Wisnioski}, E. and {Shangguan}, J. and {Lee}, M. and {Price}, S.~H. and {Eisenhauer}, F. and {Renzini}, A. and {Nestor Shachar}, A. and {Herrera-Camus}, R.},
	title = "{Galaxy morphologies at cosmic noon with JWST: A foundation for exploring gas transport with bars and spiral arms}",
	journal = {\aap},
	year = 2025,
	month = aug,
	volume = {700},
	eid = {A42},
	pages = {A42},
	doi = {10.1051/0004-6361/202554725},
	archivePrefix = {arXiv},
	eprint = {2503.21738},
	primaryClass = {astro-ph.GA},
	adsurl = {https://ui.adsabs.harvard.edu/abs/2025A&A...700A..42E}
}

@ARTICLE{Umehata2025b,
	author = {{Umehata}, Hideki and {Kubo}, Mariko and {Smail}, Ian and {Lehmer}, Bret D. and {Monson}, Erik B. and {Nakanishi}, Kouichiro and {Matsuda}, Yuichi},
	title = "{ADF22-WEB: ALMA and JWST (sub)kpc-scale views of dusty star-forming galaxies in a $z\approx$3 proto-cluster}",
	journal = {arXiv e-prints},
	year = 2025,
	month = feb,
	eid = {arXiv:2502.01868},
	pages = {arXiv:2502.01868},
	doi = {10.48550/arXiv.2502.01868},
	archivePrefix = {arXiv},
	eprint = {2502.01868},
	primaryClass = {astro-ph.GA},
	adsurl = {https://ui.adsabs.harvard.edu/abs/2025arXiv250201868U}
}

@ARTICLE{Huang2025,
	author = {{Huang}, Shuo and {Umehata}, Hideki and {Smail}, Ian and {Nakanishi}, Kouichiro and {Hatsukade}, Bunyo and {Kubo}, Mariko and {Tamura}, Yoichi and {Saito}, Tomoki and {Ikarashi}, Soh},
	title = "{ADF22+: A declining faint end in the far-infrared luminosity function in the SSA22 protocluster at z = 3.09}",
	journal = {\aap},
	year = 2025,
	month = jul,
	volume = {699},
	eid = {A324},
	pages = {A324},
	doi = {10.1051/0004-6361/202554868},
	archivePrefix = {arXiv},
	eprint = {2503.23372},
	primaryClass = {astro-ph.GA},
	adsurl = {https://ui.adsabs.harvard.edu/abs/2025A&A...699A.324H}
}

@ARTICLE{McKay2025,
	author = {{McKay}, S.~J. and {Barger}, A.~J. and {Cowie}, L.~L. and {Nicandro Rosenthal}, M.~J.},
	title = "{The Physical Properties and Morphologies of Faint Dusty Star-forming Galaxies Identified with JWST}",
	journal = {\apj},
	year = 2025,
	month = jul,
	volume = {988},
	number = {1},
	eid = {135},
	pages = {135},
	doi = {10.3847/1538-4357/ade394},
	archivePrefix = {arXiv},
	eprint = {2503.00102},
	primaryClass = {astro-ph.GA},
	adsurl = {https://ui.adsabs.harvard.edu/abs/2025ApJ...988..135M}
}

@ARTICLE{Bell2003,
	author = {{Bell}, Eric F. and {McIntosh}, Daniel H. and {Katz}, Neal and {Weinberg}, Martin D.},
	title = "{The Optical and Near-Infrared Properties of Galaxies. I. Luminosity and Stellar Mass Functions}",
	journal = {\apjs},
	year = 2003,
	month = dec,
	volume = {149},
	number = {2},
	pages = {289-312},
	doi = {10.1086/378847},
	archivePrefix = {arXiv},
	eprint = {astro-ph/0302543},
	primaryClass = {astro-ph},
	adsurl = {https://ui.adsabs.harvard.edu/abs/2003ApJS..149..289B}
}

@ARTICLE{Ren2025,
	author = {{Ren}, Jian and {Liu}, F.~S. and {Li}, Nan and {Zhao}, Pinsong and {Cui}, Qifan and {Song}, Qi and {Li}, Yubin and {Mo}, Hao and {Yesuf}, Hassen M. and {Wang}, Weichen and {An}, Fangxia and {Zheng}, Xian Zhong},
	title = "{The Evolution of the Size and Merger Fraction of Submillimeter Galaxies across 1 < z {\ensuremath{\lesssim}} 6 as Observed by JWST}",
	journal = {\apj},
	year = 2025,
	month = apr,
	volume = {982},
	number = {2},
	eid = {200},
	pages = {200},
	doi = {10.3847/1538-4357/adb961},
	archivePrefix = {arXiv},
	eprint = {2502.15569},
	primaryClass = {astro-ph.GA},
	adsurl = {https://ui.adsabs.harvard.edu/abs/2025ApJ...982..200R}
}

@ARTICLE{Venkateshwaran2024,
	author = {{Venkateshwaran}, Aparna and {Weiss}, Axel and {Sulzenauer}, Nikolaus and {Menten}, Karl and {Aravena}, Manuel and {Chapman}, Scott C. and {Gonzalez}, Anthony and {Gururajan}, Gayathri and {Hayward}, Christopher C. and {Hill}, Ryley and {Reuter}, Cassie and {Spilker}, Justin S. and {Vieira}, Joaquin D.},
	title = "{Kinematic Analysis of z = 4.3 Galaxies in the SPT2349{\textendash}56 Protocluster Core}",
	journal = {\apj},
	year = 2024,
	month = dec,
	volume = {977},
	number = {2},
	eid = {161},
	pages = {161},
	doi = {10.3847/1538-4357/ad7bb4},
	archivePrefix = {arXiv},
	eprint = {2409.13823},
	primaryClass = {astro-ph.GA},
	adsurl = {https://ui.adsabs.harvard.edu/abs/2024ApJ...977..161V}
}

@ARTICLE{Perez2024b,
	author = {{P{\'e}rez-Mart{\'\i}nez}, Jose Manuel and {Dannerbauer}, Helmut and {Koyama}, Yusei and {P{\'e}rez-Gonz{\'a}lez}, Pablo G. and {Shimakawa}, Rhythm and {Kodama}, Tadayuki and {Zhang}, Yuheng and {Daikuhara}, Kazuki and {D'Eugenio}, Chiara and {Naufal}, Abdurrahman},
	title = "{JWST/NIRCam Pa{\ensuremath{\beta}} Narrowband Imaging Reveals Ordinary Dust Extinction for H{\ensuremath{\alpha}} Emitters within the Spiderweb Protocluster at z = 2.16}",
	journal = {\apj},
	year = 2024,
	month = dec,
	volume = {977},
	number = {1},
	eid = {74},
	pages = {74},
	doi = {10.3847/1538-4357/ad8156},
	archivePrefix = {arXiv},
	eprint = {2410.03366},
	primaryClass = {astro-ph.GA},
	adsurl = {https://ui.adsabs.harvard.edu/abs/2024ApJ...977...74P}
}

@ARTICLE{Naufal2024,
	author = {{Naufal}, Abdurrahman and {Koyama}, Yusei and {D'Eugenio}, Chiara and {Dannerbauer}, Helmut and {Shimakawa}, Rhythm and {P{\'e}rez-Mart{\'\i}nez}, Jose Manuel and {Kodama}, Tadayuki and {Zhang}, Yuheng and {Daikuhara}, Kazuki},
	title = "{Revealing the Quiescent Galaxy Population in the Spiderweb Protocluster at z = 2.16 with Deep HST/WFC3 Slitless Spectroscopy}",
	journal = {\apj},
	year = 2024,
	month = dec,
	volume = {977},
	number = {1},
	eid = {58},
	pages = {58},
	doi = {10.3847/1538-4357/ad8dcf},
	archivePrefix = {arXiv},
	eprint = {2410.16643},
	primaryClass = {astro-ph.GA},
	adsurl = {https://ui.adsabs.harvard.edu/abs/2024ApJ...977...58N}
}

@ARTICLE{McKinney2025,
	author = {{McKinney}, Jed and {Casey}, Caitlin M. and {Long}, Arianna S. and {Cooper}, Olivia R. and {Manning}, Sinclaire M. and {Franco}, Maximilien and {Akins}, Hollis and {Lambrides}, Erini and {Gammon}, Elaine and {Silva}, Camila and {Gentile}, Fabrizio and {Zavala}, Jorge A. and {Amvrosiadis}, Aristeidis and {Andika}, Irham and {Brinch}, Malte and {Champagne}, Jaclyn B. and {Chartab}, Nima and {Drakos}, Nicole E. and {Faisst}, Andreas L. and {Fujimoto}, Seiji and {Gillman}, Steven and {Gozaliasl}, Ghassem and {Greve}, Thomas R. and {Harish}, Santosh and {Hayward}, Christopher C. and {Hirschmann}, Michaela and {Ilbert}, Olivier and {Kalita}, Boris S. and {Kartaltepe}, Jeyhan S. and {Koekemoer}, Anton M. and {Kokorev}, Vasily and {Liu}, Daizhong and {Magdis}, Georgios and {McCracken}, Henry Joy and {Rhodes}, Jason and {Robertson}, Brant E. and {Talia}, Margherita and {Valentino}, Francesco and {Vijayan}, Aswin P.},
	title = "{SCUBADive. I. JWST+ALMA Analysis of 289 Submillimeter Galaxies in COSMOS-web}",
	journal = {\apj},
	year = 2025,
	month = feb,
	volume = {979},
	number = {2},
	eid = {229},
	pages = {229},
	doi = {10.3847/1538-4357/ada357},
	archivePrefix = {arXiv},
	eprint = {2408.08346},
	primaryClass = {astro-ph.GA},
	adsurl = {https://ui.adsabs.harvard.edu/abs/2025ApJ...979..229M}
}

@ARTICLE{Martorano2024,
	author = {{Martorano}, Marco and {van der Wel}, Arjen and {Baes}, Maarten and {Bell}, Eric F. and {Brammer}, Gabriel and {Franx}, Marijn and {Nersesian}, Angelos},
	title = "{The Size{\textendash}Mass Relation at Rest-frame 1.5 {\ensuremath{\mu}}m from JWST/NIRCam in the COSMOS-WEB and PRIMER-COSMOS Fields}",
	journal = {\apj},
	year = 2024,
	month = sep,
	volume = {972},
	number = {2},
	eid = {134},
	pages = {134},
	doi = {10.3847/1538-4357/ad5c6a},
	archivePrefix = {arXiv},
	eprint = {2406.17756},
	primaryClass = {astro-ph.GA},
	adsurl = {https://ui.adsabs.harvard.edu/abs/2024ApJ...972..134M}
}

@ARTICLE{Lebail2024,
	author = {{Le Bail}, Aur{\'e}lien and {Daddi}, Emanuele and {Elbaz}, David and {Dickinson}, Mark and {Giavalisco}, Mauro and {Magnelli}, Benjamin and {G{\'o}mez-Guijarro}, Carlos and {Kalita}, Boris S. and {Koekemoer}, Anton M. and {Holwerda}, Benne W. and {Bournaud}, Fr{\'e}d{\'e}ric and {de la Vega}, Alexander and {Calabr{\`o}}, Antonello and {Dekel}, Avishai and {Cheng}, Yingjie and {Bisigello}, Laura and {Franco}, Maximilien and {Costantin}, Luca and {Lucas}, Ray A. and {P{\'e}rez-Gonz{\'a}lez}, Pablo G. and {Lu}, Shiying and {Wilkins}, Stephen M. and {Arrabal Haro}, Pablo and {Bagley}, Micaela B. and {Finkelstein}, Steven L. and {Kartaltepe}, Jeyhan S. and {Papovich}, Casey and {Pirzkal}, Nor and {Yung}, L.~Y. Aaron},
	title = "{JWST/CEERS sheds light on dusty star-forming galaxies: Forming bulges, lopsidedness, and outside-in quenching at cosmic noon}",
	journal = {\aap},
	year = 2024,
	month = aug,
	volume = {688},
	eid = {A53},
	pages = {A53},
	doi = {10.1051/0004-6361/202347465},
	archivePrefix = {arXiv},
	eprint = {2307.07599},
	primaryClass = {astro-ph.GA},
	adsurl = {https://ui.adsabs.harvard.edu/abs/2024A&A...688A..53L}
}

@ARTICLE{Hodge2025,
	author = {{Hodge}, J.~A. and {da Cunha}, E. and {Kendrew}, S. and {Li}, J. and {Smail}, I. and {Westoby}, B.~A. and {Nayak}, O. and {Swinbank}, A.~M. and {Chen}, C. -C. and {Walter}, F. and {van der Werf}, P. and {Cracraft}, M. and {Battisti}, A. and {Brandt}, W.~N. and {Calistro Rivera}, G. and {Chapman}, S.~C. and {Cox}, P. and {Dannerbauer}, H. and {Decarli}, R. and {Frias Castillo}, M. and {Greve}, T.~R. and {Knudsen}, K.~K. and {Leslie}, S. and {Menten}, K.~M. and {Rybak}, M. and {Schinnerer}, E. and {Wardlow}, J.~L. and {Weiss}, A.},
	title = "{ALESS-JWST: Joint (Sub)kiloparsec JWST and ALMA Imaging of z \raisebox{-0.5ex}\textasciitilde 3 Submillimeter Galaxies Reveals Heavily Obscured Bulge Formation Events}",
	journal = {\apj},
	year = 2025,
	month = jan,
	volume = {978},
	number = {2},
	eid = {165},
	pages = {165},
	doi = {10.3847/1538-4357/ad9a52},
	archivePrefix = {arXiv},
	eprint = {2407.15846},
	primaryClass = {astro-ph.GA},
	adsurl = {https://ui.adsabs.harvard.edu/abs/2025ApJ...978..165H}
}

@ARTICLE{Crespo2024,
	author = {{Crespo G{\'o}mez}, A. and {Colina}, L. and {{\'A}lvarez-M{\'a}rquez}, J. and {Bik}, A. and {Boogaard}, L. and {{\"O}stlin}, G. and {Pei{\ss}ker}, F. and {Walter}, F. and {Labiano}, A. and {P{\'e}rez-Gonz{\'a}lez}, P.~G. and {Greve}, T.~R. and {Wright}, G. and {Alonso-Herrero}, A. and {Caputi}, K.~I. and {Costantin}, L. and {Eckart}, A. and {Garc{\'\i}a-Mar{\'\i}n}, M. and {Gillman}, S. and {Hjorth}, J. and {Iani}, E. and {Langeroodi}, D. and {Pye}, J.~P. and {Rinaldi}, P. and {Tikkanen}, T. and {van der Werf}, P. and {Lagage}, P.~O. and {van Dishoeck}, E.~F.},
	title = "{JWST/MIRI unveils the stellar component of the GN20 dusty galaxy overdensity at z = 4.05}",
	journal = {\aap},
	year = 2024,
	month = nov,
	volume = {691},
	eid = {A325},
	pages = {A325},
	doi = {10.1051/0004-6361/202449750},
	archivePrefix = {arXiv},
	eprint = {2402.18672},
	primaryClass = {astro-ph.GA},
	adsurl = {https://ui.adsabs.harvard.edu/abs/2024A&A...691A.325C}
}

@ARTICLE{Amvrosiadis2025,
	author = {{Amvrosiadis}, A. and {Wardlow}, J.~L. and {Birkin}, J.~E. and {Smail}, I. and {Swinbank}, A.~M. and {Nightingale}, J. and {Bertoldi}, F. and {Brandt}, W.~N. and {Casey}, C.~M. and {Chapman}, S.~C. and {Chen}, C. -C. and {Cox}, P. and {da Cunha}, E. and {Dannerbauer}, H. and {Dudzevi{\v{c}}i{\={u}}t{\.{e}}}, U. and {Gullberg}, B. and {Hodge}, J.~A. and {Knudsen}, K.~K. and {Menten}, K. and {Walter}, F. and {van der Werf}, P.},
	title = "{The kinematics of massive high-redshift dusty star-forming galaxies}",
	journal = {\mnras},
	year = 2025,
	month = feb,
	volume = {536},
	number = {4},
	pages = {3757-3783},
	doi = {10.1093/mnras/stae2760},
	archivePrefix = {arXiv},
	eprint = {2312.08959},
	primaryClass = {astro-ph.GA},
	adsurl = {https://ui.adsabs.harvard.edu/abs/2025MNRAS.536.3757A}
}

@ARTICLE{Daikuhara2024,
	author = {{Daikuhara}, Kazuki and {Kodama}, Tadayuki and {P{\'e}rez-Mart{\'\i}nez}, Jose M. and {Shimakawa}, Rhythm and {Suzuki}, Tomoko L. and {Tadaki}, Ken-ichi and {Koyama}, Yusei and {Tanaka}, Ichi},
	title = "{Star-formation activity of low-mass galaxies at the peak epoch of galaxy formation probed by deep narrow-band imaging}",
	journal = {\mnras},
	year = 2024,
	month = jun,
	volume = {531},
	number = {2},
	pages = {2335-2355},
	doi = {10.1093/mnras/stae1243},
	archivePrefix = {arXiv},
	eprint = {2405.20663},
	primaryClass = {astro-ph.GA},
	adsurl = {https://ui.adsabs.harvard.edu/abs/2024MNRAS.531.2335D}
}

@ARTICLE{Han2024,
	author = {{Han}, Seongbong and {Jang}, J.~K. and {Contini}, Emanuele and {Dubois}, Yohan and {Jeon}, Seyoung and {Kaviraj}, Sugata and {Kimm}, Taysun and {Kraljic}, Katarina and {Oh}, Sree and {Peirani}, S{\'e}bastien and {Pichon}, Christophe and {Yi}, Sukyoung K.},
	title = "{Exploring Lenticular Galaxy Formation in Field Environments Using NewHorizon: Evidence for Counterrotating Gas Accretion as a Formation Channel}",
	journal = {\apj},
	year = 2024,
	month = dec,
	volume = {977},
	number = {1},
	eid = {116},
	pages = {116},
	doi = {10.3847/1538-4357/ad8ba7},
	archivePrefix = {arXiv},
	eprint = {2411.05910},
	primaryClass = {astro-ph.GA},
	adsurl = {https://ui.adsabs.harvard.edu/abs/2024ApJ...977..116H}
}

@ARTICLE{Simpson2017,
	author = {{Simpson}, J.~M. and {Smail}, Ian and {Swinbank}, A.~M. and {Ivison}, R.~J. and {Dunlop}, J.~S. and {Geach}, J.~E. and {Almaini}, O. and {Arumugam}, V. and {Bremer}, M.~N. and {Chen}, Chian-Chou and {Conselice}, C. and {Coppin}, K.~E.~K. and {Farrah}, D. and {Ibar}, E. and {Hartley}, W.~G. and {Ma}, C.~J. and {Micha{\l}owski}, M.~J. and {Scott}, D. and {Spaans}, M. and {Thomson}, A.~P. and {van der Werf}, P.~P.},
	title = "{The SCUBA-2 Cosmology Legacy Survey: Multi-wavelength Properties of ALMA-identified Submillimeter Galaxies in UKIDSS UDS}",
	journal = {\apj},
	year = 2017,
	month = apr,
	volume = {839},
	number = {1},
	eid = {58},
	pages = {58},
	doi = {10.3847/1538-4357/aa65d0},
	archivePrefix = {arXiv},
	eprint = {1611.03084},
	primaryClass = {astro-ph.GA},
	adsurl = {https://ui.adsabs.harvard.edu/abs/2017ApJ...839...58S}
}

@ARTICLE{Targett2013,
	author = {{Targett}, T.~A. and {Dunlop}, J.~S. and {Cirasuolo}, M. and {McLure}, R.~J. and {Bruce}, V.~A. and {Fontana}, A. and {Galametz}, A. and {Paris}, D. and {Dav{\'e}}, R. and {Dekel}, A. and {Faber}, S.~M. and {Ferguson}, H.~C. and {Grogin}, N.~A. and {Kartaltepe}, J.~S. and {Kocevski}, D.~D. and {Koekemoer}, A.~M. and {Kurczynski}, P. and {Lai}, K. and {Lotz}, J.},
	title = "{The properties of (sub-)millimetre-selected galaxies as revealed by CANDELS HST WFC3/IR imaging in GOODS-South}",
	journal = {\mnras},
	year = 2013,
	month = jul,
	volume = {432},
	number = {3},
	pages = {2012-2042},
	doi = {10.1093/mnras/stt482},
	archivePrefix = {arXiv},
	eprint = {1208.3464},
	primaryClass = {astro-ph.CO},
	adsurl = {https://ui.adsabs.harvard.edu/abs/2013MNRAS.432.2012T}
}

@ARTICLE{Hainline2011,
	author = {{Hainline}, Laura J. and {Blain}, A.~W. and {Smail}, Ian and {Alexander}, D.~M. and {Armus}, L. and {Chapman}, S.~C. and {Ivison}, R.~J.},
	title = "{The Stellar Mass Content of Submillimeter-selected Galaxies}",
	journal = {\apj},
	year = 2011,
	month = oct,
	volume = {740},
	number = {2},
	eid = {96},
	pages = {96},
	doi = {10.1088/0004-637X/740/2/96},
	archivePrefix = {arXiv},
	eprint = {1006.0238},
	primaryClass = {astro-ph.CO},
	adsurl = {https://ui.adsabs.harvard.edu/abs/2011ApJ...740...96H}
}

@ARTICLE{Cowie2018,
	author = {{Cowie}, L.~L. and {Gonz{\'a}lez-L{\'o}pez}, J. and {Barger}, A.~J. and {Bauer}, F.~E. and {Hsu}, L. -Y. and {Wang}, W. -H.},
	title = "{A Submillimeter Perspective on the GOODS Fields (SUPER GOODS). III. A Large Sample of ALMA Sources in the GOODS-S}",
	journal = {\apj},
	year = 2018,
	month = oct,
	volume = {865},
	number = {2},
	eid = {106},
	pages = {106},
	doi = {10.3847/1538-4357/aadc63},
	archivePrefix = {arXiv},
	eprint = {1805.09424},
	primaryClass = {astro-ph.GA},
	adsurl = {https://ui.adsabs.harvard.edu/abs/2018ApJ...865..106C}
}

@ARTICLE{Gomez2018,
	author = {{G{\'o}mez-Guijarro}, C. and {Toft}, S. and {Karim}, A. and {Magnelli}, B. and {Magdis}, G.~E. and {Jim{\'e}nez-Andrade}, E.~F. and {Capak}, P.~L. and {Fraternali}, F. and {Fujimoto}, S. and {Riechers}, D.~A. and {Schinnerer}, E. and {Smol{\v{c}}i{\'c}}, V. and {Aravena}, M. and {Bertoldi}, F. and {Cortzen}, I. and {Hasinger}, G. and {Hu}, E.~M. and {Jones}, G.~C. and {Koekemoer}, A.~M. and {Lee}, N. and {McCracken}, H.~J. and {Micha{\l}owski}, M.~J. and {Navarrete}, F. and {Povi{\'c}}, M. and {Puglisi}, A. and {Romano-D{\'\i}az}, E. and {Sheth}, K. and {Silverman}, J.~D. and {Staguhn}, J. and {Steinhardt}, C.~L. and {Stockmann}, M. and {Tanaka}, M. and {Valentino}, F. and {van Kampen}, E. and {Zirm}, A.},
	title = "{Starburst to Quiescent from HST/ALMA: Stars and Dust Unveil Minor Mergers in Submillimeter Galaxies at z {\ensuremath{\sim}} 4.5}",
	journal = {\apj},
	year = 2018,
	month = apr,
	volume = {856},
	number = {2},
	eid = {121},
	pages = {121},
	doi = {10.3847/1538-4357/aab206},
	archivePrefix = {arXiv},
	eprint = {1802.07751},
	primaryClass = {astro-ph.GA},
	adsurl = {https://ui.adsabs.harvard.edu/abs/2018ApJ...856..121G}
}

@ARTICLE{Swinbank2010,
	author = {{Swinbank}, A.~M. and {Smail}, Ian and {Chapman}, S.~C. and {Borys}, C. and {Alexander}, D.~M. and {Blain}, A.~W. and {Conselice}, C.~J. and {Hainline}, L.~J. and {Ivison}, R.~J.},
	title = "{A Hubble Space Telescope NICMOS and ACS morphological study of z \raisebox{-0.5ex}\textasciitilde 2 submillimetre galaxies}",
	journal = {\mnras},
	year = 2010,
	month = jun,
	volume = {405},
	number = {1},
	pages = {234-244},
	doi = {10.1111/j.1365-2966.2010.16485.x},
	archivePrefix = {arXiv},
	eprint = {1002.2518},
	primaryClass = {astro-ph.CO},
	adsurl = {https://ui.adsabs.harvard.edu/abs/2010MNRAS.405..234S}
}

@ARTICLE{Hodge2016,
	author = {{Hodge}, J.~A. and {Swinbank}, A.~M. and {Simpson}, J.~M. and {Smail}, I. and {Walter}, F. and {Alexander}, D.~M. and {Bertoldi}, F. and {Biggs}, A.~D. and {Brandt}, W.~N. and {Chapman}, S.~C. and {Chen}, C.~C. and {Coppin}, K.~E.~K. and {Cox}, P. and {Dannerbauer}, H. and {Edge}, A.~C. and {Greve}, T.~R. and {Ivison}, R.~J. and {Karim}, A. and {Knudsen}, K.~K. and {Menten}, K.~M. and {Rix}, H. -W. and {Schinnerer}, E. and {Wardlow}, J.~L. and {Weiss}, A. and {van der Werf}, P.},
	title = "{Kiloparsec-scale Dust Disks in High-redshift Luminous Submillimeter Galaxies}",
	journal = {\apj},
	year = 2016,
	month = dec,
	volume = {833},
	number = {1},
	eid = {103},
	pages = {103},
	doi = {10.3847/1538-4357/833/1/103},
	archivePrefix = {arXiv},
	eprint = {1609.09649},
	primaryClass = {astro-ph.GA},
	adsurl = {https://ui.adsabs.harvard.edu/abs/2016ApJ...833..103H}
}

@ARTICLE{Rizzo2021,
	author = {{Rizzo}, Francesca and {Vegetti}, Simona and {Fraternali}, Filippo and {Stacey}, Hannah R. and {Powell}, Devon},
	title = "{Dynamical properties of z  4.5 dusty star-forming galaxies and their connection with local early-type galaxies}",
	journal = {\mnras},
	year = 2021,
	month = nov,
	volume = {507},
	number = {3},
	pages = {3952-3984},
	doi = {10.1093/mnras/stab2295},
	archivePrefix = {arXiv},
	eprint = {2102.05671},
	primaryClass = {astro-ph.GA},
	adsurl = {https://ui.adsabs.harvard.edu/abs/2021MNRAS.507.3952R}
}

@ARTICLE{Cochrane2021,
	author = {{Cochrane}, R.~K. and {Best}, P.~N. and {Smail}, I. and {Ibar}, E. and {Cheng}, C. and {Swinbank}, A.~M. and {Molina}, J. and {Sobral}, D. and {Dudzevi{\v{c}}i{\={u}}t{\.{e}}}, U.},
	title = "{Resolving a dusty, star-forming SHiZELS galaxy at z = 2.2 with HST, ALMA, and SINFONI on kiloparsec scales}",
	journal = {\mnras},
	year = 2021,
	month = may,
	volume = {503},
	number = {2},
	pages = {2622-2638},
	doi = {10.1093/mnras/stab467},
	archivePrefix = {arXiv},
	eprint = {2102.07791},
	primaryClass = {astro-ph.GA},
	adsurl = {https://ui.adsabs.harvard.edu/abs/2021MNRAS.503.2622C}
}

@ARTICLE{Lelli2021,
	author = {{Lelli}, Federico and {Di Teodoro}, Enrico M. and {Fraternali}, Filippo and {Man}, Allison W.~S. and {Zhang}, Zhi-Yu and {De Breuck}, Carlos and {Davis}, Timothy A. and {Maiolino}, Roberto},
	title = "{A massive stellar bulge in a regularly rotating galaxy 1.2 billion years after the Big Bang}",
	journal = {Science},
	year = 2021,
	month = feb,
	volume = {371},
	number = {6530},
	pages = {713-716},
	doi = {10.1126/science.abc1893},
	archivePrefix = {arXiv},
	eprint = {2102.05957},
	primaryClass = {astro-ph.GA},
	adsurl = {https://ui.adsabs.harvard.edu/abs/2021Sci...371..713L}
}

@ARTICLE{Chen2020,
	author = {{Chen}, Chian-Chou and {Harrison}, C.~M. and {Smail}, I. and {Swinbank}, A.~M. and {Turner}, O.~J. and {Wardlow}, J.~L. and {Brandt}, W.~N. and {Calistro Rivera}, G. and {Chapman}, S.~C. and {Cooke}, E.~A. and {Dannerbauer}, H. and {Dunlop}, J.~S. and {Farrah}, D. and {Micha{\l}owski}, M.~J. and {Schinnerer}, E. and {Simpson}, J.~M. and {Thomson}, A.~P. and {van der Werf}, P.~P.},
	title = "{Extended H{\ensuremath{\alpha}} over compact far-infrared continuum in dusty submillimeter galaxies. Insights into dust distributions and star-formation rates at z {\ensuremath{\sim}} 2}",
	journal = {\aap},
	year = 2020,
	month = mar,
	volume = {635},
	eid = {A119},
	pages = {A119},
	doi = {10.1051/0004-6361/201936286},
	archivePrefix = {arXiv},
	eprint = {2002.03545},
	primaryClass = {astro-ph.GA},
	adsurl = {https://ui.adsabs.harvard.edu/abs/2020A&A...635A.119C}
}

@ARTICLE{Hodge2019,
	author = {{Hodge}, J.~A. and {Smail}, I. and {Walter}, F. and {da Cunha}, E. and {Swinbank}, A.~M. and {Rybak}, M. and {Venemans}, B. and {Brandt}, W.~N. and {Calistro Rivera}, G. and {Chapman}, S.~C. and {Chen}, Chian-Chou and {Cox}, P. and {Dannerbauer}, H. and {Decarli}, R. and {Greve}, T.~R. and {Knudsen}, K.~K. and {Menten}, K.~M. and {Schinnerer}, E. and {Simpson}, J.~M. and {van der Werf}, P. and {Wardlow}, J.~L. and {Weiss}, A.},
	title = "{ALMA Reveals Potential Evidence for Spiral Arms, Bars, and Rings in High-redshift Submillimeter Galaxies}",
	journal = {\apj},
	year = 2019,
	month = may,
	volume = {876},
	number = {2},
	eid = {130},
	pages = {130},
	doi = {10.3847/1538-4357/ab1846},
	archivePrefix = {arXiv},
	eprint = {1810.12307},
	primaryClass = {astro-ph.GA},
	adsurl = {https://ui.adsabs.harvard.edu/abs/2019ApJ...876..130H}
}

@ARTICLE{Chen2017,
	author = {{Chen}, Chian-Chou and {Hodge}, J.~A. and {Smail}, Ian and {Swinbank}, A.~M. and {Walter}, Fabian and {Simpson}, J.~M. and {Calistro Rivera}, Gabriela and {Bertoldi}, F. and {Brandt}, W.~N. and {Chapman}, S.~C. and {da Cunha}, Elisabete and {Dannerbauer}, H. and {De Breuck}, C. and {Harrison}, C.~M. and {Ivison}, R.~J. and {Karim}, A. and {Knudsen}, K.~K. and {Wardlow}, J.~L. and {Wei{\ss}}, A. and {van der Werf}, P.~P.},
	title = "{A Spatially Resolved Study of Cold Dust, Molecular Gas, H II Regions, and Stars in the z = 2.12 Submillimeter Galaxy ALESS67.1}",
	journal = {\apj},
	year = 2017,
	month = sep,
	volume = {846},
	number = {2},
	eid = {108},
	pages = {108},
	doi = {10.3847/1538-4357/aa863a},
	archivePrefix = {arXiv},
	eprint = {1708.08937},
	primaryClass = {astro-ph.GA},
	adsurl = {https://ui.adsabs.harvard.edu/abs/2017ApJ...846..108C}
}

@ARTICLE{Hodge2012,
	author = {{Hodge}, J.~A. and {Carilli}, C.~L. and {Walter}, F. and {de Blok}, W.~J.~G. and {Riechers}, D. and {Daddi}, E. and {Lentati}, L.},
	title = "{Evidence for a Clumpy, Rotating Gas Disk in a Submillimeter Galaxy at z = 4}",
	journal = {\apj},
	year = 2012,
	month = nov,
	volume = {760},
	number = {1},
	eid = {11},
	pages = {11},
	doi = {10.1088/0004-637X/760/1/11},
	archivePrefix = {arXiv},
	eprint = {1209.2418},
	primaryClass = {astro-ph.CO},
	adsurl = {https://ui.adsabs.harvard.edu/abs/2012ApJ...760...11H}
}

@ARTICLE{ZhouD2024,
	author = {{Zhou}, D. and {Greve}, T.~R. and {Gullberg}, B. and {Lee}, M.~M. and {Di Mascolo}, L. and {Dicker}, S.~R. and {Romero}, C.~E. and {Chapman}, S.~C. and {Chen}, C. -C. and {Cornish}, T. and {Devlin}, M.~J. and {Ho}, L.~C. and {Kohno}, K. and {Lagos}, C.~D.~P. and {Mason}, B.~S. and {Mroczkowski}, T. and {Wagg}, J.~F.~W. and {Wang}, Q.~D. and {Wang}, R. and {Brinch}, M. and {Dannerbauer}, H. and {Jiang}, X. -J. and {Lauritsen}, L.~R.~B. and {Vijayan}, A.~P. and {Vizgan}, D. and {Wardlow}, J.~L. and {Sarazin}, C.~L. and {Sarmiento}, K.~P. and {Serjeant}, S. and {Bhandarkar}, T.~A. and {Haridas}, S.~K. and {Moravec}, E. and {Orlowski-Scherer}, J. and {Sievers}, J.~L.~R. and {Tanaka}, I. and {Wang}, Y. -J. and {Zeballos}, M. and {Laza-Ramos}, A. and {Liu}, Y. and {Hassan}, M.~S.~R. and {Jwel}, A.~K.~M. and {Nazri}, A.~A. and {Lim}, M.~K. and {Ibrahim}, U.~F.~S.~U.},
	title = "{The RAdio Galaxy Environment Reference Survey (RAGERS): Evidence of an anisotropic distribution of submillimeter galaxies in the 4C 23.56 protocluster at z = 2.48}",
	journal = {\aap},
	year = 2024,
	month = oct,
	volume = {690},
	eid = {A196},
	pages = {A196},
	doi = {10.1051/0004-6361/202348500},
	archivePrefix = {arXiv},
	eprint = {2408.02177},
	primaryClass = {astro-ph.GA},
	adsurl = {https://ui.adsabs.harvard.edu/abs/2024A&A...690A.196Z}
}

@ARTICLE{Wang2013,
	author = {{Wang}, S.~X. and {Brandt}, W.~N. and {Luo}, B. and {Smail}, I. and {Alexander}, D.~M. and {Danielson}, A.~L.~R. and {Hodge}, J.~A. and {Karim}, A. and {Lehmer}, B.~D. and {Simpson}, J.~M. and {Swinbank}, A.~M. and {Walter}, F. and {Wardlow}, J.~L. and {Xue}, Y.~Q. and {Chapman}, S.~C. and {Coppin}, K.~E.~K. and {Dannerbauer}, H. and {De Breuck}, C. and {Menten}, K.~M. and {van der Werf}, P.},
	title = "{An ALMA Survey of Submillimeter Galaxies in the Extended Chandra Deep Field-South: The AGN Fraction and X-Ray Properties of Submillimeter Galaxies}",
	journal = {\apj},
	year = 2013,
	month = dec,
	volume = {778},
	number = {2},
	eid = {179},
	pages = {179},
	doi = {10.1088/0004-637X/778/2/179},
	archivePrefix = {arXiv},
	eprint = {1310.6364},
	primaryClass = {astro-ph.CO},
	adsurl = {https://ui.adsabs.harvard.edu/abs/2013ApJ...778..179W}
}

@ARTICLE{Swinbank2014,
	author = {{Swinbank}, A.~M. and {Simpson}, J.~M. and {Smail}, Ian and {Harrison}, C.~M. and {Hodge}, J.~A. and {Karim}, A. and {Walter}, F. and {Alexander}, D.~M. and {Brandt}, W.~N. and {de Breuck}, C. and {da Cunha}, E. and {Chapman}, S.~C. and {Coppin}, K.~E.~K. and {Danielson}, A.~L.~R. and {Dannerbauer}, H. and {Decarli}, R. and {Greve}, T.~R. and {Ivison}, R.~J. and {Knudsen}, K.~K. and {Lagos}, C.~D.~P. and {Schinnerer}, E. and {Thomson}, A.~P. and {Wardlow}, J.~L. and {Wei{\ss}}, A. and {van der Werf}, P.},
	title = "{An ALMA survey of sub-millimetre Galaxies in the Extended Chandra Deep Field South: the far-infrared properties of SMGs}",
	journal = {\mnras},
	year = 2014,
	month = feb,
	volume = {438},
	number = {2},
	pages = {1267-1287},
	doi = {10.1093/mnras/stt2273},
	archivePrefix = {arXiv},
	eprint = {1310.6362},
	primaryClass = {astro-ph.CO},
	adsurl = {https://ui.adsabs.harvard.edu/abs/2014MNRAS.438.1267S}
}

@ARTICLE{Wardlow2011,
	author = {{Wardlow}, J.~L. and {Smail}, Ian and {Coppin}, K.~E.~K. and {Alexander}, D.~M. and {Brandt}, W.~N. and {Danielson}, A.~L.~R. and {Luo}, B. and {Swinbank}, A.~M. and {Walter}, F. and {Wei{\ss}}, A. and {Xue}, Y.~Q. and {Zibetti}, S. and {Bertoldi}, F. and {Biggs}, A.~D. and {Chapman}, S.~C. and {Dannerbauer}, H. and {Dunlop}, J.~S. and {Gawiser}, E. and {Ivison}, R.~J. and {Knudsen}, K.~K. and {Kov{\'a}cs}, A. and {Lacey}, C.~G. and {Menten}, K.~M. and {Padilla}, N. and {Rix}, H. -W. and {van der Werf}, P.~P.},
	title = "{The LABOCA survey of the Extended Chandra Deep Field-South: a photometric redshift survey of submillimetre galaxies}",
	journal = {\mnras},
	year = 2011,
	month = aug,
	volume = {415},
	number = {2},
	pages = {1479-1508},
	doi = {10.1111/j.1365-2966.2011.18795.x},
	archivePrefix = {arXiv},
	eprint = {1006.2137},
	primaryClass = {astro-ph.CO},
	adsurl = {https://ui.adsabs.harvard.edu/abs/2011MNRAS.415.1479W}
}

@ARTICLE{Bouwens2020,
	author = {{Bouwens}, Rychard and {Gonz{\'a}lez-L{\'o}pez}, Jorge and {Aravena}, Manuel and {Decarli}, Roberto and {Novak}, Mladen and {Stefanon}, Mauro and {Walter}, Fabian and {Boogaard}, Leindert and {Carilli}, Chris and {Dudzevi{\v{c}}i{\={u}}t{\.{e}}}, Ugn{\.{e}} and {Smail}, Ian and {Daddi}, Emanuele and {da Cunha}, Elisabete and {Ivison}, Rob and {Nanayakkara}, Themiya and {Cortes}, Paulo and {Cox}, Pierre and {Inami}, Hanae and {Oesch}, Pascal and {Popping}, Gerg{\"o} and {Riechers}, Dominik and {van der Werf}, Paul and {Weiss}, Axel and {Fudamoto}, Yoshi and {Wagg}, Jeff},
	title = "{The ALMA Spectroscopic Survey Large Program: The Infrared Excess of z = 1.5-10 UV-selected Galaxies and the Implied High-redshift Star Formation History}",
	journal = {\apj},
	year = 2020,
	month = oct,
	volume = {902},
	number = {2},
	eid = {112},
	pages = {112},
	doi = {10.3847/1538-4357/abb830},
	archivePrefix = {arXiv},
	eprint = {2009.10727},
	primaryClass = {astro-ph.GA},
	adsurl = {https://ui.adsabs.harvard.edu/abs/2020ApJ...902..112B}
}

@ARTICLE{Shimakawa2025,
	author = {{Shimakawa}, Rhythm and {Koyama}, Yusei and {Kodama}, Tadayuki and {Dannerbauer}, Helmut and {P{\'e}rez-Mart{\'\i}nez}, J.~M. and {R{\"o}ttgering}, Huub J.~A. and {Tanaka}, Ichi and {D'Eugenio}, Chiara and {Naufal}, Abdurrahman and {Daikuhara}, Kazuki and {Zhang}, Yuheng},
	title = "{Spider-Webb: JWST Near Infrared Camera resolved galaxy star formation and nuclear activities in the Spiderweb protocluster at z = 2.16}",
	journal = {\mnras},
	year = 2025,
	month = feb,
	volume = {537},
	number = {1},
	pages = {L36-L41},
	doi = {10.1093/mnrasl/slae098},
	archivePrefix = {arXiv},
	eprint = {2410.11174},
	primaryClass = {astro-ph.GA},
	adsurl = {https://ui.adsabs.harvard.edu/abs/2025MNRAS.537L..36S}
}

@ARTICLE{Dubois2014,
	author = {{Dubois}, Y. and {Pichon}, C. and {Welker}, C. and {Le Borgne}, D. and {Devriendt}, J. and {Laigle}, C. and {Codis}, S. and {Pogosyan}, D. and {Arnouts}, S. and {Benabed}, K. and {Bertin}, E. and {Blaizot}, J. and {Bouchet}, F. and {Cardoso}, J. -F. and {Colombi}, S. and {de Lapparent}, V. and {Desjacques}, V. and {Gavazzi}, R. and {Kassin}, S. and {Kimm}, T. and {McCracken}, H. and {Milliard}, B. and {Peirani}, S. and {Prunet}, S. and {Rouberol}, S. and {Silk}, J. and {Slyz}, A. and {Sousbie}, T. and {Teyssier}, R. and {Tresse}, L. and {Treyer}, M. and {Vibert}, D. and {Volonteri}, M.},
	title = "{Dancing in the dark: galactic properties trace spin swings along the cosmic web}",
	journal = {\mnras},
	year = 2014,
	month = oct,
	volume = {444},
	number = {2},
	pages = {1453-1468},
	doi = {10.1093/mnras/stu1227},
	archivePrefix = {arXiv},
	eprint = {1402.1165},
	primaryClass = {astro-ph.CO},
	adsurl = {https://ui.adsabs.harvard.edu/abs/2014MNRAS.444.1453D}
}

@ARTICLE{Aumer2013,
	author = {{Aumer}, Michael and {White}, Simon D.~M.},
	title = "{Idealized models for galactic disc formation and evolution in `realistic' {\ensuremath{\Lambda}}CDM haloes}",
	journal = {\mnras},
	year = 2013,
	month = jan,
	volume = {428},
	number = {2},
	pages = {1055-1076},
	doi = {10.1093/mnras/sts083},
	archivePrefix = {arXiv},
	eprint = {1203.1190},
	primaryClass = {astro-ph.GA},
	adsurl = {https://ui.adsabs.harvard.edu/abs/2013MNRAS.428.1055A}
}

@ARTICLE{Sales2012,
	author = {{Sales}, Laura V. and {Navarro}, Julio F. and {Theuns}, Tom and {Schaye}, Joop and {White}, Simon D.~M. and {Frenk}, Carlos S. and {Crain}, Robert A. and {Dalla Vecchia}, Claudio},
	title = "{The origin of discs and spheroids in simulated galaxies}",
	journal = {\mnras},
	year = 2012,
	month = jun,
	volume = {423},
	number = {2},
	pages = {1544-1555},
	doi = {10.1111/j.1365-2966.2012.20975.x},
	archivePrefix = {arXiv},
	eprint = {1112.2220},
	primaryClass = {astro-ph.CO},
	adsurl = {https://ui.adsabs.harvard.edu/abs/2012MNRAS.423.1544S}
}

@ARTICLE{Scannapieco2009,
	author = {{Scannapieco}, Cecilia and {White}, Simon D.~M. and {Springel}, Volker and {Tissera}, Patricia B.},
	title = "{The formation and survival of discs in a {\ensuremath{\Lambda}}CDM universe}",
	journal = {\mnras},
	year = 2009,
	month = jun,
	volume = {396},
	number = {2},
	pages = {696-708},
	doi = {10.1111/j.1365-2966.2009.14764.x},
	archivePrefix = {arXiv},
	eprint = {0812.0976},
	primaryClass = {astro-ph},
	adsurl = {https://ui.adsabs.harvard.edu/abs/2009MNRAS.396..696S}
}

@ARTICLE{Ikarashi2015,
	author = {{Ikarashi}, Soh and {Ivison}, R.~J. and {Caputi}, Karina I. and {Aretxaga}, Itziar and {Dunlop}, James S. and {Hatsukade}, Bunyo and {Hughes}, David H. and {Iono}, Daisuke and {Izumi}, Takuma and {Kawabe}, Ryohei and {Kohno}, Kotaro and {Lagos}, Claudia D.~P. and {Motohara}, Kentaro and {Nakanishi}, Kouichiro and {Ohta}, Kouji and {Tamura}, Yoichi and {Umehata}, Hideki and {Wilson}, Grant W. and {Yabe}, Kiyoto and {Yun}, Min S.},
	title = "{Compact Starbursts in z {\ensuremath{\sim}} 3-6 Submillimeter Galaxies Revealed by ALMA}",
	journal = {\apj},
	year = 2015,
	month = sep,
	volume = {810},
	number = {2},
	eid = {133},
	pages = {133},
	doi = {10.1088/0004-637X/810/2/133},
	archivePrefix = {arXiv},
	eprint = {1411.5038},
	primaryClass = {astro-ph.GA},
	adsurl = {https://ui.adsabs.harvard.edu/abs/2015ApJ...810..133I}
}

@ARTICLE{Toft2014,
	author = {{Toft}, S. and {Smol{\v{c}}i{\'c}}, V. and {Magnelli}, B. and {Karim}, A. and {Zirm}, A. and {Michalowski}, M. and {Capak}, P. and {Sheth}, K. and {Schawinski}, K. and {Krogager}, J. -K. and {Wuyts}, S. and {Sanders}, D. and {Man}, A.~W.~S. and {Lutz}, D. and {Staguhn}, J. and {Berta}, S. and {Mccracken}, H. and {Krpan}, J. and {Riechers}, D.},
	title = "{Submillimeter Galaxies as Progenitors of Compact Quiescent Galaxies}",
	journal = {\apj},
	year = 2014,
	month = feb,
	volume = {782},
	number = {2},
	eid = {68},
	pages = {68},
	doi = {10.1088/0004-637X/782/2/68},
	archivePrefix = {arXiv},
	eprint = {1401.1510},
	primaryClass = {astro-ph.GA},
	adsurl = {https://ui.adsabs.harvard.edu/abs/2014ApJ...782...68T}
}

@ARTICLE{Gonzalez2011,
	author = {{Gonz{\'a}lez}, Juan E. and {Lacey}, C.~G. and {Baugh}, C.~M. and {Frenk}, C.~S.},
	title = "{The role of submillimetre galaxies in hierarchical galaxy formation}",
	journal = {\mnras},
	year = 2011,
	month = may,
	volume = {413},
	number = {2},
	pages = {749-762},
	doi = {10.1111/j.1365-2966.2010.18169.x},
	archivePrefix = {arXiv},
	eprint = {1006.0230},
	primaryClass = {astro-ph.CO},
	adsurl = {https://ui.adsabs.harvard.edu/abs/2011MNRAS.413..749G}
}

@ARTICLE{Jian2012,
	author = {{Jian}, Hung-Yu and {Lin}, Lihwai and {Chiueh}, Tzihong},
	title = "{Environmental Dependence of the Galaxy Merger Rate in a {\ensuremath{\Lambda}}CDM Universe}",
	journal = {\apj},
	year = 2012,
	month = jul,
	volume = {754},
	number = {1},
	eid = {26},
	pages = {26},
	doi = {10.1088/0004-637X/754/1/26},
	archivePrefix = {arXiv},
	eprint = {1205.1588},
	primaryClass = {astro-ph.CO},
	adsurl = {https://ui.adsabs.harvard.edu/abs/2012ApJ...754...26J}
}

@ARTICLE{Hopkins2009,
	author = {{Hopkins}, Philip F. and {Cox}, Thomas J. and {Younger}, Joshua D. and {Hernquist}, Lars},
	title = "{How do Disks Survive Mergers?}",
	journal = {\apj},
	year = 2009,
	month = feb,
	volume = {691},
	number = {2},
	pages = {1168-1201},
	doi = {10.1088/0004-637X/691/2/1168},
	archivePrefix = {arXiv},
	eprint = {0806.1739},
	primaryClass = {astro-ph},
	adsurl = {https://ui.adsabs.harvard.edu/abs/2009ApJ...691.1168H}
}

@ARTICLE{Stewart2017,
	author = {{Stewart}, Kyle R. and {Maller}, Ariyeh H. and {O{\~n}orbe}, Jose and {Bullock}, James S. and {Joung}, M. Ryan and {Devriendt}, Julien and {Ceverino}, Daniel and {Kere{\v{s}}}, Du{\v{s}}an and {Hopkins}, Philip F. and {Faucher-Gigu{\`e}re}, Claude-Andr{\'e}},
	title = "{High Angular Momentum Halo Gas: A Feedback and Code-independent Prediction of LCDM}",
	journal = {\apj},
	year = 2017,
	month = jul,
	volume = {843},
	number = {1},
	eid = {47},
	pages = {47},
	doi = {10.3847/1538-4357/aa6dff},
	archivePrefix = {arXiv},
	eprint = {1606.08542},
	primaryClass = {astro-ph.GA},
	adsurl = {https://ui.adsabs.harvard.edu/abs/2017ApJ...843...47S}
}

@ARTICLE{McAlpine2019,
	author = {{McAlpine}, Stuart and {Smail}, Ian and {Bower}, Richard G. and {Swinbank}, A.~M. and {Trayford}, James W. and {Theuns}, Tom and {Baes}, Maarten and {Camps}, Peter and {Crain}, Robert A. and {Schaye}, Joop},
	title = "{The nature of submillimetre and highly star-forming galaxies in the EAGLE simulation}",
	journal = {\mnras},
	year = 2019,
	month = sep,
	volume = {488},
	number = {2},
	pages = {2440-2454},
	doi = {10.1093/mnras/stz1692},
	archivePrefix = {arXiv},
	eprint = {1901.05467},
	primaryClass = {astro-ph.GA},
	adsurl = {https://ui.adsabs.harvard.edu/abs/2019MNRAS.488.2440M}
}

@ARTICLE{Wang2025,
	author = {{Wang}, Weichen and {Cantalupo}, Sebastiano and {Pensabene}, Antonio and {Galbiati}, Marta and {Travascio}, Andrea and {Steidel}, Charles C. and {Maseda}, Michael V. and {Pezzulli}, Gabriele and {de Beer}, Stephanie and {Fossati}, Matteo and {Fumagalli}, Michele and {Gallego}, Sofia G. and {Lazeyras}, Titouan and {Mackenzie}, Ruari and {Matthee}, Jorryt and {Nanayakkara}, Themiya and {Quadri}, Giada},
	title = "{A giant disk galaxy two billion years after the Big Bang}",
	journal = {Nature Astronomy},
	year = 2025,
	month = may,
	volume = {9},
	pages = {710-719},
	doi = {10.1038/s41550-025-02500-2},
	archivePrefix = {arXiv},
	eprint = {2409.17956},
	primaryClass = {astro-ph.GA},
	adsurl = {https://ui.adsabs.harvard.edu/abs/2025NatAs...9..710W}
}

@ARTICLE{Umehata2025a,
	author = {{Umehata}, Hideki and {Steidel}, Charles C. and {Smail}, Ian and {Swinbank}, Mark and {Monson}, Erik B. and {Rosario}, David and {Lehmer}, Bret D. and {Nakanishi}, Kouichiro and {Kubo}, Mariko and {Iono}, Daisuke and {Alexander}, David M. and {Kohno}, Kotaro and {Tamura}, Yoichi and {Ivison}, Rob J. and {Saito}, Toshiki and {Mitsuhashi}, Ikki and {Huang}, Shuo and {Matsuda}, Yuichi},
	title = "{ADF22-WEB: A giant barred spiral starburst galaxy in the z = 3.1 SSA22 protocluster core}",
	journal = {\pasj},
	year = 2025,
	month = apr,
	volume = {77},
	number = {2},
	pages = {432-445},
	doi = {10.1093/pasj/psaf010},
	archivePrefix = {arXiv},
	eprint = {2410.22155},
	primaryClass = {astro-ph.GA},
	adsurl = {https://ui.adsabs.harvard.edu/abs/2025PASJ...77..432U}
}

@ARTICLE{Chen2015,
	author = {{Chen}, Chian-Chou and {Smail}, Ian and {Swinbank}, A.~M. and {Simpson}, J.~M. and {Ma}, Cheng-Jiun and {Alexander}, D.~M. and {Biggs}, A.~D. and {Brandt}, W.~N. and {Chapman}, S.~C. and {Coppin}, K.~E.~K. and {Danielson}, A.~L.~R. and {Dannerbauer}, H. and {Edge}, A.~C. and {Greve}, T.~R. and {Ivison}, R.~J. and {Karim}, A. and {Menten}, Karl M. and {Schinnerer}, E. and {Walter}, F. and {Wardlow}, J.~L. and {Wei{\ss}}, A. and {van der Werf}, P.~P.},
	title = "{An ALMA Survey of Submillimeter Galaxies in the Extended Chandra Deep Field South: Near-infrared Morphologies and Stellar Sizes}",
	journal = {\apj},
	year = 2015,
	month = feb,
	volume = {799},
	number = {2},
	eid = {194},
	pages = {194},
	doi = {10.1088/0004-637X/799/2/194},
	archivePrefix = {arXiv},
	eprint = {1412.0668},
	primaryClass = {astro-ph.GA},
	adsurl = {https://ui.adsabs.harvard.edu/abs/2015ApJ...799..194C}
}

@ARTICLE{Speagle2014,
       author = {{Speagle}, J.~S. and {Steinhardt}, C.~L. and {Capak}, P.~L. and {Silverman}, J.~D.},
        title = "{A Highly Consistent Framework for the Evolution of the Star-Forming ``Main Sequence'' from z \raisebox{-0.5ex}\textasciitilde 0-6}",
      journal = {\apjs},
         year = 2014,
        month = oct,
       volume = {214},
       number = {2},
          eid = {15},
        pages = {15},
          doi = {10.1088/0067-0049/214/2/15},
archivePrefix = {arXiv},
       eprint = {1405.2041},
 primaryClass = {astro-ph.GA},
       adsurl = {https://ui.adsabs.harvard.edu/abs/2014ApJS..214...15S}
}

@ARTICLE{Shimakawa2024b,
	author = {{Shimakawa}, Rhythm and {P{\'e}rez-Mart{\'\i}nez}, J.~M. and {Dannerbauer}, Helmut and {Koyama}, Yusei and {Kodama}, Tadayuki and {P{\'e}rez-Gonz{\'a}lez}, Pablo G. and {D'Eugenio}, Chiara and {Zhang}, Yuheng and {Naufal}, Abdurrahman and {Daikuhara}, Kazuki},
	title = "{JWST/NIRCam Narrowband Survey of Pa{\ensuremath{\beta}} Emitters in the Spiderweb Protocluster at z = 2.16}",
	journal = {\apj},
	year = 2024,
	month = dec,
	volume = {977},
	number = {1},
	eid = {73},
	pages = {73},
	doi = {10.3847/1538-4357/ad8155},
	archivePrefix = {arXiv},
	eprint = {2410.03362},
	primaryClass = {astro-ph.GA},
	adsurl = {https://ui.adsabs.harvard.edu/abs/2024ApJ...977...73S}
}

@ARTICLE{Zhang2024,
	author = {{Zhang}, Y.~H. and {Dannerbauer}, H. and {P{\'e}rez-Mart{\'\i}nez}, J.~M. and {Koyama}, Y. and {Zheng}, X.~Z. and {D'Eugenio}, C. and {Emonts}, B.~H.~C. and {Calvi}, R. and {Chen}, Z. and {Daikuhara}, K. and {De Breuck}, C. and {Jin}, S. and {Kodama}, T. and {Lehnert}, M.~D. and {Naufal}, A. and {Shimakawa}, R.},
	title = "{ASW$^{2}$DF: Census of the obscured star formation in a galaxy cluster in formation at z = 2.2}",
	journal = {\aap},
	year = 2024,
	month = dec,
	volume = {692},
	eid = {A22},
	pages = {A22},
	doi = {10.1051/0004-6361/202451379},
	archivePrefix = {arXiv},
	eprint = {2410.10169},
	primaryClass = {astro-ph.GA},
	adsurl = {https://ui.adsabs.harvard.edu/abs/2024A&A...692A..22Z}
}

@ARTICLE{Tadaki2020,
       author = {{Tadaki}, Ken-ichi and {Belli}, Sirio and {Burkert}, Andreas and {Dekel}, Avishai and {F{\"o}rster Schreiber}, Natascha M. and {Genzel}, Reinhard and {Hayashi}, Masao and {Herrera-Camus}, Rodrigo and {Kodama}, Tadayuki and {Kohno}, Kotaro and {Koyama}, Yusei and {Lee}, Minju M. and {Lutz}, Dieter and {Mowla}, Lamiya and {Nelson}, Erica J. and {Renzini}, Alvio and {Suzuki}, Tomoko L. and {Tacconi}, Linda J. and {{\"U}bler}, Hannah and {Wisnioski}, Emily and {Wuyts}, Stijn},
        title = "{Structural Evolution in Massive Galaxies at z {\ensuremath{\sim}} 2}",
      journal = {\apj},
         year = 2020,
        month = sep,
       volume = {901},
       number = {1},
          eid = {74},
        pages = {74},
          doi = {10.3847/1538-4357/abaf4a},
archivePrefix = {arXiv},
       eprint = {2009.01976},
 primaryClass = {astro-ph.GA},
       adsurl = {https://ui.adsabs.harvard.edu/abs/2020ApJ...901...74T}
}

@ARTICLE{Rivera2018,
       author = {{Calistro Rivera}, Gabriela and {Hodge}, J.~A. and {Smail}, Ian and {Swinbank}, A.~M. and {Weiss}, A. and {Wardlow}, J.~L. and {Walter}, F. and {Rybak}, M. and {Chen}, Chian-Chou and {Brandt}, W.~N. and {Coppin}, K. and {da Cunha}, E. and {Dannerbauer}, H. and {Greve}, T.~R. and {Karim}, A. and {Knudsen}, K.~K. and {Schinnerer}, E. and {Simpson}, J.~M. and {Venemans}, B. and {van der Werf}, P.~P.},
        title = "{Resolving the ISM at the Peak of Cosmic Star Formation with ALMA: The Distribution of CO and Dust Continuum in z {\ensuremath{\sim}} 2.5 Submillimeter Galaxies}",
      journal = {\apj},
         year = 2018,
        month = aug,
       volume = {863},
       number = {1},
          eid = {56},
        pages = {56},
          doi = {10.3847/1538-4357/aacffa},
archivePrefix = {arXiv},
       eprint = {1804.06852},
 primaryClass = {astro-ph.GA},
       adsurl = {https://ui.adsabs.harvard.edu/abs/2018ApJ...863...56C}
}

@ARTICLE{Cochrane2019,
       author = {{Cochrane}, R.~K. and {Hayward}, C.~C. and {Angl{\'e}s-Alc{\'a}zar}, D. and {Lotz}, J. and {Parsotan}, T. and {Ma}, X. and {Kere{\v{s}}}, D. and {Feldmann}, R. and {Faucher-Gigu{\`e}re}, C.~A. and {Hopkins}, P.~F.},
        title = "{Predictions for the spatial distribution of the dust continuum emission in 1 < z < 5 star-forming galaxies}",
      journal = {\mnras},
         year = 2019,
        month = sep,
       volume = {488},
       number = {2},
        pages = {1779-1789},
          doi = {10.1093/mnras/stz1736},
archivePrefix = {arXiv},
       eprint = {1905.13234},
 primaryClass = {astro-ph.GA},
       adsurl = {https://ui.adsabs.harvard.edu/abs/2019MNRAS.488.1779C}
}

@ARTICLE{Popping2022,
       author = {{Popping}, Gerg{\"o} and {Pillepich}, Annalisa and {Calistro Rivera}, Gabriela and {Schulz}, Sebastian and {Hernquist}, Lars and {Kaasinen}, Melanie and {Marinacci}, Federico and {Nelson}, Dylan and {Vogelsberger}, Mark},
        title = "{The dust-continuum size of TNG50 galaxies at z = 1-5: a comparison with the distribution of stellar light, stars, dust, and H$_{2}$}",
      journal = {\mnras},
         year = 2022,
        month = mar,
       volume = {510},
       number = {3},
        pages = {3321-3334},
          doi = {10.1093/mnras/stab3312},
archivePrefix = {arXiv},
       eprint = {2101.12218},
 primaryClass = {astro-ph.GA},
       adsurl = {https://ui.adsabs.harvard.edu/abs/2022MNRAS.510.3321P}
}

@ARTICLE{Fujimoto2018,
       author = {{Fujimoto}, Seiji and {Ouchi}, Masami and {Kohno}, Kotaro and {Yamaguchi}, Yuki and {Hatsukade}, Bunyo and {Ueda}, Yoshihiro and {Shibuya}, Takatoshi and {Inoue}, Shigeki and {Oogi}, Taira and {Toft}, Sune and {G{\'o}mez-Guijarro}, Carlos and {Wang}, Tao and {Espada}, Daniel and {Nagao}, Tohru and {Tanaka}, Ichi and {Ao}, Yiping and {Umehata}, Hideki and {Taniguchi}, Yoshiaki and {Nakanishi}, Kouichiro and {Rujopakarn}, Wiphu and {Ivison}, R.~J. and {Wang}, Wei-hao and {Lee}, Minju M. and {Tadaki}, Ken-ichi and {Tamura}, Yoichi and {Dunlop}, J.~S.},
        title = "{ALMA 26 Arcmin$^{2}$ Survey of GOODS-S at One Millimeter (ASAGAO): Average Morphology of High-z Dusty Star-forming Galaxies in an Exponential Disk (n ≃ 1)}",
      journal = {\apj},
         year = 2018,
        month = jul,
       volume = {861},
       number = {1},
          eid = {7},
        pages = {7},
          doi = {10.3847/1538-4357/aac6c4},
archivePrefix = {arXiv},
       eprint = {1802.02136},
 primaryClass = {astro-ph.GA},
       adsurl = {https://ui.adsabs.harvard.edu/abs/2018ApJ...861....7F}
}

@INPROCEEDINGS{Perrin2012,
       author = {{Perrin}, Marshall D. and {Soummer}, R{\'e}mi and {Elliott}, Erin M. and {Lallo}, Matthew D. and {Sivaramakrishnan}, Anand},
        title = "{Simulating point spread functions for the James Webb Space Telescope with WebbPSF}",
    booktitle = {Space Telescopes and Instrumentation 2012: Optical, Infrared, and Millimeter Wave},
         year = 2012,
       editor = {{Clampin}, Mark C. and {Fazio}, Giovanni G. and {MacEwen}, Howard A. and {Oschmann}, Jacobus M., Jr.},
       series = {Society of Photo-Optical Instrumentation Engineers (SPIE) Conference Series},
       volume = {8442},
        month = sep,
          eid = {84423D},
        pages = {84423D},
          doi = {10.1117/12.925230},
       adsurl = {https://ui.adsabs.harvard.edu/abs/2012SPIE.8442E..3DP}
}

@INPROCEEDINGS{Perrin2014,
       author = {{Perrin}, Marshall D. and {Sivaramakrishnan}, Anand and {Lajoie}, Charles-Philippe and {Elliott}, Erin and {Pueyo}, Laurent and {Ravindranath}, Swara and {Albert}, Lo{\"\i}c.},
        title = "{Updated point spread function simulations for JWST with WebbPSF}",
    booktitle = {Space Telescopes and Instrumentation 2014: Optical, Infrared, and Millimeter Wave},
         year = 2014,
       editor = {{Oschmann}, Jacobus M., Jr. and {Clampin}, Mark and {Fazio}, Giovanni G. and {MacEwen}, Howard A.},
       series = {Society of Photo-Optical Instrumentation Engineers (SPIE) Conference Series},
       volume = {9143},
        month = aug,
          eid = {91433X},
        pages = {91433X},
          doi = {10.1117/12.2056689},
       adsurl = {https://ui.adsabs.harvard.edu/abs/2014SPIE.9143E..3XP}
}

@ARTICLE{Finkelstein2023,
       author = {{Finkelstein}, Steven L. and {Bagley}, Micaela B. and {Ferguson}, Henry C. and {Wilkins}, Stephen M. and {Kartaltepe}, Jeyhan S. and {Papovich}, Casey and {Yung}, L.~Y. Aaron and {Arrabal Haro}, Pablo and {Behroozi}, Peter and {Dickinson}, Mark and {Kocevski}, Dale D. and {Koekemoer}, Anton M. and {Larson}, Rebecca L. and {Le Bail}, Aur{\'e}lien and {Morales}, Alexa M. and {P{\'e}rez-Gonz{\'a}lez}, Pablo G. and {Burgarella}, Denis and {Dav{\'e}}, Romeel and {Hirschmann}, Michaela and {Somerville}, Rachel S. and {Wuyts}, Stijn and {Bromm}, Volker and {Casey}, Caitlin M. and {Fontana}, Adriano and {Fujimoto}, Seiji and {Gardner}, Jonathan P. and {Giavalisco}, Mauro and {Grazian}, Andrea and {Grogin}, Norman A. and {Hathi}, Nimish P. and {Hutchison}, Taylor A. and {Jha}, Saurabh W. and {Jogee}, Shardha and {Kewley}, Lisa J. and {Kirkpatrick}, Allison and {Long}, Arianna S. and {Lotz}, Jennifer M. and {Pentericci}, Laura and {Pierel}, Justin D.~R. and {Pirzkal}, Nor and {Ravindranath}, Swara and {Ryan}, Russell E. and {Trump}, Jonathan R. and {Yang}, Guang and {Bhatawdekar}, Rachana and {Bisigello}, Laura and {Buat}, V{\'e}ronique and {Calabr{\`o}}, Antonello and {Castellano}, Marco and {Cleri}, Nikko J. and {Cooper}, M.~C. and {Croton}, Darren and {Daddi}, Emanuele and {Dekel}, Avishai and {Elbaz}, David and {Franco}, Maximilien and {Gawiser}, Eric and {Holwerda}, Benne W. and {Huertas-Company}, Marc and {Jaskot}, Anne E. and {Leung}, Gene C.~K. and {Lucas}, Ray A. and {Mobasher}, Bahram and {Pandya}, Viraj and {Tacchella}, Sandro and {Weiner}, Benjamin J. and {Zavala}, Jorge A.},
        title = "{CEERS Key Paper. I. An Early Look into the First 500 Myr of Galaxy Formation with JWST}",
      journal = {\apjl},
         year = 2023,
        month = mar,
       volume = {946},
       number = {1},
          eid = {L13},
        pages = {L13},
          doi = {10.3847/2041-8213/acade4},
archivePrefix = {arXiv},
       eprint = {2211.05792},
 primaryClass = {astro-ph.GA},
       adsurl = {https://ui.adsabs.harvard.edu/abs/2023ApJ...946L..13F}
}

@ARTICLE{Casey2024,
       author = {{Casey}, Caitlin M. and {Akins}, Hollis B. and {Shuntov}, Marko and {Ilbert}, Olivier and {Paquereau}, Louise and {Franco}, Maximilien and {Hayward}, Christopher C. and {Finkelstein}, Steven L. and {Boylan-Kolchin}, Michael and {Robertson}, Brant E. and {Allen}, Natalie and {Brinch}, Malte and {Cooper}, Olivia R. and {Ding}, Xuheng and {Drakos}, Nicole E. and {Faisst}, Andreas L. and {Fujimoto}, Seiji and {Gillman}, Steven and {Harish}, Santosh and {Hirschmann}, Michaela and {Jin}, Shuowen and {Kartaltepe}, Jeyhan S. and {Koekemoer}, Anton M. and {Kokorev}, Vasily and {Liu}, Daizhong and {Long}, Arianna S. and {Magdis}, Georgios and {Maraston}, Claudia and {Martin}, Crystal L. and {McCracken}, Henry Joy and {McKinney}, Jed and {Mobasher}, Bahram and {Rhodes}, Jason and {Rich}, R. Michael and {Sanders}, David B. and {Silverman}, John D. and {Toft}, Sune and {Vijayan}, Aswin P. and {Weaver}, John R. and {Wilkins}, Stephen M. and {Yang}, Lilan and {Zavala}, Jorge A.},
        title = "{COSMOS-Web: Intrinsically Luminous z {\ensuremath{\gtrsim}} 10 Galaxy Candidates Test Early Stellar Mass Assembly}",
      journal = {\apj},
         year = 2024,
        month = apr,
       volume = {965},
       number = {1},
          eid = {98},
        pages = {98},
          doi = {10.3847/1538-4357/ad2075},
archivePrefix = {arXiv},
       eprint = {2308.10932},
 primaryClass = {astro-ph.GA},
       adsurl = {https://ui.adsabs.harvard.edu/abs/2024ApJ...965...98C}
}

@ARTICLE{Peng2002,
       author = {{Peng}, Chien Y. and {Ho}, Luis C. and {Impey}, Chris D. and {Rix}, Hans-Walter},
        title = "{Detailed Structural Decomposition of Galaxy Images}",
      journal = {\aj},
         year = 2002,
        month = jul,
       volume = {124},
       number = {1},
        pages = {266-293},
          doi = {10.1086/340952},
archivePrefix = {arXiv},
       eprint = {astro-ph/0204182},
 primaryClass = {astro-ph},
       adsurl = {https://ui.adsabs.harvard.edu/abs/2002AJ....124..266P}
}

@ARTICLE{Ding2021,
       author = {{Ding}, Xuheng and {Birrer}, Simon and {Treu}, Tommaso and {Silverman}, John D.},
        title = "{Galaxy shapes of Light (GaLight): a 2D modeling of galaxy images}",
      journal = {arXiv e-prints},
         year = 2021,
        month = nov,
          eid = {arXiv:2111.08721},
        pages = {arXiv:2111.08721},
          doi = {10.48550/arXiv.2111.08721},
archivePrefix = {arXiv},
       eprint = {2111.08721},
 primaryClass = {astro-ph.GA},
       adsurl = {https://ui.adsabs.harvard.edu/abs/2021arXiv211108721D}
}

@ARTICLE{Snyder2015b,
       author = {{Snyder}, Gregory F. and {Torrey}, Paul and {Lotz}, Jennifer M. and {Genel}, Shy and {McBride}, Cameron K. and {Vogelsberger}, Mark and {Pillepich}, Annalisa and {Nelson}, Dylan and {Sales}, Laura V. and {Sijacki}, Debora and {Hernquist}, Lars and {Springel}, Volker},
        title = "{Galaxy morphology and star formation in the Illustris Simulation at z = 0}",
      journal = {\mnras},
         year = 2015,
        month = dec,
       volume = {454},
       number = {2},
        pages = {1886-1908},
          doi = {10.1093/mnras/stv2078},
archivePrefix = {arXiv},
       eprint = {1502.07747},
 primaryClass = {astro-ph.GA},
       adsurl = {https://ui.adsabs.harvard.edu/abs/2015MNRAS.454.1886S}
}

@ARTICLE{Snyder2015a,
       author = {{Snyder}, Gregory F. and {Lotz}, Jennifer and {Moody}, Christopher and {Peth}, Michael and {Freeman}, Peter and {Ceverino}, Daniel and {Primack}, Joel and {Dekel}, Avishai},
        title = "{Diverse structural evolution at z > 1 in cosmologically simulated galaxies}",
      journal = {\mnras},
         year = 2015,
        month = aug,
       volume = {451},
       number = {4},
        pages = {4290-4310},
          doi = {10.1093/mnras/stv1231},
archivePrefix = {arXiv},
       eprint = {1409.1583},
 primaryClass = {astro-ph.GA},
       adsurl = {https://ui.adsabs.harvard.edu/abs/2015MNRAS.451.4290S}
}

@ARTICLE{Lotz2008b,
       author = {{Lotz}, Jennifer M. and {Jonsson}, Patrik and {Cox}, T.~J. and {Primack}, Joel R.},
        title = "{Galaxy merger morphologies and time-scales from simulations of equal-mass gas-rich disc mergers}",
      journal = {\mnras},
         year = 2008,
        month = dec,
       volume = {391},
       number = {3},
        pages = {1137-1162},
          doi = {10.1111/j.1365-2966.2008.14004.x},
archivePrefix = {arXiv},
       eprint = {0805.1246},
 primaryClass = {astro-ph},
       adsurl = {https://ui.adsabs.harvard.edu/abs/2008MNRAS.391.1137L}
}

@ARTICLE{Lotz2004,
       author = {{Lotz}, Jennifer M. and {Primack}, Joel and {Madau}, Piero},
        title = "{A New Nonparametric Approach to Galaxy Morphological Classification}",
      journal = {\aj},
         year = 2004,
        month = jul,
       volume = {128},
       number = {1},
        pages = {163-182},
          doi = {10.1086/421849},
archivePrefix = {arXiv},
       eprint = {astro-ph/0311352},
 primaryClass = {astro-ph},
       adsurl = {https://ui.adsabs.harvard.edu/abs/2004AJ....128..163L}
}

@ARTICLE{Conselice2003,
       author = {{Conselice}, Christopher J.},
        title = "{The Relationship between Stellar Light Distributions of Galaxies and Their Formation Histories}",
      journal = {\apjs},
         year = 2003,
        month = jul,
       volume = {147},
       number = {1},
        pages = {1-28},
          doi = {10.1086/375001},
archivePrefix = {arXiv},
       eprint = {astro-ph/0303065},
 primaryClass = {astro-ph},
       adsurl = {https://ui.adsabs.harvard.edu/abs/2003ApJS..147....1C}
}

@ARTICLE{Abraham2003,
       author = {{Abraham}, Roberto G. and {van den Bergh}, Sidney and {Nair}, Preethi},
        title = "{A New Approach to Galaxy Morphology. I. Analysis of the Sloan Digital Sky Survey Early Data Release}",
      journal = {\apj},
         year = 2003,
        month = may,
       volume = {588},
       number = {1},
        pages = {218-229},
          doi = {10.1086/373919},
archivePrefix = {arXiv},
       eprint = {astro-ph/0301239},
 primaryClass = {astro-ph},
       adsurl = {https://ui.adsabs.harvard.edu/abs/2003ApJ...588..218A}
}

@ARTICLE{Bershady2000,
       author = {{Bershady}, Matthew A. and {Jangren}, Anna and {Conselice}, Christopher J.},
        title = "{Structural and Photometric Classification of Galaxies. I. Calibration Based on a Nearby Galaxy Sample}",
      journal = {\aj},
         year = 2000,
        month = jun,
       volume = {119},
       number = {6},
        pages = {2645-2663},
          doi = {10.1086/301386},
archivePrefix = {arXiv},
       eprint = {astro-ph/0002262},
 primaryClass = {astro-ph},
       adsurl = {https://ui.adsabs.harvard.edu/abs/2000AJ....119.2645B}
}

@ARTICLE{Rodriguez-Gomez2019,
       author = {{Rodriguez-Gomez}, Vicente and {Snyder}, Gregory F. and {Lotz}, Jennifer M. and {Nelson}, Dylan and {Pillepich}, Annalisa and {Springel}, Volker and {Genel}, Shy and {Weinberger}, Rainer and {Tacchella}, Sandro and {Pakmor}, R{\"u}diger and {Torrey}, Paul and {Marinacci}, Federico and {Vogelsberger}, Mark and {Hernquist}, Lars and {Thilker}, David A.},
        title = "{The optical morphologies of galaxies in the IllustrisTNG simulation: a comparison to Pan-STARRS observations}",
      journal = {\mnras},
         year = 2019,
        month = mar,
       volume = {483},
       number = {3},
        pages = {4140-4159},
          doi = {10.1093/mnras/sty3345},
archivePrefix = {arXiv},
       eprint = {1809.08239},
 primaryClass = {astro-ph.GA},
       adsurl = {https://ui.adsabs.harvard.edu/abs/2019MNRAS.483.4140R}
}

@ARTICLE{Polletta2024,
	author = {{Polletta}, M. and {Frye}, B.~L. and {Garuda}, N. and {Willner}, S.~P. and {Berta}, S. and {Kneissl}, R. and {Dole}, H. and {Jansen}, R.~A. and {Lehnert}, M.~D. and {Cohen}, S.~H. and {Summers}, J. and {Windhorst}, R.~A. and {D'Silva}, J.~C.~J. and {Koekemoer}, A.~M. and {Coe}, D. and {Conselice}, C.~J. and {Driver}, S.~P. and {Grogin}, N.~A. and {Marshall}, M.~A. and {Nonino}, M. and {Ortiz}, III, R. and {Pirzkal}, N. and {Robotham}, A. and {Ryan}, R.~E. and {Willmer}, C.~N.~A. and {Yan}, H. and {Arumugam}, V. and {Cheng}, C. and {Gim}, H.~B. and {Hathi}, N.~P. and {Holwerda}, B. and {Kamieneski}, P. and {Keel}, W.~C. and {Li}, J. and {Pascale}, M. and {Rottgering}, H. and {Smith}, B.~M. and {Yun}, M.~S.},
	title = "{JWST's PEARLS: Resolved study of the stellar and dust components in starburst galaxies at cosmic noon}",
	journal = {\aap},
	year = 2024,
	month = oct,
	volume = {690},
	eid = {A285},
	pages = {A285},
	doi = {10.1051/0004-6361/202450671},
	archivePrefix = {arXiv},
	eprint = {2405.07986},
	primaryClass = {astro-ph.GA},
	adsurl = {https://ui.adsabs.harvard.edu/abs/2024A&A...690A.285P}
}

@ARTICLE{Ito2024,
       author = {{Ito}, Kei and {Valentino}, Francesco and {Brammer}, Gabriel and {Faisst}, Andreas L. and {Gillman}, Steven and {G{\'o}mez-Guijarro}, Carlos and {Gould}, Katriona M.~L. and {Heintz}, Kasper E. and {Ilbert}, Olivier and {Jespersen}, Christian Kragh and {Kokorev}, Vasily and {Kubo}, Mariko and {Magdis}, Georgios E. and {McPartland}, Conor J.~R. and {Onodera}, Masato and {Rizzo}, Francesca and {Tanaka}, Masayuki and {Toft}, Sune and {Vijayan}, Aswin P. and {Weaver}, John R. and {Whitaker}, Katherine E. and {Wright}, Lillian},
        title = "{Size{\textendash}Stellar Mass Relation and Morphology of Quiescent Galaxies at z {\ensuremath{\geq}} 3 in Public JWST Fields}",
      journal = {\apj},
         year = 2024,
        month = apr,
       volume = {964},
       number = {2},
          eid = {192},
        pages = {192},
          doi = {10.3847/1538-4357/ad2512},
archivePrefix = {arXiv},
       eprint = {2307.06994},
 primaryClass = {astro-ph.GA},
       adsurl = {https://ui.adsabs.harvard.edu/abs/2024ApJ...964..192I}
}

@ARTICLE{Brammer2011,
       author = {{Brammer}, Gabriel B. and {Whitaker}, K.~E. and {van Dokkum}, P.~G. and {Marchesini}, D. and {Franx}, M. and {Kriek}, M. and {Labb{\'e}}, I. and {Lee}, K. -S. and {Muzzin}, A. and {Quadri}, R.~F. and {Rudnick}, G. and {Williams}, R.},
        title = "{The Number Density and Mass Density of Star-forming and Quiescent Galaxies at 0.4 <= z <= 2.2}",
      journal = {\apj},
         year = 2011,
        month = sep,
       volume = {739},
       number = {1},
          eid = {24},
        pages = {24},
          doi = {10.1088/0004-637X/739/1/24},
archivePrefix = {arXiv},
       eprint = {1104.2595},
 primaryClass = {astro-ph.CO},
       adsurl = {https://ui.adsabs.harvard.edu/abs/2011ApJ...739...24B}
}

@MISC{Dannerbauer2021,
       author = {{Dannerbauer}, Helmut and {Koyama}, Yusei and {Jin}, Shuowen and {Kodama}, Tadayuki and {Perez Martinez}, Jose Manuel and {Shimakawa}, Rhythm},
        title = "{Mapping, resolving and penetrating into the dusty Spiderweb protocluster with unqiue Pa-beta imaging}",
 howpublished = {JWST Proposal. Cycle 1, ID. \#1572},
         year = 2021,
        month = mar,
        pages = {1572},
       adsurl = {https://ui.adsabs.harvard.edu/abs/2021jwst.prop.1572D}
}

@MISC{Koyama2022,
       author = {{Koyama}, Yusei and {Dannerbauer}, Helmut and {Calvi}, Rosa and {Chen}, Zhengyi and {Kodama}, Tadayuki and {Naufal}, Abdurrahman and {Perez Martinez}, Jose Manuel and {Shimakawa}, Rhythm},
        title = "{A complete census of quiescent galaxies in the dense core of the Spiderweb protocluster at z=2.16}",
 howpublished = {HST Proposal. Cycle 30, ID. \#17117},
         year = 2022,
        month = jun,
        pages = {17117},
       adsurl = {https://ui.adsabs.harvard.edu/abs/2022hst..prop17117K}
}

@INPROCEEDINGS{Rieke2005,
       author = {{Rieke}, Marcia J. and {Kelly}, Douglas and {Horner}, Scott},
        title = "{Overview of James Webb Space Telescope and NIRCam's Role}",
    booktitle = {Cryogenic Optical Systems and Instruments XI},
         year = 2005,
       editor = {{Heaney}, James B. and {Burriesci}, Lawrence G.},
       series = {Society of Photo-Optical Instrumentation Engineers (SPIE) Conference Series},
       volume = {5904},
        month = aug,
        pages = {1-8},
          doi = {10.1117/12.615554},
       adsurl = {https://ui.adsabs.harvard.edu/abs/2005SPIE.5904....1R}
}

@ARTICLE{Ren2024,
       author = {{Ren}, Jian and {Liu}, F.~S. and {Li}, Nan and {Cui}, Qifan and {Zhao}, Pinsong and {Li}, Yubin and {Song}, Qi and {Yesuf}, Hassen M. and {Zheng}, Xian Zhong},
        title = "{Calibrating Nonparametric Morphological Indicators from JWST Images for Galaxies over 0.5 < z < 3}",
      journal = {\apj},
         year = 2024,
        month = jul,
       volume = {969},
       number = {1},
          eid = {4},
        pages = {4},
          doi = {10.3847/1538-4357/ad4117},
archivePrefix = {arXiv},
       eprint = {2404.16686},
 primaryClass = {astro-ph.GA},
       adsurl = {https://ui.adsabs.harvard.edu/abs/2024ApJ...969....4R}
}

@ARTICLE{Perez-Gonzalez2024,
       author = {{P{\'e}rez-Gonz{\'a}lez}, Pablo G. and {Rinaldi}, Pierluigi and {Caputi}, Karina I. and {{\'A}lvarez-M{\'a}rquez}, Javier and {Annunziatella}, Marianna and {Langeroodi}, Danial and {Moutard}, Thibaud and {Boogaard}, Leindert and {Iani}, Edoardo and {Melinder}, Jens and {Costantin}, Luca and {{\"O}stlin}, G{\"o}ran and {Colina}, Luis and {Greve}, Thomas R. and {Wright}, Gillian and {Alonso-Herrero}, Almudena and {Bik}, Arjan and {Bosman}, Sarah E.~I. and {Crespo G{\'o}mez}, Alejandro and {Dicken}, Daniel and {Eckart}, Andreas and {Garc{\'\i}a-Mar{\'\i}n}, Macarena and {Gillman}, Steven and {G{\"u}del}, Manuel and {Henning}, Thomas and {Hjorth}, Jens and {Jermann}, Iris and {Labiano}, {\'A}lvaro and {Meyer}, Romain A. and {Pei{\ensuremath{\beta}}ker}, Florian and {Pye}, John P. and {Ray}, Thomas P. and {Tikkanen}, Tuomo and {Walter}, Fabian and {van der Werf}, Paul P.},
        title = "{A NIRCam-dark Galaxy Detected with the MIRI/F1000W Filter in the MIDIS/JADES Hubble Ultra Deep Field}",
      journal = {\apjl},
         year = 2024,
        month = jul,
       volume = {969},
       number = {1},
          eid = {L10},
        pages = {L10},
          doi = {10.3847/2041-8213/ad517b},
archivePrefix = {arXiv},
       eprint = {2402.16942},
 primaryClass = {astro-ph.GA},
       adsurl = {https://ui.adsabs.harvard.edu/abs/2024ApJ...969L..10P}
}

@ARTICLE{Boogaard2024,
       author = {{Boogaard}, Leindert A. and {Gillman}, Steven and {Melinder}, Jens and {Walter}, Fabian and {Colina}, Luis and {{\"O}stlin}, G{\"o}ran and {Caputi}, Karina I. and {Iani}, Edoardo and {P{\'e}rez-Gonz{\'a}lez}, Pablo and {van der Werf}, Paul and {Greve}, Thomas R. and {Wright}, Gillian and {Alonso-Herrero}, Almudena and {{\'A}lvarez-M{\'a}rquez}, Javier and {Annunziatella}, Marianna and {Bik}, Arjan and {Bosman}, Sarah and {Costantin}, Luca and {Crespo G{\'o}mez}, Alejandro and {Dicken}, Dan and {Eckart}, Andreas and {Hjorth}, Jens and {Jermann}, Iris and {Labiano}, Alvaro and {Langeroodi}, Danial and {Meyer}, Romain A. and {Moutard}, Thibaud and {Pei{\ss}ker}, Florian and {Pye}, John P. and {Rinaldi}, Pierluigi and {Tikkanen}, Tuomo V. and {Topinka}, Martin and {Henning}, Thomas},
        title = "{MIDIS: JWST/MIRI Reveals the Stellar Structure of ALMA-selected Galaxies in the Hubble Ultra Deep Field at Cosmic Noon}",
      journal = {\apj},
         year = 2024,
        month = jul,
       volume = {969},
       number = {1},
          eid = {27},
        pages = {27},
          doi = {10.3847/1538-4357/ad43e5},
archivePrefix = {arXiv},
       eprint = {2308.16895},
 primaryClass = {astro-ph.GA},
       adsurl = {https://ui.adsabs.harvard.edu/abs/2024ApJ...969...27B}
}

@ARTICLE{Tan2024b,
	author = {{Tan}, Qing-Hua and {Daddi}, Emanuele and {Magnelli}, Benjamin and {Correa}, Camila A. and {Bournaud}, Fr{\'e}d{\'e}ric and {Adscheid}, Sylvia and {Zhang}, Shao-Bo and {Elbaz}, David and {G{\'o}mez-Guijarro}, Carlos and {Kalita}, Boris S. and {Liu}, Daizhong and {Liu}, Zhaoxuan and {Pety}, J{\'e}r{\^o}me and {Puglisi}, Annagrazia and {Schinnerer}, Eva and {Silverman}, John D. and {Valentino}, Francesco},
	title = "{In situ spheroid formation in distant submillimetre-bright galaxies}",
	journal = {\nat},
	year = 2024,
	month = dec,
	volume = {636},
	number = {8041},
	pages = {69-74},
	doi = {10.1038/s41586-024-08201-6},
	archivePrefix = {arXiv},
	eprint = {2407.16578},
	primaryClass = {astro-ph.GA},
	adsurl = {https://ui.adsabs.harvard.edu/abs/2024Natur.636...69T}
}

@ARTICLE{Gillman2024,
	author = {{Gillman}, Steven and {Smail}, Ian and {Gullberg}, Bitten and {Swinbank}, A.~M. and {Vijayan}, Aswin P. and {Lee}, Minju and {Brammer}, Gabe and {Dudzevi{\v{c}}i{\={u}}t{\.{e}}}, Ugn{\.{e}} and {Greve}, Thomas R. and {Almaini}, Omar and {Brinch}, Malte and {Chapman}, Scott C. and {Chen}, Chian-Chou and {Ikarashi}, Soh and {Matsuda}, Yuichi and {Wang}, Wei-Hao and {Walter}, Fabian and {van der Werf}, Paul P.},
	title = "{The structure of massive star-forming galaxies from JWST and ALMA: Dusty, high-redshift disc galaxies}",
	journal = {\aap},
	year = 2024,
	month = nov,
	volume = {691},
	eid = {A299},
	pages = {A299},
	doi = {10.1051/0004-6361/202451006},
	archivePrefix = {arXiv},
	eprint = {2406.03544},
	primaryClass = {astro-ph.GA},
	adsurl = {https://ui.adsabs.harvard.edu/abs/2024A&A...691A.299G}
}

@ARTICLE{Liu2024,
       author = {{Liu}, Zhaoxuan and {Silverman}, John D. and {Daddi}, Emanuele and {Puglisi}, Annagrazia and {Renzini}, Alvio and {Kalita}, Boris S. and {Kartaltepe}, Jeyhan S. and {Kashino}, Daichi and {Rodighiero}, Giulia and {Rujopakarn}, Wiphu and {Suzuki}, Tomoko L. and {Tanaka}, Takumi S. and {Valentino}, Francesco and {Andika}, Irham Taufik and {Casey}, Caitlin M. and {Faisst}, Andreas and {Franco}, Maximilien and {Gozaliasl}, Ghassem and {Gillman}, Steven and {Hayward}, Christopher C. and {Koekemoer}, Anton M. and {Kokorev}, Vasily and {Lambrides}, Erini and {Lee}, Minju M. and {Magdis}, Georgios E. and {Harish}, Santosh and {McCracken}, Henry Joy and {Rhodes}, Jason and {Shuntov}, Marko and {Ding}, Xuheng},
        title = "{JWST and ALMA Discern the Assembly of Structural and Obscured Components in a High-redshift Starburst Galaxy}",
      journal = {\apj},
         year = 2024,
        month = jun,
       volume = {968},
       number = {1},
          eid = {15},
        pages = {15},
          doi = {10.3847/1538-4357/ad4096},
archivePrefix = {arXiv},
       eprint = {2311.14809},
 primaryClass = {astro-ph.GA},
       adsurl = {https://ui.adsabs.harvard.edu/abs/2024ApJ...968...15L}
}

@ARTICLE{Zhuang2024,
       author = {{Zhuang}, Ming-Yang and {Shen}, Yue},
        title = "{Characterization of JWST NIRCam PSFs and Implications for AGN+host Image Decomposition}",
      journal = {\apj},
         year = 2024,
        month = feb,
       volume = {962},
       number = {2},
          eid = {139},
        pages = {139},
          doi = {10.3847/1538-4357/ad1183},
archivePrefix = {arXiv},
       eprint = {2304.13776},
 primaryClass = {astro-ph.GA},
       adsurl = {https://ui.adsabs.harvard.edu/abs/2024ApJ...962..139Z}
}

@ARTICLE{Ward2024,
       author = {{Ward}, Ethan and {de la Vega}, Alexander and {Mobasher}, Bahram and {McGrath}, Elizabeth J. and {Iyer}, Kartheik G. and {Calabr{\`o}}, Antonello and {Costantin}, Luca and {Dickinson}, Mark and {Holwerda}, Benne W. and {Huertas-Company}, Marc and {Hirschmann}, Michaela and {Lucas}, Ray A. and {Pandya}, Viraj and {Wilkins}, Stephen M. and {Yung}, L.~Y. Aaron and {Arrabal Haro}, Pablo and {Bagley}, Micaela B. and {Finkelstein}, Steven L. and {Kartaltepe}, Jeyhan S. and {Koekemoer}, Anton M. and {Papovich}, Casey and {Pirzkal}, Nor},
        title = "{Evolution of the Size{\textendash}Mass Relation of Star-forming Galaxies Since z = 5.5 Revealed by CEERS}",
      journal = {\apj},
         year = 2024,
        month = feb,
       volume = {962},
       number = {2},
          eid = {176},
        pages = {176},
          doi = {10.3847/1538-4357/ad20ed},
archivePrefix = {arXiv},
       eprint = {2311.02162},
 primaryClass = {astro-ph.GA},
       adsurl = {https://ui.adsabs.harvard.edu/abs/2024ApJ...962..176W}
}

@ARTICLE{Ji2024,
       author = {{Ji}, Zhiyuan and {Williams}, Christina C. and {Suess}, Katherine A. and {Tacchella}, Sandro and {Johnson}, Benjamin D. and {Robertson}, Brant and {Alberts}, Stacey and {Baker}, William M. and {Baum}, Stefi and {Bhatawdekar}, Rachana and {Bonaventura}, Nina and {Boyett}, Kristan and {Bunker}, Andrew J. and {Carniani}, Stefano and {Charlot}, Stephane and {Chen}, Zuyi and {Chevallard}, Jacopo and {Curtis-Lake}, Emma and {D'Eugenio}, Francesco and {de Graaff}, Anna and {DeCoursey}, Christa and {Egami}, Eiichi and {Eisenstein}, Daniel J. and {Hainline}, Kevin and {Hausen}, Ryan and {Helton}, Jakob M. and {Looser}, Tobias J. and {Lyu}, Jianwei and {Maiolino}, Roberto and {Maseda}, Michael V. and {Nelson}, Erica and {Rieke}, George and {Rieke}, Marcia and {Rix}, Hans-Walter and {Sandles}, Lester and {Sun}, Fengwu and {{\"U}bler}, Hannah and {Willmer}, Christopher N.~A. and {Willott}, Chris and {Witstok}, Joris},
        title = "{JADES: Rest-frame UV-to-NIR Size Evolution of Massive Quiescent Galaxies from Redshift z=5 to z=0.5}",
      journal = {arXiv e-prints},
         year = 2024,
        month = jan,
          eid = {arXiv:2401.00934},
        pages = {arXiv:2401.00934},
          doi = {10.48550/arXiv.2401.00934},
archivePrefix = {arXiv},
       eprint = {2401.00934},
 primaryClass = {astro-ph.GA},
       adsurl = {https://ui.adsabs.harvard.edu/abs/2024arXiv240100934J}
}

 \begin{appendix} 

\noindent\begin{minipage}{\textwidth}
	\section{Gallery of the eight ALMA sources undetected by JWST in the Spiderweb field.}
	\label{sec:appendix_undetections}
	\begin{center}
		\includegraphics[width=0.49\textwidth]{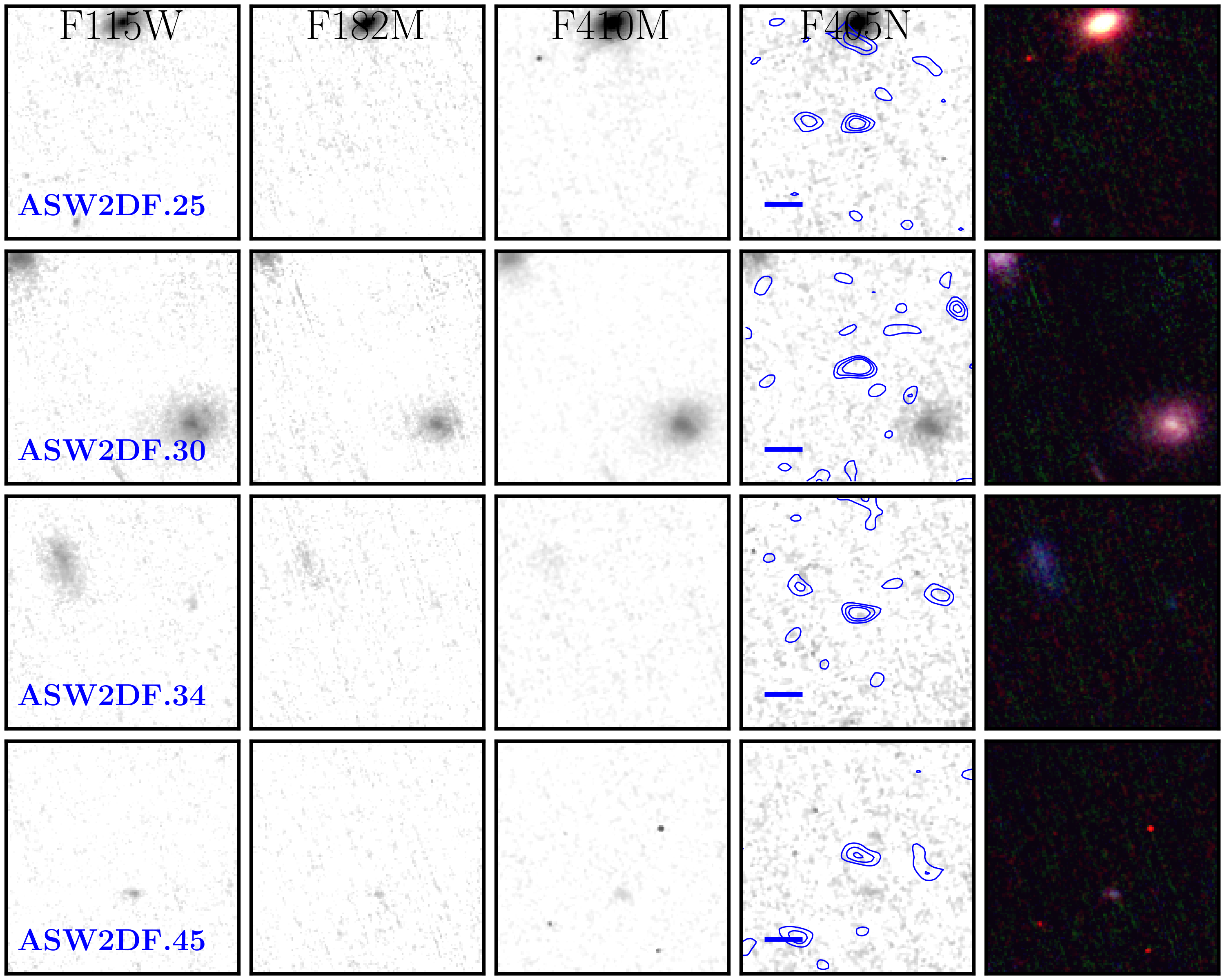}
		\includegraphics[width=0.49\textwidth]{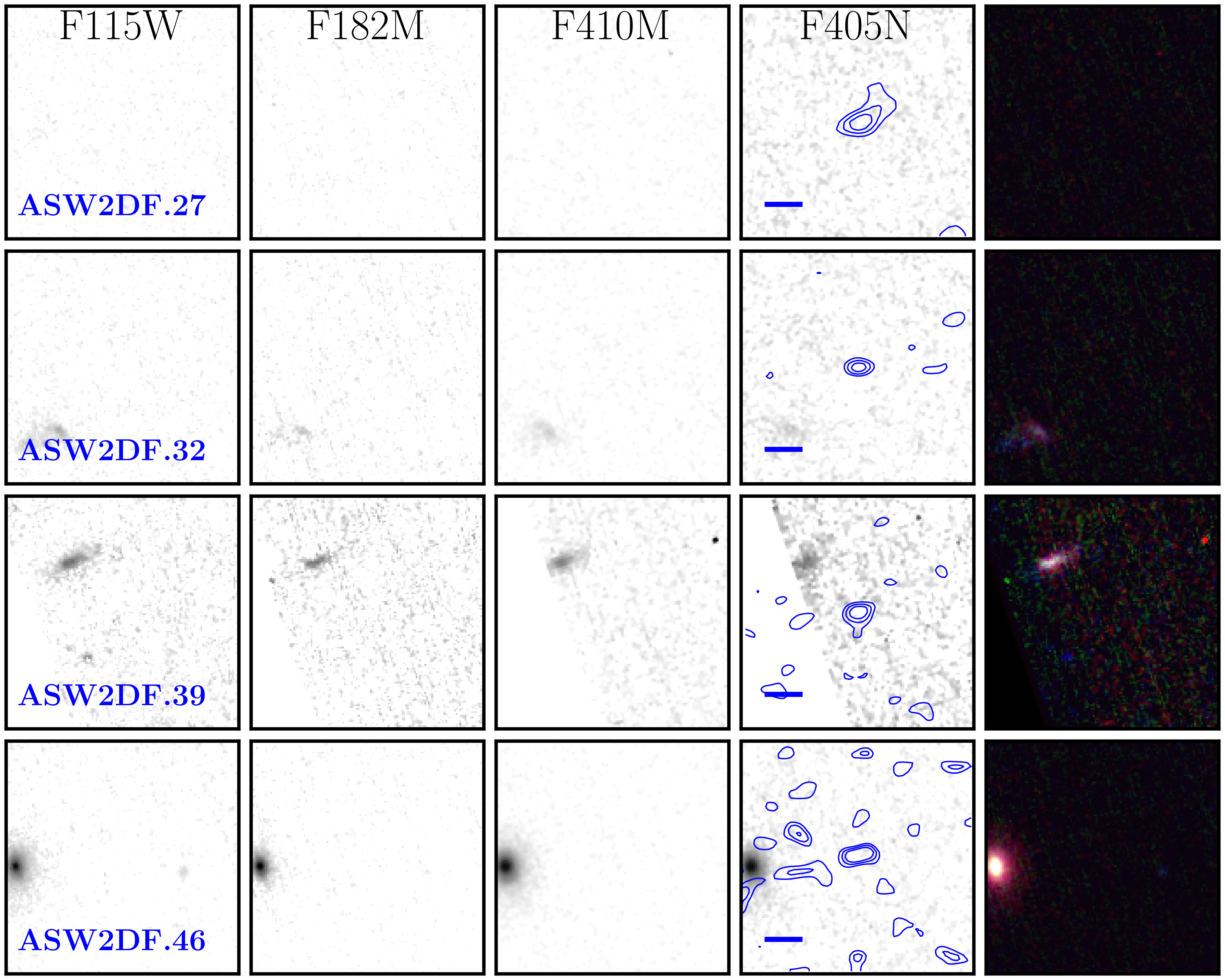}
	\end{center}
	\smallskip
	\textbf{Figure A.1.} The gallery of 8 ALMA sources which are not detected by JWST/NIRCam. The cutout settings and layout are identical to those in Figure~\ref{fig: cutouts}.
\end{minipage}

	\vspace{1cm}
	\noindent\begin{minipage}{\textwidth}
		\section{Gallery of the 12 ALMA sources lack of redshift information.}
		\label{sec:appendix_nonmembers}
		\begin{center}
			\includegraphics[width=0.49\textwidth]{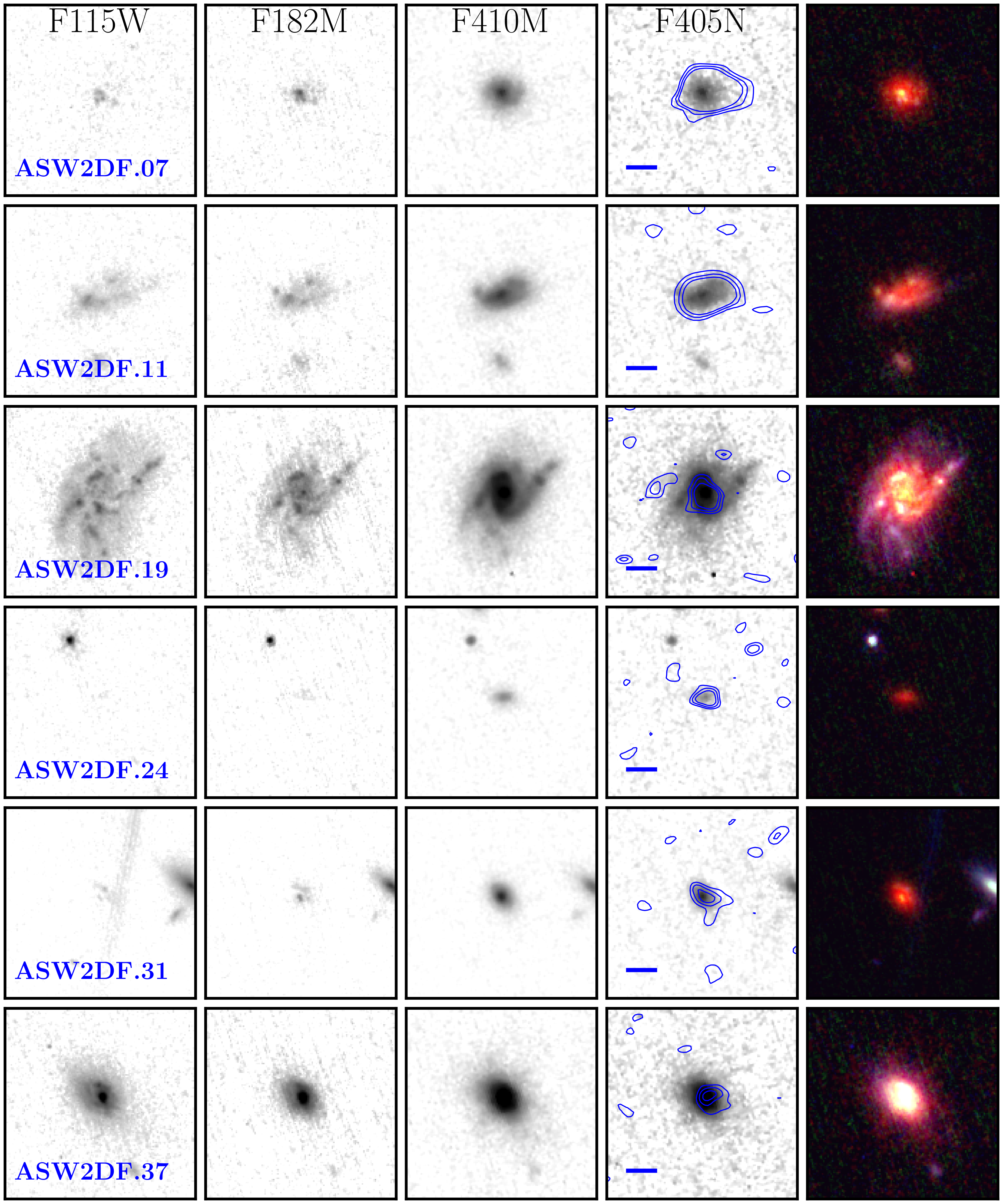}
			\includegraphics[width=0.49\textwidth]{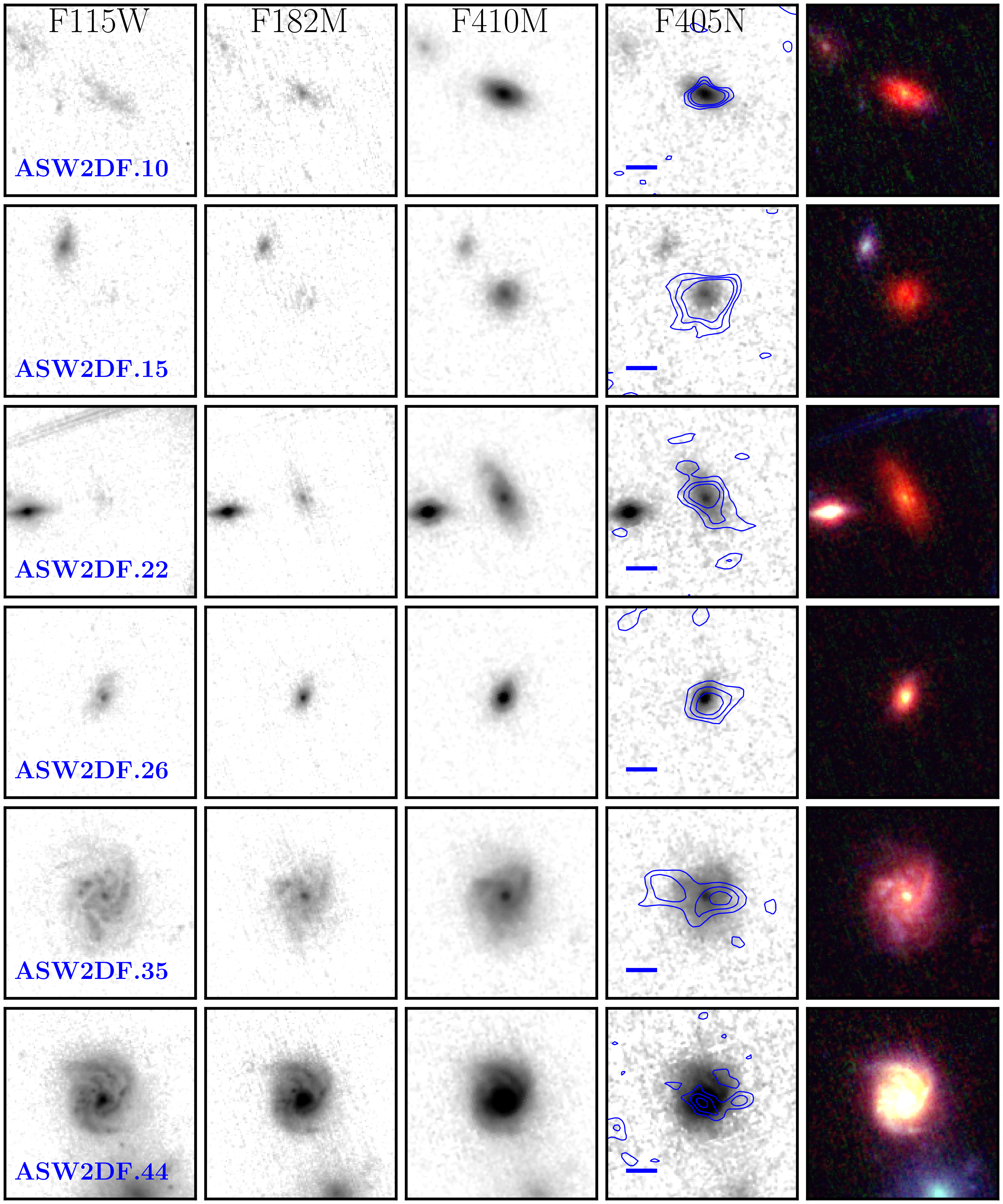}
		\end{center}
		\smallskip
		\textbf{Figure A.1.} The gallery of 12 sources detected by both ALMA and JWST/NIRCam, but are not confirmed to be members. The cutout settings and layout are identical to those in Figure~\ref{fig: cutouts}.
	\end{minipage}

\begin{sidewaystable*}
	
\section{Morphological analysis of the 22 ALMA sources detected by JWST in the Spiderweb field.}
	
\centering
\caption{Morphological properties of the nine DSFG members of the Spiderweb protocluster.} 
\label{tab: dsfgs_morph}
	\begin{tabular}{cccccccccccc}
		\toprule
		ID  & Gini$/M_{\rm 20, F410M}$ & R$_{\rm e, F115W}$ & n$_{\rm F115W}$ & R$_{\rm e, F182M}$ & n$_{\rm F182M}$ & R$_{\rm e, F405N}$ & n$_{\rm F405N}$ & R$_{\rm e, F410M}$ & n$_{\rm F410M}$ &  R$_{\rm e, 1.2\,mm}$ & Notes \\
		& & (kpc) & & (kpc) & & (kpc) & & (kpc) & & (kpc) & \\
		\midrule
04 & 0.45/-1.75 & $5.43\pm0.14$ & $0.54\pm0.05$ & $4.39\pm0.07$ & $0.30\pm0.01$ & $3.06\pm0.01$ & $0.61\pm0.01$ & $2.93\pm0.00$ & $0.60\pm0.00$ & $1.01\pm0.21$ & dusty, disk \\
05 & 0.45/-1.69 & $8.56\pm0.21$ & $1.64\pm0.04$ & $6.04\pm0.16$ & $1.75\pm0.05$ & $3.76\pm0.02$ & $1.16\pm0.01$ & $3.89\pm0.01$ & $1.31\pm0.00$ & $2.44\pm0.32$ & spirals, core \\
12 & 0.54/-1.86 & $1.93\pm0.07$ & $8.99\pm0.09$ & $5.06\pm0.27$ & $8.96\pm0.15$ & $1.54\pm0.01$ & $1.86\pm0.03$ & $1.72\pm0.01$ & $2.75\pm0.02$ & $2.20\pm0.65$ & compact AGN \\
14 & 0.49/-1.16 & $7.47\pm0.29$ & $2.20\pm0.27$ & $5.51\pm0.49$ & $1.22\pm0.24$ & $3.28\pm0.02$ & $0.30\pm0.00$ & $3.39\pm0.03$ & $1.16\pm0.02$ & -- & merger \\
20 & 0.53/-1.76 & $5.26\pm0.08$ & $0.52\pm0.02$ & $6.41\pm0.25$ & $1.74\pm0.07$ & $3.94\pm0.03$ & $1.21\pm0.01$ & $3.75\pm0.01$ & $1.33\pm0.01$ & $2.20\pm0.70$ & extended disk \\
23 & 0.42/-1.60 & $6.16\pm0.15$ & $3.21\pm0.05$ & $3.15\pm0.04$ & $0.75\pm0.02$ & $3.09\pm0.01$ & $0.84\pm0.01$ & $3.03\pm0.01$ & $0.87\pm0.00$ & -- & clumps, merger \\
28 & 0.47/-1.71 & $4.54\pm0.14$ & $0.32\pm0.03$ & $5.08\pm0.21$ & $0.67\pm0.05$ & $3.94\pm0.02$ & $0.71\pm0.01$ & $3.91\pm0.01$ & $0.94\pm0.01$ & -- & dusty, disk \\
29 & 0.53/-1.75 & $8.90\pm0.25$ & $1.96\pm0.05$ & $8.92\pm0.79$ & $3.31\pm0.20$ & $4.20\pm0.23$ & $2.77\pm0.14$ & $5.59\pm0.14$ & $3.86\pm0.08$ & -- & clumps, merger \\
47 & 0.49/-1.81 & $8.20\pm2.02$ & $0.80\pm0.32$ & $4.48\pm0.39$ & $0.49\pm0.11$ & $2.95\pm0.04$ & $1.08\pm0.04$ & $2.95\pm0.02$ & $1.27\pm0.02$ & -- & dusty, disk \\
		\bottomrule
	\end{tabular}
\tablefoot{Column 1 marks the IDs of the nine DSFG members. Column 2 lists the Gini and $M_{20}$ parameters measured in F410M filter, the errors of these measurements are extremely small and below the last decimal place of the values, and not shown in the table; columns 3–10 provide the effective radii and Sérsic indices in F115W, F182M, F405N and F410M, respectively. Column 11 shows dust sizes derived from ALMA 1.2\,mm observations. The final column lists the brief descriptions on individual source based on their morphologies.}

\vspace{1cm}
\centering
\caption{Morphological properties of the 12 ALMA sources lack of redshift information.} 
\label{tab: other_morph}
	\begin{tabular}{cccccccccccc}
		\toprule
		ID  & Gini$/M_{\rm 20, F410M}$ & R$_{\rm e, F115W}$ & n$_{\rm F115W}$ & R$_{\rm e, F182M}$ & n$_{\rm F182M}$ & R$_{\rm e, F405N}$ & n$_{\rm F405N}$ & R$_{\rm e, F410M}$ & n$_{\rm F410M}$ &  R$_{\rm e, 1.2\,mm}$ & Notes \\
		& & (arcsec) & & (arcsec) & & (arcsec) & & (arcsec) & & (arcsec) & \\
		\midrule
07 & 0.46/-1.71 & $0.25\pm0.04$ & $1.59\pm0.27$ & $0.37\pm0.03$ & $3.53\pm0.24$ & $0.23\pm0.00$ & $1.22\pm0.03$ & $0.24\pm0.00$ & $1.46\pm0.01$ & $0.21\pm0.07$ & disk, core \\
10 & 0.48/-1.79 & $0.73\pm0.04$ & $1.03\pm0.09$ & $0.69\pm0.12$ & $2.70\pm0.32$ & $0.23\pm0.00$ & $1.16\pm0.01$ & $0.23\pm0.00$ & $1.31\pm0.01$ & $0.12\pm0.04$ & disk, core \\
11 & 0.47/-1.59 & $0.64\pm0.01$ & $1.23\pm0.05$ & $0.50\pm0.01$ & $0.38\pm0.02$ & $0.36\pm0.00$ & $0.66\pm0.01$ & $0.38\pm0.00$ & $0.89\pm0.01$ & -- & disk, spirals \\
15 & 0.48/-1.65 & $0.64\pm0.24$ & $2.13\pm0.53$ & $0.36\pm0.03$ & $0.65\pm0.12$ & $0.21\pm0.00$ & $0.91\pm0.03$ & $0.22\pm0.00$ & $1.07\pm0.01$ & $0.30\pm0.10$ & compact disk \\
19 & 0.51/-1.71 & $0.88\pm0.00$ & $0.52\pm0.01$ & $0.63\pm0.00$ & $0.74\pm0.01$ & $0.48\pm0.00$ & $1.42\pm0.01$ & $0.45\pm0.00$ & $1.28\pm0.00$ & $0.23\pm0.09$ & spirals, clumps \\
22 & 0.50/-1.83 & $5.09\pm0.03$ & $4.35\pm0.18$ & $1.27\pm0.21$ & $3.73\pm0.44$ & $0.48\pm0.01$ & $1.68\pm0.03$ & $0.43\pm0.00$ & $1.54\pm0.01$ & -- & disk, core \\
24 & 0.50/-1.70 & $1.08\pm0.26$ & $4.19\pm1.42$ & $0.27\pm0.04$ & $0.71\pm0.23$ & $0.19\pm0.01$ & $1.15\pm0.11$ & $0.17\pm0.00$ & $1.14\pm0.07$ & -- & dusty, compact \\
26 & 0.52/-1.71 & $0.38\pm0.01$ & $2.78\pm0.06$ & $0.23\pm0.01$ & $3.50\pm0.10$ & $0.14\pm0.00$ & $2.12\pm0.04$ & $0.12\pm0.00$ & $3.20\pm0.04$ & $0.18\pm0.14$ & disk, core \\
31 & 0.49/-1.78 & $0.40\pm0.02$ & $0.59\pm0.06$ & $1.45\pm0.20$ & $6.45\pm0.60$ & $0.14\pm0.00$ & $1.60\pm0.04$ & $0.14\pm0.00$ & $1.55\pm0.01$ & $0.18\pm0.10$ & dusty, compact \\
35 & 0.47/-1.76 & $0.51\pm0.00$ & $0.50\pm0.01$ & $0.62\pm0.01$ & $1.30\pm0.02$ & $0.51\pm0.00$ & $1.17\pm0.01$ & $0.52\pm0.00$ & $1.16\pm0.00$ & $0.40\pm0.12$ & spirals, clumps \\
37 & 0.51/-1.88 & $0.16\pm0.00$ & $6.17\pm0.07$ & $0.24\pm0.01$ & $8.59\pm0.16$ & $0.15\pm0.00$ & $2.84\pm0.01$ & $0.17\pm0.00$ & $3.63\pm0.01$ & $0.21\pm0.12$ & disk, bulge \\
44 & 0.50/-1.80 & $0.68\pm0.00$ & $3.05\pm0.02$ & $0.51\pm0.00$ & $3.10\pm0.01$ & $0.34\pm0.00$ & $1.99\pm0.01$ & $0.31\pm0.00$ & $2.01\pm0.00$ & $0.44\pm0.15$ & disk, bulge \\
		\bottomrule
	\end{tabular}
\tablefoot{Columns are the same as those in Table \ref{tab: dsfgs_morph}, while the effective radii are in the unit of arcsec due to the lack of redshift information.}
\end{sidewaystable*}

\clearpage
\noindent\begin{minipage}{\textwidth}
	\section{Validation of the mass-to-light ratio stellar-sass estimates}
	\label{sec:stellar masses}
    
    \begin{center}	
    \includegraphics[width=0.6\textwidth]{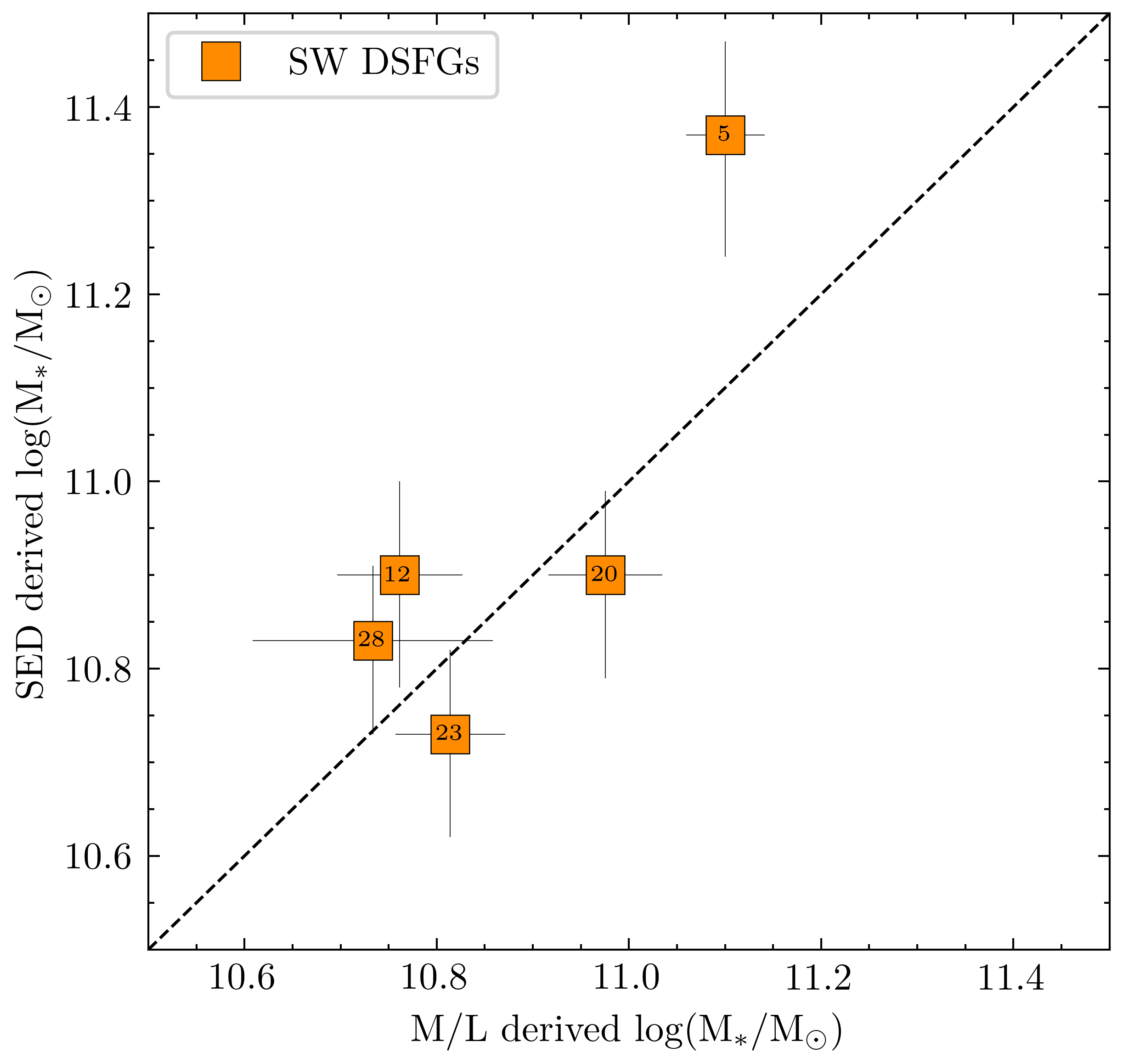}
	\label{fig:stellar_masses}
    \end{center}
	\smallskip
	\textbf{Figure D.1.} Comparison of UV–to–NIR SED-based stellar masses with the new mass-to-light estimates for the five ALMA-detected members with full multiwavelength coverage. Source IDs from Table~\ref{tab: dsfgs_phys} are indicated within each data point.
    \\
    \\
    Here we evaluate the robustness of the mass-to-light ratio approach used to derive the stellar masses for four of our nine spectroscopically confirmed ALMA-detected sources (excluding the Spiderweb Galaxy; see Sect.~\ref{subsubsec: mass_size}). We do so by comparing the M/L-based stellar masses with those obtained from full SED fitting \citep{Perez2023} for the five sources detected in up to ten photometric bands. Figure D.1 presents this comparison, together with a 1:1 reference line. The five sources (ASW2DF.05, 12, 20, 23, 28) lie close to this relation, with a mean offset of 0.13 dex and a 1$\sigma$ scatter of 0.17 dex. This level of agreement demonstrates that our approach provides consistent stellar-mass estimates and supports its application to the remaining four sources (ASW2DF.04, 14, 29, 47) in our analysis (Sect.~\ref{subsubsec: mass_size}).

\end{minipage}

\end{appendix}

\end{document}